\documentclass[%
 aps,
twocolumn,preprintnumbers,floatfix,
prev,
nofootinbib,
 amsmath,amssymb,
]{revtex4-1}

\newcommand{\kahlerij}{\mathcal{K}_{ij}}
\newcommand{\mij}{\mathcal{M}_{ij}}
\usepackage{graphicx}
\usepackage{dcolumn}
\usepackage{bm}
\usepackage{xcolor}
\usepackage{graphicx}
\usepackage{tabularx}
\usepackage{yfonts}
\usepackage{float}
\usepackage{mathtools}
\usepackage{subcaption}
\usepackage{enumerate}
\usepackage{hyperref}
\usepackage{bbold}
\usepackage{cleveref}
\usepackage{mathrsfs}
\newcolumntype{K}[1]{>{\centering\arraybackslash}p{#1}}
\usepackage{cancel}
\usepackage{nth}
\usepackage{caption}
\usepackage{xfrac}
\newcommand{\beq}{\begin{equation}}
\newcommand{\eeq}{\end{equation}}
\newcommand{\beqa}{\begin{eqnarray}}
\newcommand{\eeqa}{\end{eqnarray}}

\usepackage[T1]{fontenc}

\begin{document}
\newcommand{\tkDM}[1]{\textcolor{red}{#1}}  

\preprint{APS/123-QED}
\begin{flushleft}
KCL-PH-TH/\today
\end{flushleft}


\title{The Spectrum of the Axion Dark Sector}


\author{Matthew J. Stott$^a$}
\email{matthew.stott@kcl.ac.uk}
\author{David J. E. Marsh$^a$}
\email{david.marsh@kcl.ac.uk}
\author{Chakrit Pongkitivanichkul$^{a,b,c}$}
\email{chakpo@kku.ac.th}
\author{Layne C. Price$^{d}$}
\email{laynep@andrew.cmu.edu}
\author{Bobby S. Acharya$^{a,e}$}
\email{bobby.acharya@kcl.ac.uk}

\affiliation{
$^a$ Theoretical Particle Physics and Cosmology Group, Department of Physics, King's College London, University of London, Strand, London, WC2R 2LS, United Kingdom \\
$^b$ The Center for Future High Energy Physics, Institute of High Energy Physics, Beijing, China \\
$^c$ Department of Physics, Khon Kaen University, Khon Kaen, Thailand \\
$^d$ McWilliams Center for Cosmology, Department of Physics, Carnegie Mellon University, Pittsburgh, PA 15213, USA\\
$^e$ The Abdus Salam International Centre for Theoretical Physics, Strada Costiera 11, Trieste, Italy}

\date{\today}


\begin{abstract}
Axions arise in many theoretical extensions of the Standard Model of particle physics, in particular the ``string axiverse''. If the axion masses, $m_a$, and (effective) decay constants, $f_a$, lie in specific ranges, then axions contribute to the cosmological dark matter and dark energy densities. We compute the background cosmological (quasi-)observables for models with a large number of axion fields, $n_{\rm ax}\sim \mathcal{O}(10-100)$, with the masses and decay constants drawn from statistical distributions. This reduces the number of parameters from $2n_{\rm ax}$ to a small number of ``hyperparameters''. We consider a number of distributions, from those motivated purely by statistical considerations, to those where the structure is specified according to a class of M-theory models. Using Bayesian methods we are able to constrain the hyperparameters of the distributions. In some cases the hyperparameters can be related to string theory, e.g. constraining the number ratio of axions to moduli, or the typical decay constant scale needed to provide the correct relic densities. Our methodology incorporates the use of both random matrix theory and Bayesian networks.

\end{abstract}


\maketitle


\section{Introduction}
\label{sec:intro}

The Standard Model of particle physics is an overwhelming triumph of \nth{20} century physics. Combined with the general theory of relativity (and a model for neutrino masses), it is able to describe all terrestrial phenomena over a vast range of energy scales, and it has been verified with exquisite precision in the \nth{21} century by the work conducted at the Large Hadron Collider~\cite{pdg}. The Standard Model fails spectacularly, however, when applied on cosmological scales. Observations of the cosmic microwave background (CMB) temperature and polarisation anisotropies, for example, imply that the present-day energy density of the Universe is dominated by Dark Matter (DM) and Dark Energy (DE)~\cite{Ade:2015xua}. The particle content of the Standard Model contains no candidate for DM~\cite{2005PhR...405..279B}, and the value of the DE density, if assumed to be solely due to the cosmological constant, $\Lambda$, cannot be explained~\cite{1989RvMP...61....1W}.\footnote{Cosmology, of course, also presents another two huge problems for the Standard Model: the baryon asymmetry, and the generation of initial conditions (inflation). We will not discuss these problems further.} 

\begin{figure*}[t]
\hspace*{-2cm}
\centering
    \includegraphics[width=1.2\linewidth]{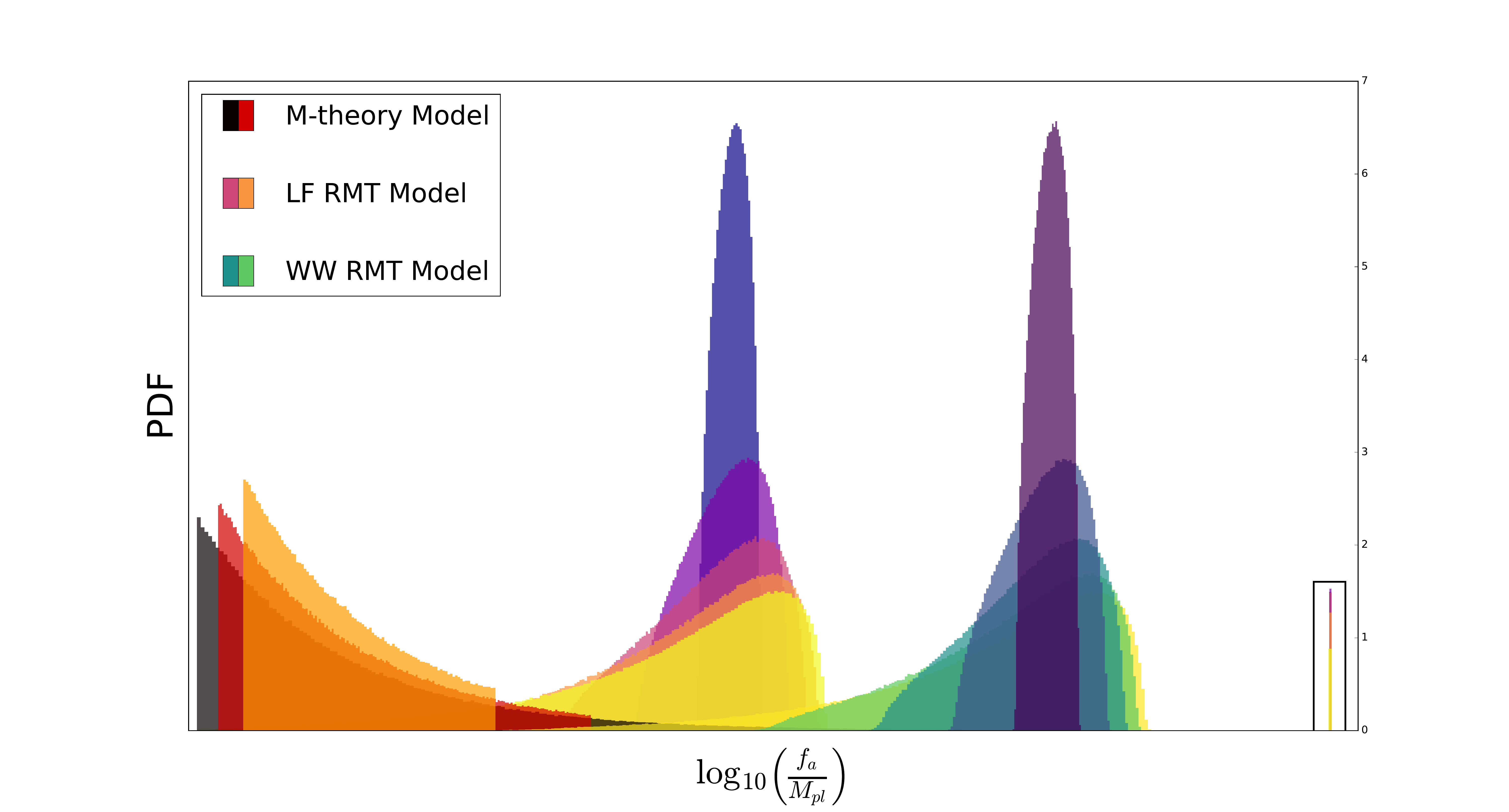}
\caption{{\bf String axiverse RMT model axion decay constant spectra:} Probability density plots displaying the spectra for the axion decay constants, $f_a$, defined as the eigenvalues of the kinetic matrix in Eq.~(\ref{eq:fbasis}) constructed using 7500 iterations with $n_{\rm ax}=100$. The shape of the spectrum determines the initial axion field range as well as effecting the axion mass distribution after rotating to the canonical basis. The highlighted (black rectangle) values demonstrate the enhancement of the eigenvalue spectra width when using non-zero mean, non-gaussian distributions (LF RMT model, Section~\ref{sec:logflat}) for the kinetic matrix. For visual clarity we include an arbitrary normalisation offset on the distribution mean. In practice the normalisation is given by the Planck scale, $M_{pl}$, and the mean is determined by a free model parameter of the order of the fundamental scale.}
\label{decayf}
\end{figure*}

These problems at the heart of particle and cosmological physics today force us to explore a wide range of theories beyond the Standard Model (BSM). Many such theories invoke ideas combining a combination of extra dimensions of spacetime and supersymmetry (SUSY), with the leading such theory being string/M-theory (e.g. Ref.~\cite{2007stmt.book.....B}). The extra dimensions are compact in these models which in turn leads, in the low energy, (3+1)-dimensional description, to the existence of massless pseudoscalar axion-like fields (which, for simplicity, we now refer to as simply ``\emph{axions}'')~\cite{1984PhLB..149..351W,Svrcek:2006yi}.\footnote{There is also the presence of scalar moduli to account for. We discuss moduli stabilisation in due course.}

The number of axions depends on the topology of the compact dimensions. In realistic compactifications of string theory, this can easily be in the range of $\mathcal{O}(10)$ to $\mathcal{O}(100)$, or more (e.g. Ref.~\cite{Kreuzer:2000xy}). The axions generically acquire masses, $m_{a}$, due to non-perturbative quantum effects (e.g. instantons~\cite{1988NuPhB.306..890G,1988assy.book.....C}), and as such the masses depend exponentially on parameters of the UV theory, such as the size of extra dimensions. In the context of string theory there are many effects which can be used to generate potentials for the axion fields such as worldsheet or brane instantons. On the other hand the axion ``decay constants'', $f_{a}$, are expected to be of order of the UV scale~\cite{2003JCAP...06..001B}. Large decay constants lead to suppressed couplings between axions and the Standard Model. This leads to the theoretical expectation that there should exist some large number of light, stable, axions given the potential complexity of the extra-dimensional manifold: an idea known as the ``\emph{string axiverse}''~\cite{Arvanitaki:2009fg}. 

\emph{Light, stable axions are excellent DM candidates, and can also contribute to the DE density, with a rich phenomenology} (for a review of axion cosmology, see Ref.~\cite{Marsh:2015xka}). However, a large number of axion fields brings with it, $2n_{\rm ax}^2$ parameters coming from the kinetic and mass matrices present in an effective field description, making a brute force treatment of the cosmology difficult. Natural questions which arise are: what is the typical DM and DE cosmology which emerges from a string axiverse model? Under what conditions do string axiverse models give rise to realistic cosmologies? In order to address these problems, we present an initial study in the context of of string axiverse cosmology for simplified axiverse models relating to both the problems of DM and DE, utilising the frameworks of Bayesian Networks and random matrix theory (RMT). In this study we present five different models, characterised by their corresponding distributions for the elements of both the kinetic matrix (which is related to the K\"{a}hler metric) and mass matrix of a multi-axion field theory. The distributions for $f_{a}$ and $m_{a}$, after rotation of the matrices to the canonical diagonal basis, determine the cosmology of string axiverse theories, and we present constraints on the hyperparameters of these distributions from the DM and DE densities.

One of our models, inspired by the Jeffreys prior, incorperating scale invariance of the physical quantities, is a statistical straw man: log-flat eigenvalue distributions, ``maximally ignorant'' of any underlying fundamental theory. Another straw man assumes a trivial kinetic matrix, with the mass matrix eigenvalue distribution derived from the Mar\v{c}henko-Pastur law for random matrices (loosely related to axion models~\cite{Dimopoulos:2005ac,Easther:2005zr}). The other three models assume non-trivial distributions in the kinetic matrix, giving rise to non-trivial distributions for the axion decay constants, $f_a$, in the diagonal basis. Our most physically motivated model for the matrix distributions is derived from considering the string axiverse arising in M-theory compactified on $G_2$-manifolds~\cite{Acharya:2010zx}. The distributions of the decay constants for these models are shown (in arbitrary units) in Fig.~\ref{decayf} (we define the decay constants \emph{before} accounting for ``alignment'' [see Section~\ref{sec:initial}]). The form of the resulting mass distributions after rotation of the matrices differ from the straw-man models, and are discussed throughout this paper. Table~\ref{tab:models} describes each of the models we consider in this study, and their associated location in the paper.    

We make no discussion in this work of the possible couplings between axions and the Standard Model, or any production modes for axions other than vacuum realignment. This is the simplest possible model-independent approach to the axiverse in a cosmological context. See Refs.~\cite{Marsh:2015xka,2015ARNPS..65..485G} for discussion of other axion production modes and detection of axions through non-gravitational interactions.

\begin{table}
\begin{center} 	
\caption{String axiverse models used throughout this study with their corresponding short hand notation. Also detailed are their sections of appearance in the text giving the properties of their construction as well as their DM/DE cosmology considerations.}
\label{tab:models}
\begin{tabularx}{0.46\textwidth}{K{0.21\textwidth}K{0.08\textwidth}K{0.15\textwidth}} 
\toprule
Model    & Label & Section \\
\hline
\textbf{I. Scale Invariant} & \textbf{SI} & \textbf{Sec.~\ref{sec:haarmeasure}}\\ 
 \ \ i. {\small \textit{Dark Matter}} & SI-DM& \textquotedbl     \\ 
  \ \ ii. {\small \textit{Dark Energy}} & SI-DE& \textquotedbl     \\ \hline
\textbf{II. Mar\v{c}enko-Pastur} & \textbf{MP} & \textbf{Sec.~\ref{sec:mpmodel}/\ref{sec:mplaw}}\\ 
 \ \ i. {\small \textit{Dark Matter}} & MP-DM& Sec.~\ref{sec:mpexdm}/ \ref{sec:mpmcmc}     \\
  \ \ ii. {\small \textit{Dark Energy}} & MP-DE& Sec.~\ref{sec:mpexde}/ \ref{sec:mpmcmc}     \\ \hline
\textbf{III. Wishart/Wishart} & \textbf{WW} & \textbf{Sec.~\ref{sec:WWmat}/\ref{appendix:rmt}}\\ 
 \ \ i. {\small \textit{Dark Matter}} & WW-DM& Sec.~\ref{sec:wwexdm}     \\
  \ \ ii. {\small \textit{Dark Energy}} & WW-DE& Sec.~\ref{sec:wwexde}      \\ \hline
\textbf{IV. Log-Flat/Log-Flat} & \textbf{LF} & \textbf{Sec.~\ref{sec:logflat}/\ref{sec:eigspect}}\\ 
 \ \ i. {\small \textit{Dark Matter}} & LF-DM& Sec.~\ref{sec:lfdmex}      \\
  \ \ ii. {\small \textit{Dark Energy}} & LF-DE& Sec.~\ref{sec:lfdeex}      \\ \hline
  \textbf{V. M-Theory} & \textbf{MT} & \textbf{Sec.~\ref{sec:mthethe}/\ref{sec:mthethe2}}\\ 
 \ \ i. {\small \textit{Dark Matter}} & MT-DM& Sec.~\ref{sec:mtdm}/\ref{sec:mtmcmcdm}     \\
  \ \ ii. {\small \textit{Dark Energy}} & MT-DE& Sec.~\ref{sec:mtde}      \\
\botrule
\end{tabularx}
\end{center}

\end{table} 
The paper is organised with the following structure. Section~\ref{sec:theory} presents an initial look at axions in string theory as well as detailing our effective model for string axiverse cosmology, introducing the key concepts of the kinetic matrix, $\kahlerij$, and mass matrix, $\mij$ along with the initial field conditions. Section~\ref{sec:axiverse_rmt} presents a set of random matrix theory models for $\kahlerij$ and $\mij$. We also present in this section a random matrix approach to $G_2$ compactifications of M-theory. Our results begin in Section~\ref{sec:results1}, where we present example cosmologies for all of our models with either fixed values of the underlying parameters or gridded scans of multidimensional parameter space. Section~\ref{sec:results2} presents constraints on the random matrix parameters from a Markov Chain Monte Carlo (MCMC) analysis of the quasi-observables from the CMB using Bayesian networks; we cover only a subset of the possible models, with a complete treatment left for future work. We conclude with discussions of our study in Section~\ref{sec:conclusions}. 

Appendix~\ref{appendix:numsim} presents details of our scheme for the numerical solutions to the equations of motion, and more details about the assumed cosmology. Appendix~\ref{appendix:stringmot} reviews the form of the superpotential arising in both M-theory and Type-IIB string theory along with details of the possible connection between random matrix theory and Type-IIB string theory on Calabi-Yau manifolds. Appendix~\ref{appendix:rmt} introduces the principle concepts of random matrix theory we incorporate in our RMT models as well as the basics of the Mar\v{c}henko-Pastur density function for sample covariance matrices and potential extensions/deviations from this law for different matrix ensembles. Finally, Appendix~\ref{sec:rejec} contains some novel examples of outlier cosmologies. 

Our numerical code, \textsc{AxionNet}, is written in \textsc{python} and is available to download from \url{https://github.com/DoddyPhysics/AxionNet}.

     

\section{Axions}

\begin{table*}[t]
\begin{center}
\caption{The full range of parameters used in this study including the cosmological input parameters along with the model dependant RMT parameters and theoretical M-theory parameters. Our cosmological density and parameter data comes from the \textit {Planck 2015 TT+lowP} likelihood's in Ref.~\cite{Ade:2015xua} with our CMB temperature defined using COBE data in Ref.~\cite{1996ApJ...473..576F}.}
\label{tab:cosmopar}
\begin{tabular*}{\textwidth}{K{0.10\textwidth}K{0.55\textwidth}K{0.20\textwidth}K{0.12\textwidth}}
\toprule
    \text{$Parameter$}        & \text{$Definition$}       & \text{$Prior/Value$} & \text{$Eq./Ref.$}  \\ 
    \hline
    \rule{0pt}{2.6ex}
  &  {\bf Cosmological}   & 
      \rule[-1.2ex]{0pt}{0pt}
\\ 
   \hline 
     $n_{\rm ax}$ & Number of axion fields & $\mathcal{O}(1-100)$ & -
\\
     $f_a$ & Axion decay constant & $\mathcal{O}(10^{-4}M_{pl}-M_{pl})$ & Eq.~(\ref{eq:fbasis})
\\
     $m_a$ & Axion mass & $[10^{-35}{\rm eV},10^{-15}{\rm eV}]$ & Eq.~(\ref{eq:massm})
\\
     $\theta_i$ & Initial field misalignment& $\mathcal{U}[0,\pi]$ & Eq.~(\ref{eq:phiini})
\\
 $\phi_i$ & Initial axion field conditions& $\mathcal{F}_{ij}\theta_j$ & Eq.~(\ref{eq:phif})
\\
$\dot{\phi}_i$  & Initial field derivative & $0$ & Eq.~(\ref{eq:hubbleeq})
\\
$\mathcal{F}_{ij}$ & Decay constant matrix & Model dependent & Eq.~(\ref{eq:phif})
\\
$a$  & Cosmic scale factor & $(10^{-8} \rightarrow 1)$ & -
\\ 
$H_0$ & Present day Hubble rate & $h M_H$ & -
\\
$M_H$ & Hubble mass scale, 100 km s${-1}$ Mpc$^{-1}$ & $2.13\times 10^{-33}\text{ eV}$ & -
\\
$M_{pl}$ & Reduced Planck mass, $1/\sqrt{8\pi G}$ & $2.435\times 10^{27}\text{ eV}$ & -
\\
$\Omega_{\rm DM}$ & Axion dark matter density parameter& $(0,1)$ & -
\\
$\Omega_{\rm DE}$ & Axion dark energy density parameter& $(0,1)$ & -
\\

    \hline
    \rule{0pt}{2.6ex}
  &  {\bf \textit{Planck 2015 TT+lowP} Parameters}   & 
      \rule[-1.2ex]{0pt}{0pt}
      \\
      \hline

& \textit{Used as quasi-observable data} & \\
\hline
$h$  & Present day Hubble rate & $0.6731 \pm 0.0096$ & \cite{Ade:2015xua}       \\
 $\Omega_{m}$  & Total matter fraction &$0.315 \pm 0.013$ & \textquotedbl      \\
  $z_{eq}$ & Redshift of matter-radiation equality &$3393 \pm 49$  & \textquotedbl       \\
\hline
& \textit{Fixed in a given model} & \\
\hline
$\Omega_{b}h^2$  & Physical baryon density (all) &$0.022$ & \cite{Ade:2015xua}       \\
$\Omega_{c}h^2$  & Physical dark matter density (DE models) &$0.12$ & \textquotedbl     \\
 $\Omega_{\Lambda} h^2$  & Physical dark energy density (DM models) &  $0.31$ & \textquotedbl    \\
 $T_{\rm CMB}$ & CMB temperature (COBE, all) & 2.725 K & \cite{1996ApJ...473..576F} \\

    \hline
    \rule{0pt}{2.6ex}
  &  {\bf Random Matrix Theory Models}   & 
      \rule[-1.2ex]{0pt}{0pt}
      \\
      \hline
      $\sigma^2_{\mathcal{K}}$  &Kinetic matrix distribution scale& $[10^{-3}M_{pl},1M_{pl}]$ & Eq.~(\ref{eq:sigmak})
\\
      $\sigma^2_{\mathcal{M}}$  &Mass matrix distribution scale& $[10^{-4}M_H,10^{36}M_H]$ & Eq.~(\ref{eq:sigmam})
\\
$\beta_{\mathcal{K}, \mathcal{M}}$  & Sub-matrix dimension parameter& $(0.0,1.0]$ & Eq.~(\ref{eq:betak})/(\ref{eq:betam}) 
\\
$\bar{f}$  & MP RMT model equal field condition scale& $[10^{-9}M_{pl},1M_{pl}]$ & Eq.~(\ref{eq:mpvev})
\\    
  $k_{\rm min}$  & LF RMT model kinetic matrix element distribution lower bound& $-5.0$ & Eq.~(\ref{eq:lfel1})
  \\
   $k_{\rm min}$  & LF RMT model kinetic matrix element distribution upper bound& $[-3.0,0.0]$ & \textquotedbl 
\\
$m_{\rm min}$  & LF RMT model mass matrix element distribution lower bound& $-5.0$ (DE), $4.0$ (DM) & Eq.~(\ref{eq:lfel2})
\\
$m_{\rm max}$  & LF RMT model mass matrix element distribution upper bound& $[-1.0,8.5]$ & \textquotedbl 
\\

\hline
    \rule{0pt}{2.6ex}
  &  {\bf M-theory Model}   & 
      \rule[-1.2ex]{0pt}{0pt}
\\
\hline
  $F/(M_{H}^2)$ & SUSY order parameter, $m_{3/2}M_{pl}$ & $5.4 \times 10^{104} (m_{3/2}/1\text{ TeV})$ & Eq.~(\ref{eq:fdim})
\\
$m_{3/2}$ & Gravitino mass & $10$ TeV & -
\\
$\Lambda$  & Instanton Mass scale, string units & [$10^{-5}$,1] & Eq.~(\ref{eq:lamdim})
\\
$s$  & Averaged value for Moduli vevs, string units & $\mathcal{U}[10,100]$/$\mathcal{N}(\bar{s},\sigma_{s})$ & Eq.~(\ref{eq:modvev})/(\ref{eq:modvev2})
\\
$\widetilde{N}_{\rm max}$  & Instanton Index Parameter &  [0.6,1.6] & Eq.~(\ref{eq:nmax})
\\ 
$a_0$ & Axion decay constant scale & 1 & Eq.~(\ref{eq:mthea})
\\  
\botrule  
\end{tabular*}
\end{center}
\label{tab:params}
\end{table*}

\label{sec:theory}
\subsection{String Axions: A Single Field Example}

Axions respect a perturbative shift symmetry, $\theta\rightarrow \theta+{\rm const.}$, of Goldstone bosons. For geometric axions, this symmetry comes from the higher dimensional gauge symmetries of supergravity. Non-perturbative effects generically break this shift symmetry down to a discrete subgroup. Axions are characterised using two parameters: the axion decay constant, $f_{a}$,
and the energy scale of the associated non-perturbative physics, $\Lambda_{a}$. The effective four dimensional Lagrangian for the dimensionless axion field with a spacetime metric signature, $(-,+,+,+)$, is
\beq
\mathscr{L} = -\frac{f_{a}^2}{2}\partial_{\mu}\theta \partial^{\mu}\theta -  \Lambda_{a}^4 U(\theta)\,,
\eeq 
where $U(\theta)$ is some periodic potential of the dimensionless fields, $\theta$. In the dilute instanton gas approximation, the field potential is given by,
\beq
V(\theta) =\Lambda_{a}^4 U(\theta)=\Lambda_{a}^4 \left(1-\cos\theta \right) \label{eq:b1}\,.
\eeq
The non-perturbative physics present an exponential dependance on the instanton action S,
\beq
\Lambda^4_a = \mu^4e^{-S}\,.
\label{eq:exposen}
\eeq
The parameter $\mu$ is a mass scale determined by the geometric mean of the SUSY breaking scale and the ``fundamental'' scale such as the String or Planck scale. The canonically normalised axion field is, 
\beq
\label{eq:phiini}
\phi = f_{a} \theta\,,
\eeq 
from which we see that the axion decay constant, $f_{a}$, sets the scale of periodicity in the potential. For small field displacements $\theta< 1$, performing a local Taylor expansion about the vacuum $\theta=0$ up to quadratic order yields the axion mass term,
\beq
m_{a} = \frac{\Lambda_{a}^2}{f_{a}}\,.
\eeq
For small field displacements, $f_{a}$ disappears as an explicit parameter in the Lagrangian. However, because of its role in the periodicity of the potential it still appears as the natural range of field values for $\phi$. In the ensuing discussion, we use $f_{a}$ as the scale of the initial conditions.


\subsection{The String Axiverse: An Effective Theory} \label{sec:effmodel}

For multiple fields arising in typical string axiverse models we must consider cross couplings in the field kinetic terms present in the non-trivial axion field space metric $\mathcal{K}_{ij}$. In SUSY theories, this is related to the K\"{a}hler metric, which, for axions paired with K\"{a}hler moduli is given by $\frac{\partial^2 K}{\partial \tau_i \partial \tau_j}$, where $K$ is the K\"{a}hler potential and $\tau_i$ represent the moduli fields (see Ref.~\cite{2007stmt.book.....B} for a more general description). In supergravity the basis for the axion fields is such that the kinetic matrix is both non-diagonal and not canonically normalised, where the general Lagrangian takes the form:
\begin{equation}
\mathscr{L} = -M_{pl}^2\mathcal{K}_{ij}\partial_{\mu}\theta_i\partial^{\mu}\theta_j -M_{pl}^2 \mathcal{M}_{ij}\theta_i\theta_j	\, .\label{eq:effL}
\end{equation}
The mass matrix is determined as usual from the K\"{a}hler potential and the superpotential, $W$. For simplicity we expand the potential to the mass term, and will not use the full general form of the cosine potential, which expresses the entries of $\mij$ in terms of the instanton charge matrix, $\mathcal{Q}$. We discuss this briefly later, and a full treatment will be the subject of future work. 

We diagonalise the Lagrangian by beginning with the diagonalisation of $\kahlerij$:
\beq
\mathcal{K} = U^T{\rm diag}(\mathcal{K}) U = \frac{1}{2}U^T{\rm diag}(f_a){\rm diag}(f_a)U \, ,
\eeq
where we \emph{define the axion decay constants, $f_a$, from the eigenvalues of $\kahlerij$ in the original (non-diagonal) basis.} We discuss how this choice relates to the axion initial conditions in the next subsection. The decay constants thus defined are (in Planck units):
\beq
\vec{f_a} = \sqrt{2 {\rm eig}(\mathcal{K})}\, .
\label{eq:fbasis}
\eeq
We next define the canonically normalised field:
\beq
\tilde{\phi}=M_{pl}{\rm diag}(f_a) U \theta \, .
\eeq
Inserting this definition we find the Lagrangian for the canonical fields:
\beq
\label{eq:nflatbasis}
\mathscr{L} = -\frac{1}{2}\partial_\mu\tilde{\phi}_i\partial^\mu\tilde{\phi}_j - \frac{1}{2}\tilde{\phi}_i\tilde{\mathcal{M}}_{ij}\tilde{\phi}_j \, .
\eeq
The new mass matrix is given by:
\beq
\tilde{\mathcal{M}}=2{\rm diag}(1/f_a)U\mathcal{M}U^T{\rm diag}(1/f_a) \, .
\eeq
The new mass matrix is diagonalised by,
\beq
\label{eq:massm}
\tilde{\mathcal{M}}=V^T{\rm diag}(m^2_a)V \, .
\eeq
Defining the mass eigenstate fields,
\beq
\phi = V\tilde{\phi} = M_{pl}V{\rm diag}(f_a)U\theta \, .
\eeq
The fully diagonalised Lagrangian is:
\begin{equation}
\mathscr{L} = -\frac{1}{2}\partial_{\mu}\phi_i\partial^{\mu}\phi_i - \frac{1}{2}{\rm diag}(m^2_a)\phi_i \phi_i\,.
\label{eq:finall}
\end{equation}
Eq.~(\ref{eq:finall}) is the canonical mass eigenstate basis with the mass spectrum dependance coming from the initial forms of $\kahlerij$, $\mathcal{M}_{ij}$, and the various rotations in field space. As is the case in the single axion example, the axion decay constants coming from diagonalisation of $\kahlerij$ now only play a role in setting the natural initial displacements of the axion fields.


\subsection{Axion Cosmology} \label{sec:eom}
We work in a homogeneous, and isotropic Universe with a flat Friedmann-Lema\^{i}tre-Robertson-Walker (FLRW) geometry:
\begin{equation}
 {\rm d}s^2 = -{\rm d}t^2 + a^2(t){\rm d}\vec{x}^2\,,
\end{equation}
where $a(t)$ is the cosmological scale factor, normalised to unity today, defining the cosmological redshift $a(z) = 1/(1+z)$. The equations of motion for the axion fields follow from the canonical action for matter,
\beq
S_m=\int {\rm d}^4 x\sqrt{-g}\mathscr{L} \, ,
\eeq
with $g$ the FLRW metric determinant. Axions obey the Klein-Gordon equation of motion: 
\begin{equation}
\label{eq:hubbleeq}
\ddot{\phi}_{i} + 3H\dot{\phi}_{i} + m_{a,i}^2\phi_i = 0\,,  
\end{equation}
where the dot denotes the derivative with respect to the cosmic time. The Friedmann constraint for the Hubble parameter, $H = \sfrac{\dot{a}}{a}$, is:
\begin{equation}
3H^2M_{pl}^2 =\sum_i \rho_i \, ,
\label{Eq:Freid}
\end{equation}
where the sum over $i$ extends over all axions, ordinary matter, dark matter, radiation, and the cosmological constant. See Appendix~\ref{appendix:numsim} for more details. 

We solve the axion field equations in cosmic time, and use the Friedmann constraint to find $a(t)$, which determines the evolution of the standard fluid components via their equation of state. The combined equation of state for the axions is given by: 
\begin{equation}
w_a = \frac{P_a}{\rho_a} = \frac{\frac{1}{2}\sum_{i}^N \dot{\phi}^2_i - V}{\frac{1}{2}\sum_{i}^N \dot{\phi}^2_i + V}\,.
\end{equation}
The total equation of state today determines the acceleration parameter, $\ddot{a}$.
\begin{figure}
\includegraphics[width=0.5\textwidth]{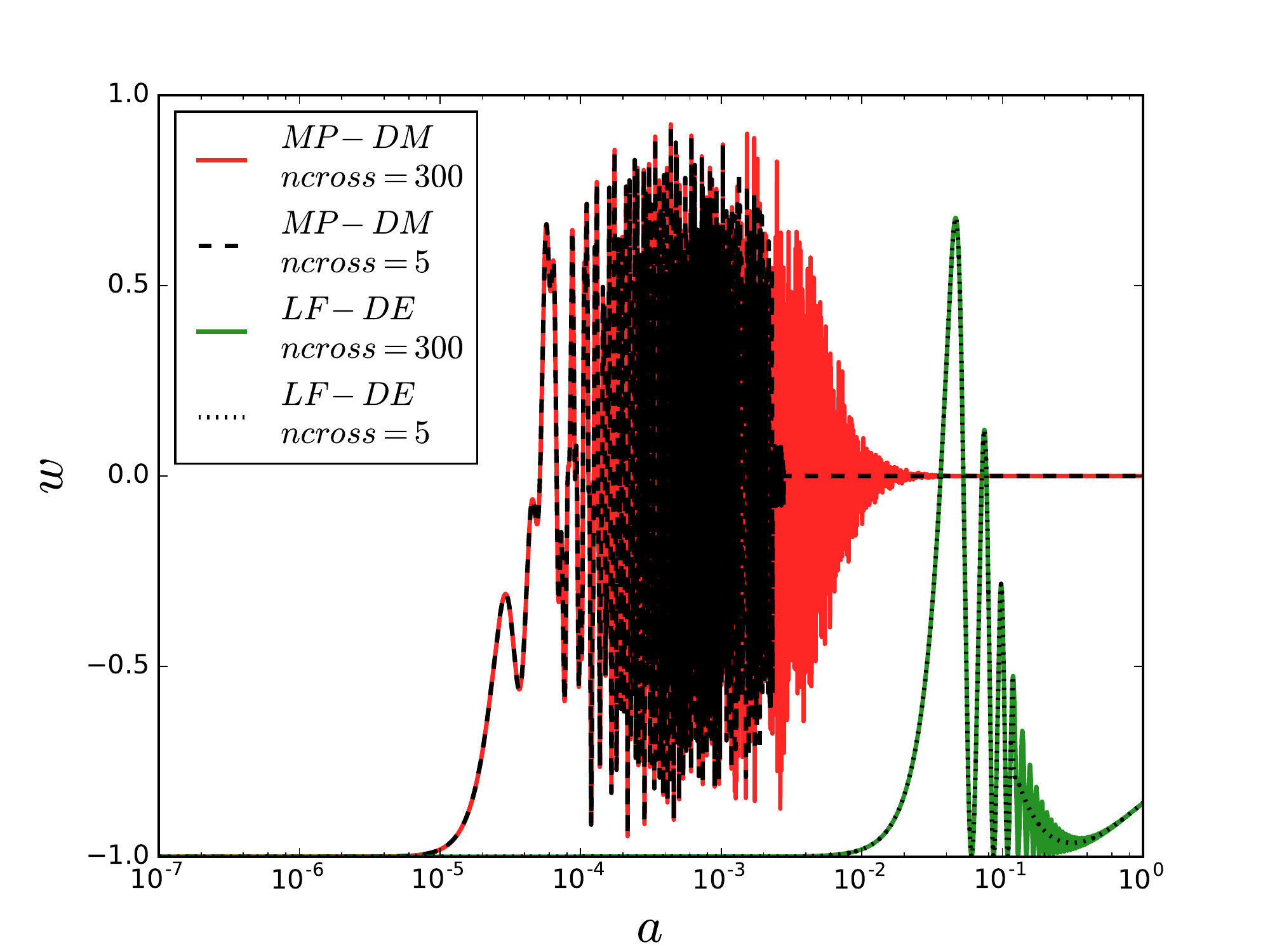}
\caption{{\bf Evolution of the collective axion equation of state:} The collective axion equation of state, $w_a$ as a function of the cosmic scale factor, $a$ for axions behaving as either the total dark matter or total dark energy in different RMT models. $n_{\rm cross}$ referrers to the numerical precision, see Appendix~\ref{appendix:numsim}.}
\label{fig:eosex2}	
\end{figure}
\begin{figure*}[t]
\centering
\begin{tabular}{cc}
    \includegraphics[width=0.49\linewidth]{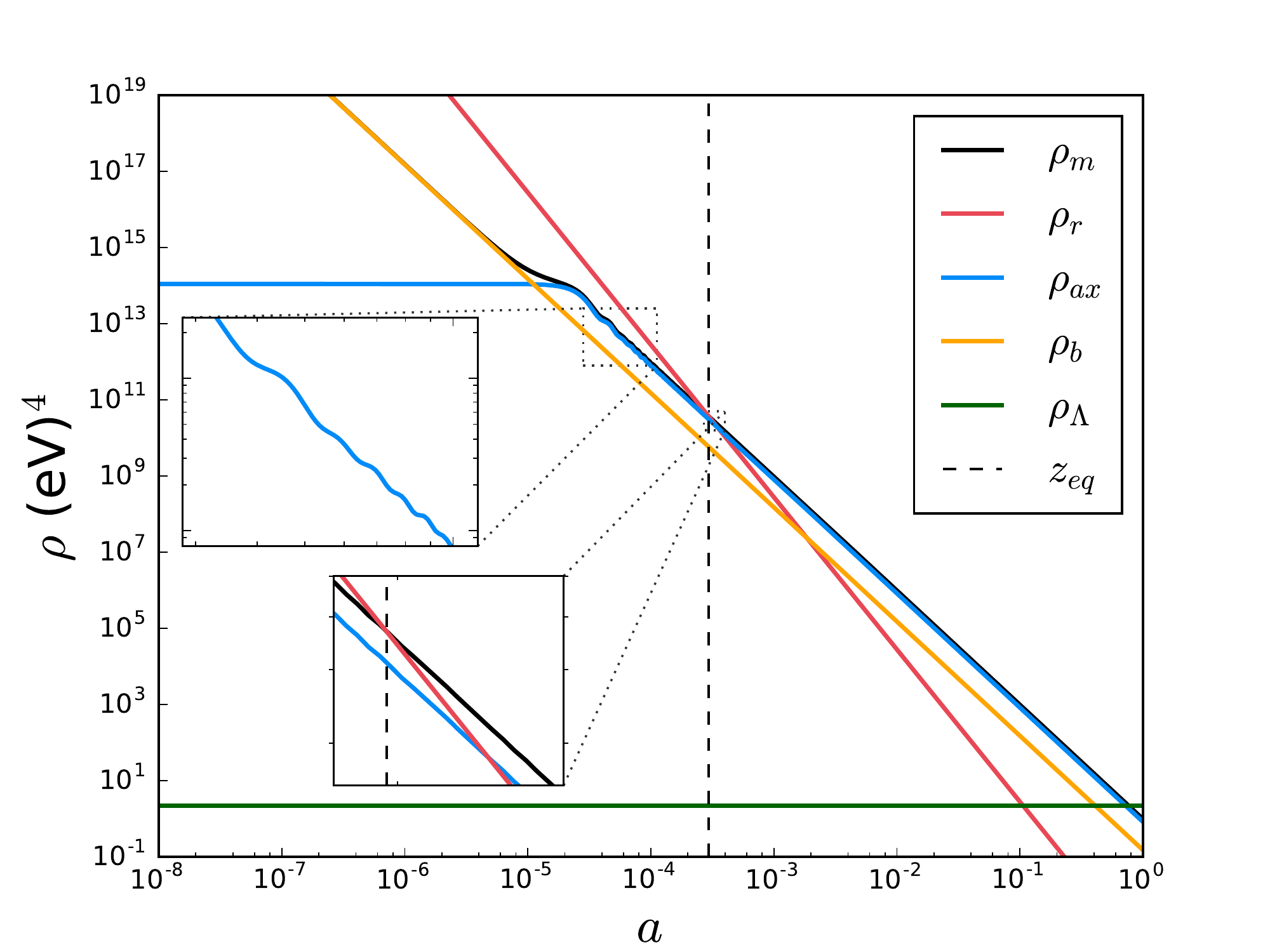}&
    \includegraphics[width=0.49\linewidth]{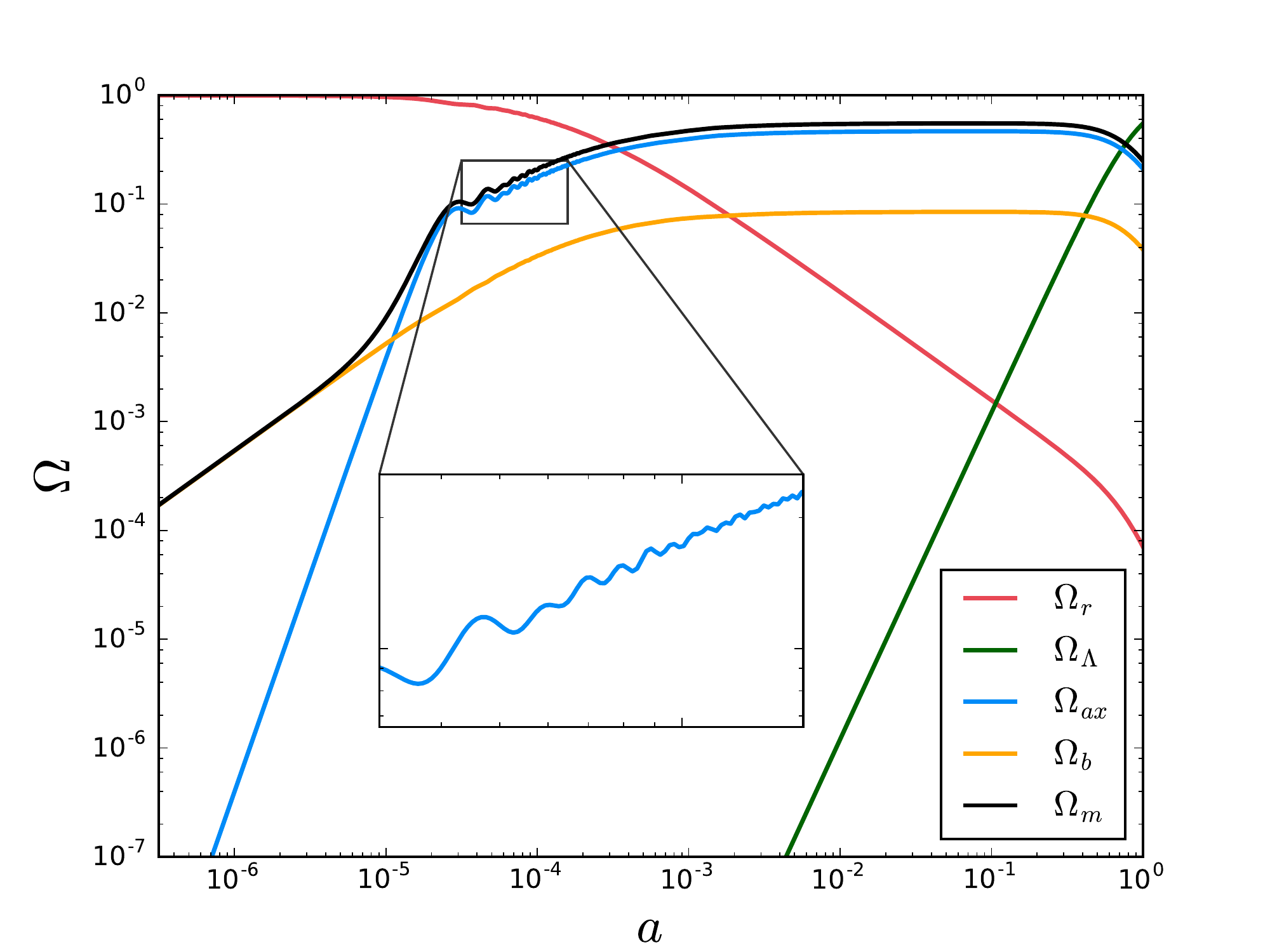}\\
\end{tabular}
\caption{{\bf Evolution of the cosmological densities and cosmological density parameters:} \emph{Left panel}: Plot for the evolution of cosmological densities, $\rho$ as a function of the cosmic scale factor, $a$ for $n_{\rm ax} = 10$ axions behaving as the total dark matter in the MP RMT model. \emph{Right panel}: Plot for the evolution of the contributions to the critical density, $\Omega_i = \sfrac{\rho_i}{3H^2}$ as a function of the cosmic scale factor, $a(t)$. Each panel details the evolution of the MP RMT axions plus the standard $\Lambda$CDM parameters $\Omega_{r}$, $\Omega_{\Lambda}$ and $\Omega_{b}$. \emph{Left figure upper inset}: Enhanced view of the effect of multi-field oscillations on the total axion density, $\rho_{ax}$. \emph{Left figure lower inset}: Comparative matter-radiation equality with crossings of $\rho_m = \rho_b+\rho_{ax}$ and $\rho_r$ at $z_{eq} = 3393$ defined in Tab.~\ref{tab:cosmopar}. \emph{Right figure inset}: Enhanced view of the effect of multi-field oscillations on the axion density parameter $\Omega_{\rm DM}$ contributing to the critical density.}
\label{fig:cosmoex}
\end{figure*}

Fig.~\ref{fig:eosex2} shows the collective equation of state for example multi-field evolutions involving $n_{\rm ax}=10$ axions for both dark matter and dark energy cosmologies in different RMT models. The dashed and dotted lines detail our approximations where we show the effect on the collective equation of state for the axion population when we restrict the individual equations of state for each field to a fixed number of oscillatory crossings used as an accuracy parameter we denote as $n_{\rm cross}$. The amplitude of the total equation of state is damped from the effects of multiple fields with non-degenerate associated scales in the population, oscillating between the values of $\leq 1$ and $\geq -1$. In the limit $n_{\rm ax} = 1$ the equation of state will continue to oscillate between -1 and 1. We find that $n_{\rm cross} = 5$ captures a significant proportion of the total field behaviour as compared to increased values of $n_{\rm cross}$. See Appendix~\ref{sec:axosc} and \ref{sec:comquasi} for details of our process used and choice of approximation. 

The axion fields are initially over damped setting the fields in slow roll, $\dot{\phi_{i}} \approx 0$, with an almost constant equation of state, $w_a \approx -1$. This type of field evolution demonstrates the ability of axions to behave as candidates in quintessence or inflationary models. As the Hubble rate, $H$ decreases the fields overcome the Hubble friction present as a damping term in their equations of motion, at a time $t_{\rm osc}^i \approx H^{-1}$ satisfying the condition $m_{a,i} \approx H$. The $i$th axion field now begins to coherently oscillate about the minimum of its potential with an amplitude determined by its initial misalignment angle. In this phase the axions will begin to dilute slower and scale as pressure-less matter where the equation of state begins to oscillate about $w_a = 0$ and a phase of underdamping begins. The axion pressure now averages to zero and the energy density begins to scale as $\rho_a \propto a^{-3}$, leaving the axion as a suitable dark matter candidate. The left hand panel of Fig.~\ref{fig:cosmoex} details an example evolution of the components of the energy density through numerical integration of the equations of motion for $n_{\rm ax}=10$ fields in the Universe as well as the remaining standard $\Lambda$CDM parameters. The evolution of the associated density parameters is plotted in the right hand panel.   

At any given time, fields with $H\gtrsim m_{a,i}$ will behave as a contribution to the total effective dark energy density, $\Omega_{\rm DE}$ and fields with $H\lesssim m_{a,i}$ behave as contribution to the total dark matter density, $\Omega_{\rm DM}$.  We classify axions as either DM or DE components of the energy density of the Universe according to the description in Appendix~\ref{appendix:numsim}. We use this to determine $\Omega_{m}=\Omega_b+\Omega_{\rm DM}$ and $\Omega_{\rm DE,tot}=\Omega_\Lambda+\Omega_{\rm DE}$. The evolution of $\rho_{m}$ with redshift determines the redshift of matter radiation equality, $z_{\rm eq}$.

\subsection{Initial Conditions}
\label{sec:initial}
The role of the axion decay constants, for our purposes, is to fix the natural initial field displacements, and thus the axion relic density from vacuum realignment~\cite{1983PhLB..120..127P,1983PhLB..120..133A,1983PhLB..120..137D}. In the (generic) case of multiple axions where the number of instantons providing the axion masses is larger than the number of axions, the notion of a single ``axion decay constant'' is not well defined.\footnote{We thank Thomas Bachlechner for discussion on this point.} 

Expanding the potential to the mass term alone, the dimensionful scales that control the evolution and relic densities are the initial displacements of the canonical fields. In all cases we set our initial conditions on the axion fields as
\beq
\label{eq:phif}
\phi^{\rm ini}_i =\mathcal{F}_{ij}\vartheta_j \, ,
\eeq
for some (random) matrix $\mathcal{F}_{ij}$, and where $\vartheta$ is a random vector with elements in the range $[0,\pi]$ (as expected for an initially massless field with a discrete shift symmetry and a symmetric potential). 

\emph{We set the initial conditions on $\vartheta_i$ to uniformly sample the field space in some basis}. We do this by noting that  \emph{there is some basis where the $\vartheta_i$ forms a cubic lattice}. We uniformly sample in this cubic basis, since this is operationally very simple. However, we note that this is not a uniform sampling of the field space in the ``charge basis'' defined by the charge matrix, $\mathcal{Q}$, an integer 
matrix whose entries reside in a charge lattice in the cosine potential, $V(\theta)=\sum_{X,i}\Lambda_X \left[ 1-\cos \left(\mathcal{Q}^{X}_i\theta_i\right)\right]$. We leave investigations of this interesting question, which is intimately related to the notions of alignment and charge quantisation for future work. Other discussions of this point, and sampling of initial conditions in general, see Refs.~\cite{Easther:2005zr,Bachlechner:2014hsa,Bachlechner:2015qja,2013JCAP...07..027E,2012PhRvD..86b3508M}.
\begin{figure*}
\centering
\hspace*{-2cm}
    \includegraphics[width=1.2\linewidth]{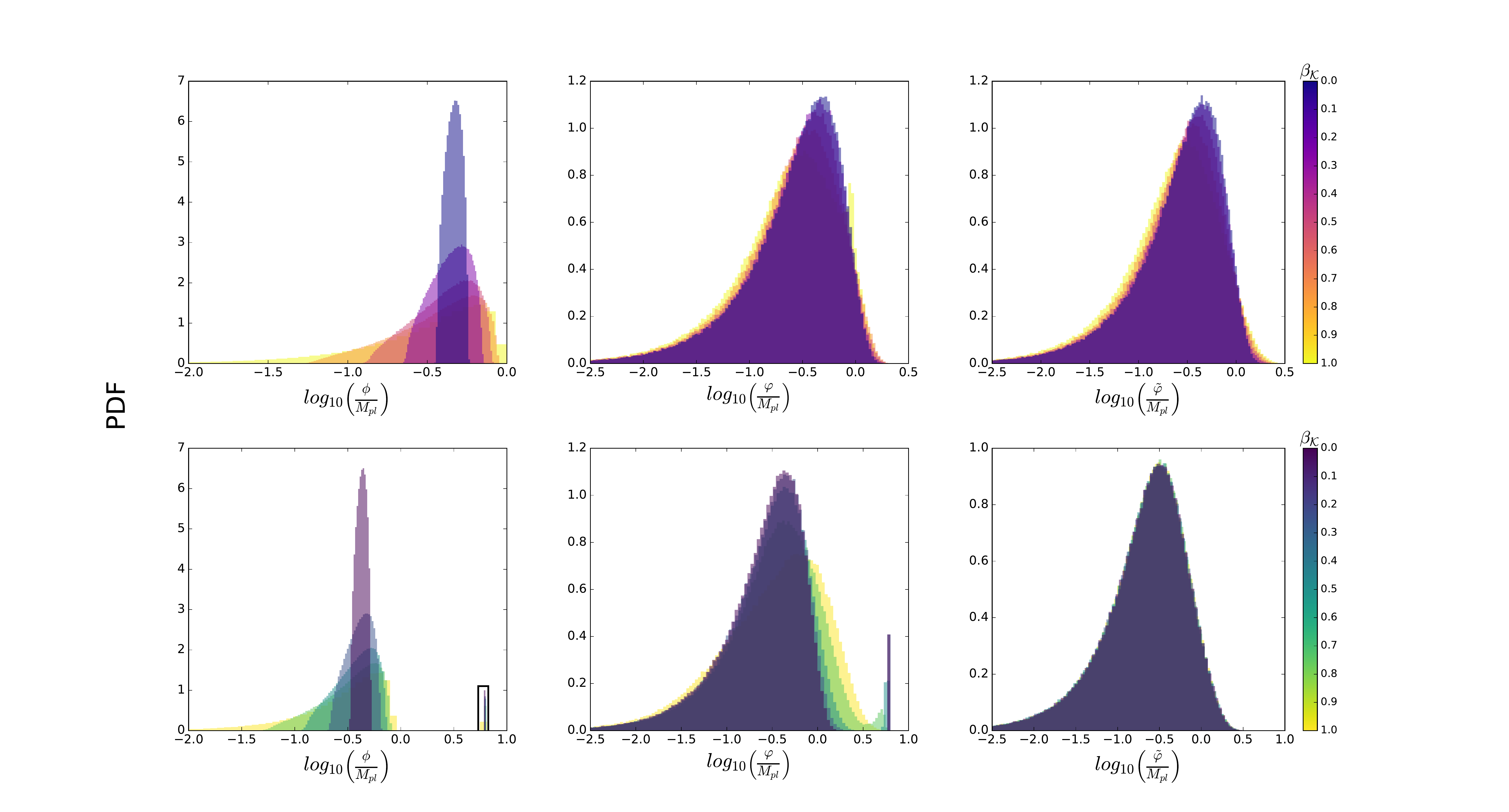}
\caption{{\bf String axiverse RMT model initial field displacement spectra:} 
Probability density plots for the initial axion field displacements defined in each basis outlined in Eqs.~(\ref{eq:fbasis}), (\ref{eq:efff1}) and (\ref{eq:efff2}) for 5000 iterations using $n_{\rm ax}=75$. \emph{Upper panels}: Zero centred mean, gaussian distributions used for the elements of the kinetic matrix $\mathcal{K}_{ij}$ (WW RMT model (Section~\ref{sec:WWmat})). \emph{Lower panels}: Non-zero centred mean, non-Gaussian distributions used for the elements of the kinetic matrix $\mathcal{K}_{ij}$ (LF RMT model (Section~\ref{sec:logflat})). The highlighted (black rectangle) values demonstrate the enhancement of the spectral width in the LF RMT model.}
\label{decayf2}
\end{figure*}
We define the matrix $\mathcal{F}_{ij}$ for two different possibilities for the cubic basis. Consider the set of transformations that turn the initial fields, $\theta$, into the canonically normalised fields, $\phi$, in index notation:
\beq
\frac{\phi_i}{M_{pl}} = V_{ij}{\rm diag}(f_a)_{jk}U_{kl}\theta_l \, .
\label{eqn:field_transform}
\eeq
In general we should expect that in the cubic basis both $\mathcal{K}_{ij}$ and $\mathcal{M}_{ij}$ are off-diagonal, and so $\vartheta_i=\theta_i$. On the other hand, it could be the case that the cubic basis is the same basis as the one in which $\mathcal{K}_{ij}$ is diagonal. In that case, it is natural to set $\vartheta_i=U_{ij}\theta_j$. We allow for both possibilities in our numerical explorations (though for the MP and MT models, where $\mathcal{K}_{ij}$ is diagonal by construction, the two choices are the same).

For completeness of discussion, we still seek to define a measure on the initial field displacements that is somewhat equivalent to the usual notion of a ``decay constant''. We define such a measure by the following vector for the general case:
\beq
\tilde{\varphi_i} := |V_{ij}{\rm diag}(f_a)_{jk}U_{kl}\langle \vartheta \rangle _l| \, ,
\label{eq:efff1}
\eeq
where $\langle \vartheta \rangle$ is the vector of $\sfrac{\pi}{2}$ values representing the average of $\vartheta$. For the case of the cubic basis with diagonal $\mathcal{K}_{ij}$, we define our measure as,
\beq
\varphi_i = |V_{ij}{\rm diag}(f_a)_{jk}\langle \vartheta \rangle_k| \, .
\label{eq:efff2}
\eeq

The overall scale of our initial conditions is set by the eigenvalues of $\mathcal{K}_{ij}$ giving the elements $f_a$. However, rotations can shift these values onto different canonical fields, allowing for N-flation type enhancement by the pythagorean sum. The initial field conditions coming from $f_a$ to $\phi$, $\tilde{\varphi}$ and $\varphi$ are shown in Fig.~\ref{decayf2}. In the upper and lower left-hand panels we show the initial field displacements of the form in Eq.~(\ref{eq:phiini}) for both the WW RMT and LF RMT models where the bulk of the spectrum is initially limited to sub Planck scale values (upper and lower left panels). $\phi$ is defined using Eq.~(\ref{eq:phif}) where $\mathcal{F}_{ij} = {\rm diag}(\sqrt{2f^2_a})_{ij}$ such that, $\phi_i = {\rm diag}(\sqrt{2}f_a)_{ij}\langle \vartheta \rangle_j$. In the upper panels we see that the initial field displacements quickly converge to a negatively skewed distribution on a logarithmic scale when using a \emph{white} Wishart matrix for $\mathcal{K}_{ij}$ (see Section~\ref{sec:WWmat}). Selecting a new basis identified by a further rotation acting on $\mathcal{F}_{ij}$ does not alter the initial field displacements where we observe a degeneracy across all values of $\beta_{\mathcal{K}}$.

When a \emph{spiked} Wishart matrix is used for $\mathcal{K}_{ij}$ (see Section~\ref{sec:logflat}) the repulsed eigenvalues shown for $\phi$ ``enhance'' the potential initial field conditions when selecting a new basis for sampling. Said alternatively the convergence of the spectra via the unitary rotations is ``slower'' in this model maintaining features of the initial matrix spectra for $\mathcal{K}_{ij}$. The spectra for each choice of basis is distinct in its output as shown in the central and right lower panels. In the basis for $\varphi$ lower values of $\beta_{\mathcal{K}}$ maintain the hard edge of the non-rotated spectra (lower left panel) with values of $\beta_{\mathcal{K}} \rightarrow 1$ providing larger probability densities for field displacement transcending the $M_{pl}$ limit. The two models converge when finally selecting $\tilde{\varphi}$ as the choice of basis. 
      
\section{The String Axiverse}
\label{sec:axiverse_rmt}

\subsection{A Random Matrix Approach to the String Axiverse}
\label{sec:rmtmodels}
\subsubsection{Generalities}

A simplified approach to modelling the \emph{string axiverse} is to use random matrix theory to encode the structure of the kinetic matrix and mass matrix appearing in the effective model description in the Lagrangian of Eq.~(\ref{eq:effL}), without detailed knowledge of the underlying K\"{a}hler potential and superpotential. The power of random matrix theory is the notion that large, complicated systems present the properties of universality, depending only on the symmetry classes of these systems. A principle observation occurs as the dimensional order of these matrices increases their spectra stabilise, their properties determined by several limiting laws such as Wigner's celebrated semicircle law. At a very basic level, random matrix theory and the universality that emerges from it can be considered a generalisation of the central limit theorem. See Appendix~\ref{appendix:rmt} for further discussion on the generalities of random matrix theory. Accessible introductions to these topics can be found on Terry Tao's blog,\footnote{\url{https://terrytao.wordpress.com/}} and in the book by Mehta~\cite{mehta}. 

In each class, the matrices we consider will all have elements drawn from the same statistical distribution. Our matrices are not block-diagonal, with blocks containing different scales. Physically therefore, there are no separate sectors: all the axions we consider receive their masses from effects of the same order. Universality then dictates that our distributions will, up to outliers, be classified by a single (mean) scale, and spread (variance, and other moments). The lack of bimodality means that the mass distributions are unlikely to furnish us simultaneously with axions classified as DM ($m_a\gtrsim H(z_{\rm eq})\sim 10^{-27}\text{ eV}$) and DE ($m_a\lesssim H(z_0) \sim 10^{-33}\text{ eV}$), while at the same time having no cosmologically problematic axions at the intermediate scale~\cite{Hlozek:2014lca}.

Given these considerations we will restrict ourselves to only considering two classes of random matrices constructed in the form of Eq.~(\ref{eq:generalmatrix}) without any loss of generality for our concerns. First, the well motivated case of matrices residing in the Wishart ensemble of real sample covariance matrices. The limiting spectrum of normalised Wishart matrices, $W = \frac{1}{p}X^T X$ where $X$ is a $(n\times p)$ rectangular matrix and $p\geq n$ is given by the Mar\v{c}henko-Pastur law (see Appendix~\ref{sec:mplaw}) with spectral properties determined by an aspect ratio, $\sfrac{n}{p} \in (0,1]$ (see below). The universality of the Mar\v{c}henko-Pastur law deems it will hold for arbitrary distributions of zero mean and unit variance. When constructing our kinetic and mass matrices in this form we shall designate them as a \emph{white} Wishart matrix parameterised by $\beta=1$ in the the standard \emph{beta} ensemble for random matrices (see Section~\ref{appendix:maten}). (\emph{White}) Wishart matrices often occur in many applications of random matrix theory and can play a key role in areas such as multidimensional Bayesian analysis \cite{doi:10.1080/03610929508831629,10.2307/2984196}. The generalised construction of Wishart matrices via higher order convulsions have spectra described by the \emph{Fuss-Catalan} distributions which could prove an interesting extension in future work \cite{2015PhRvE..92a2121M}. See Appendix~\ref{appendix:rmt} for further discussion.   

Secondly, we will investigate the properties of non-universality and extremal fluctuations in the asymptotic behaviour of singular values in random matrix models using a non-gaussian distribution for the entries of the sub matrices, $X$. Matrices constructed in this manner are subject to an eigenvalue repulsion in the form of singular eigenvalues away from a bulk region of the distribution. The bulk of these distributions is governed by the Mar\v{c}henko-Pastur density function. Further discussion can be found in Appendix~\ref{appendix:rmt} or Refs.\cite{2008PhDT.......217W,2010arXiv1011.5404M,2011arXiv1109.3704B,2012arXiv1209.3394C} for discussion of \emph{spiked} Wishart covariance models with these properties. We will not consider in detail the finer properties of the analysis associated to the largest eigenvalues for sample covariances matrices with spiked populations through there moments or the nature of the Baik, Ben Arous and P\`{e}ch\`{e} (BBP) phase transition which can lead to such phenomena \cite{2004math......3022B}. We will treat our models at the level of the statistical distributions used to construct our sub-matrices only highlighting the features and spectral properties their eigenvalue distributions may exhibit.  We will designate a matrix constructed in this way as a \emph{spiked} Wishart matrix. 

To summarise, for any given random matrix model, we construct both $\kahlerij$ and $\mij$ as normalised positive-definite matrices in the following way:
\begin{align}
A_{hj},B_{hj}  &\in \mathbb{R}^{n \times p} \label{eq:ransubm}\,, \\
\label{eq:ransubmk}
\mathcal{K}_{ij}&= \frac{1}{p} A_{ih}^TA_{hj} \ \in \mathbb{R}^{n \times n}\, , \\
\label{eq:ransubmm}
\mathcal{M}_{ij}&= \frac{1}{p} B^T_{ih}B_{hj}\ \in \mathbb{R}^{n \times n} \,,
\end{align}
where the entries of the sub-matrices $A_{hj}$ and $B_{hj}$ in Eq.~(\ref{eq:ransubm}) are random entries drawn from a given statistical distribution, $\Omega(\mu,\sigma)$. 

In our models of the string axiverse, by construction, $\mathcal{K}_{ij}$ and $\mathcal{M}_{ij}$ are square matrices with a dimension determined by the number of axions, $(n_{\rm ax},n_{\rm ax})$. By definition the sub-matrices in our RMT models, $A_{hj}$ and $B_{hj}$ need not be square.  This defines the incorporation of our aspect ratio shaping parameters $\beta_{\mathcal{K}}$ and $\beta_{\mathcal{M}}$ where the sub-matrices $A_{hj}$ and $B_{hj}$ are both rectangular with the defined dimensions $(n_{\rm ax}, n_{\rm ax}/\beta_{\mathcal{K}})$ and $(n_{\rm ax}, n_{\rm ax}/\beta_{\mathcal{M}})$ respectively. The shaping parameters are explicitly  defined as,

\begin{align}
\label{eq:betak}
\beta_{\mathcal{K}} &= \sfrac{n_{\rm ax}}{p_{\mathcal{K}}}\,,\\
\label{eq:betam}
\beta_{\mathcal{M}}	&= \sfrac{n_{\rm ax}}{p_{\mathcal{M}}}\,,
\end{align}
 
where $p_{\mathcal{K},\mathcal{M}} \geq n_{\rm ax}$. When we select the two shaping parameters to be the same value determined by $p_\mathcal{K} = p_{\mathcal{M}}$ (which we will in general through this study) we shall refer to this using the notation, $\beta_{\mathcal{K},\mathcal{M}}$. See Appendix~\ref{sec:axmod} for discussion on the role of these parameters in the context of realisations of the axiverse in string theory along with the likely values they take. 

\subsection{Scale Invariant Measure on Eigenvalues}
\label{sec:haarmeasure}

As a straw-man model, and as a baseline with which to compare our physically motivated models, we consider a log flat prior using the motivations of scale invariance on the positive, real, \emph{physical} and dimensionful parameters coming from $\mathcal{K}_{ij}$ and $\mathcal{M}_{ij}$: that is, on the decay constants in the diagonal basis along with normalisation factors of $\sqrt{2}$ and masses in the canonical diagonal basis. Such a prior is well motivated in the in context of axiverse literature~\cite{Arvanitaki:2009fg} and is inspired by the Jeffreys prior. The axion decay constants could also span several decades~\cite{Honda:2016jnd,Banks:2003sx}. We use the log-flat prior for both of these unknown dimensionful quantities as a ``\emph{maximally ignorant}'' approach.

We begin in the mass eigenstate basis (Eq.~(\ref{eq:finall})) where both $\mathcal{K}_{ij}$ and $\mathcal{M}_{ij}$ are diagonal, and consider only the eigenvalues of both the kinetic and mass matrix in this basis.
\begin{figure*}[t]
\centering
\begin{tabular}{cc}
    \includegraphics[width=0.49\linewidth]{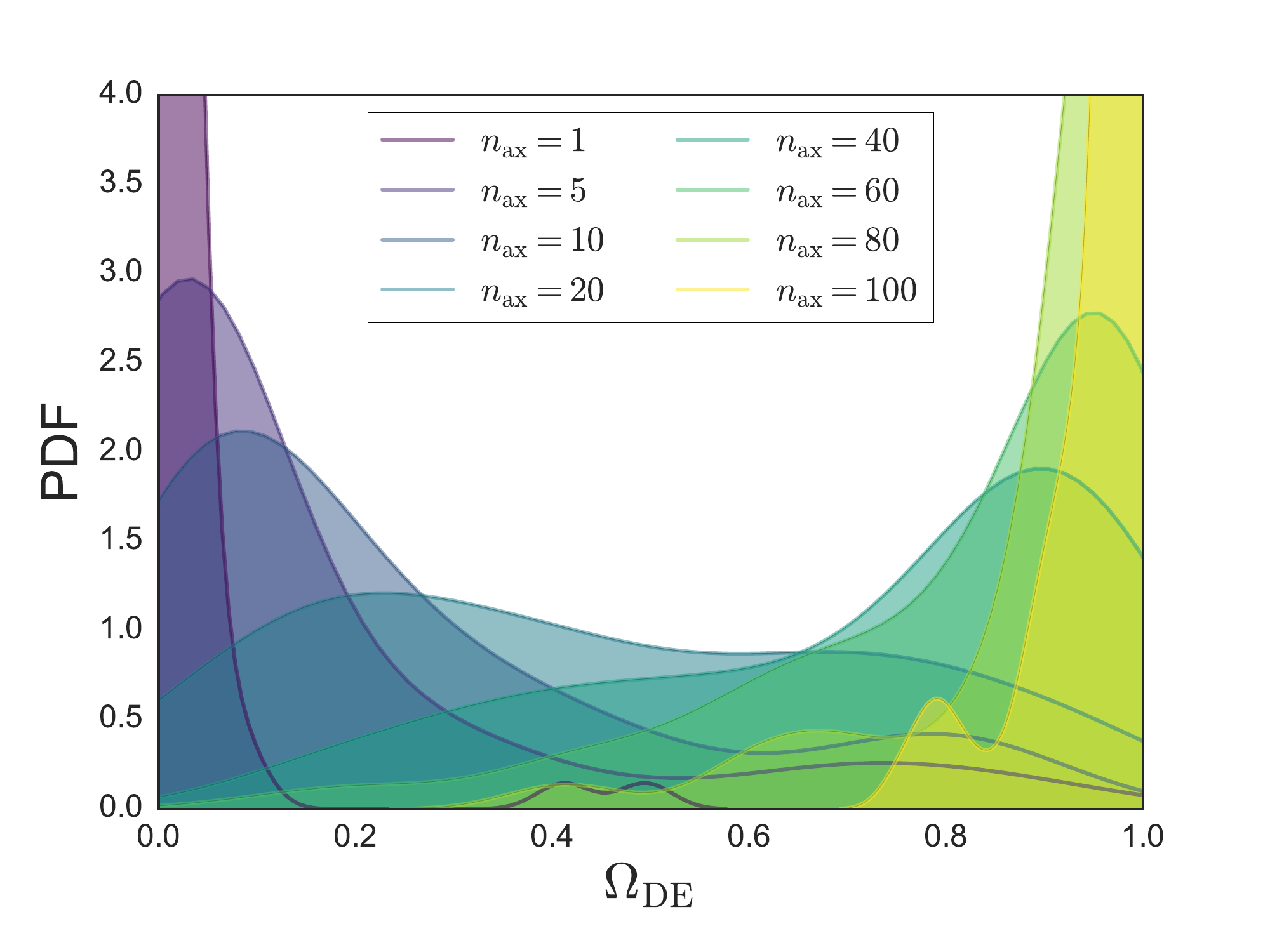}&
    \includegraphics[width=0.49\linewidth]{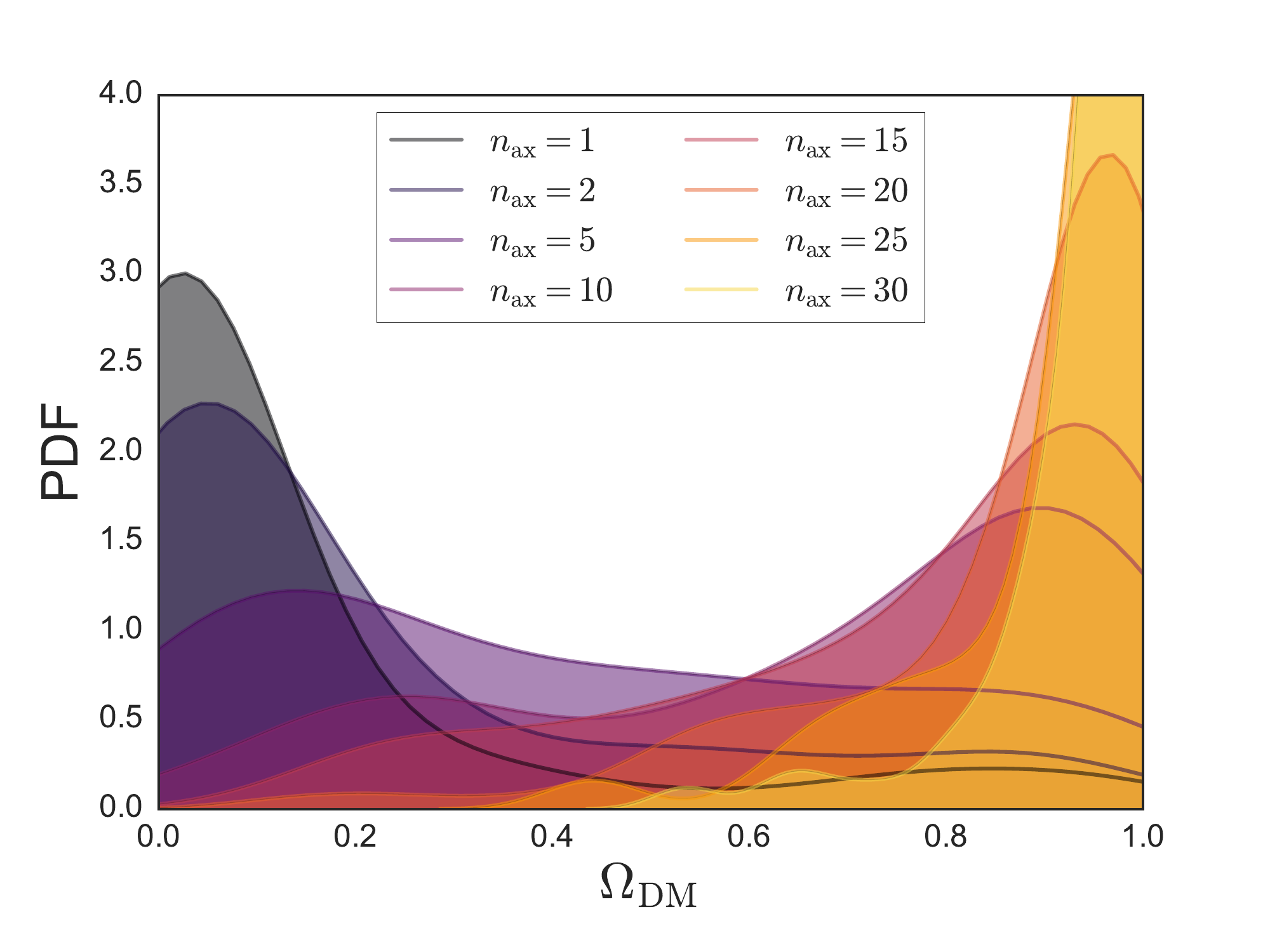}\\
\end{tabular}
    \centering
    \caption[]%
{{\bf Dark energy and dark matter cosmologies with scale invariant measure on physical quantities:} \emph{Left panel}: KDE plot for the axion dark energy density parameter, $\Omega_{{\rm DE}}$ with $n_{\rm ax}= \mathcal{O}(1)\rightarrow \mathcal{O}(100)$ with log-flat priors on both the physical parameters, $m^2_a$ and $f^2_a$ sampled in the window detailed in Eqs.~(\ref{eq:hmde1}) to (\ref{eq:hmde3}). \emph{Right panel}: KDE plot for the axion dark matter density parameter, $\Omega_{\rm DM}$ with $n_{\rm ax}= \mathcal{O}(1)\rightarrow \mathcal{O}(10)$  with log-flat priors on both the physical parameters, $m^2_a$ and $f^2_a$ sampled in the window detailed in Eqs.~(\ref{eq:hmdm1}) to (\ref{eq:hmdm3}).}
\label{fig:haarex}
\end{figure*}
The axion parameters are drawn from, 
\begin{align}
	\log_{10}({\rm eig}(\mathcal{{K}}_{ij})) &\in \mathcal{U}[k_{\rm min},k_{\rm max}] \label{eq:haar1}\,,\\
	\log_{10}({\rm eig}(\mathcal{{M}}_{ij})) &\in \mathcal{U}[m_{\rm min},m_{\rm max}]\label{eq:haar2}\,.
\end{align}
The uniform distribution is unnormalised, and is only a proper prior for our considerations once the end points of the distribution are fixed by the controlling limits. By definition this breaks the scale invariance of our prior however we retain motivations for bounded limits in concordance with the literature. The values, 
\begin{align}
{\rm eig}(\mathcal{{K}}_{ij}) = f_{a,i}^2\,,\\
{\rm eig}(\mathcal{{M}}_{ij}) = m^2_{a,i}\,, 
\end{align}
represent the elements of the diagonalised kinetic and mass matrix respectively. The limits $k_{\rm min}$ and $k_{\rm max}$ in general are associated with lower and upper bounds on non-perturbative physics scales. The upper and lower bounds, $m_{\rm min}$ and $m_{\rm max}$ represent a portion of the axion mass window suited for extracting fields behaving as either DE or DM. In the left hand panel of Fig.~\ref{fig:haarex} we show the KDE plot in the context of axions behaving as DE with the following parameter priors,

\begin{align}
	n_{\rm ax} = \mathcal{O}(1) &\rightarrow \mathcal{O}(100) \,, \label{eq:hmde1}\\
	\log_{10}(\sfrac{{\rm eig}(\mathcal{{K}}_{ij})}{M_{pl}}) &\in \mathcal{U}[-4.0,-0.5]\,,\label{eq:hmde2}\\
	\log_{10}(\sfrac{{\rm eig}(\mathcal{{M}}_{ij})}{M_H}) &\in \mathcal{U}[-2.0,2.0]\,.\label{eq:hmde3}
\end{align}
Correspondingly, the right hand panel of Fig.~\ref{fig:haarex} shows the KDE plot for axions behaving as DM using the following priors,
\begin{align}
	n_{\rm ax} = \mathcal{O}(1) &\rightarrow \mathcal{O}(10) \,, \label{eq:hmdm1}\\
	\log_{10}(\sfrac{{\rm eig}(\mathcal{{K}}_{ij})}{M_{pl}}) &\in \mathcal{U}[-4.0,-0.5]\,,\label{eq:hmdm2}\\
	\log_{10}(\sfrac{{\rm eig}(\mathcal{{M}}_{ij})}{M_H}) &\in \mathcal{U}[6.0,16.0]\,. \label{eq:hmdm3}
\end{align}
 
The requirement for axion population sizes with at least $n_{\rm ax} \approx \mathcal{O}(10)$ in order to give a realistic chance of finding cosmologies returning values of $\Omega_{\rm DE}$ sitting in the rough window $\Omega_{\rm DE} = (0.6 \rightarrow 0.8)$, is evident in the left hand panel of Fig.~\ref{fig:haarex}. The right hand panel of Fig.~\ref{fig:haarex} shows that an increase in the field population size quickly leads to the domination of axion DM when utilising a significant mass window.  
  We use the information in both panels of Fig.\ref{fig:haarex} to indicate the potential for multiple axions giving the required values of $\Omega_{\rm DM}$ and $\Omega_{\rm DE}$ whilst maximising the size of the population. In general our RMT models will consider more localised scale windows and as such we select a population size of $n_{\rm ax}=20$ to serve as a good common ground between both types of cosmology.

\subsection{Random Matrix Theory Models}

For a more physically realistic approach we should expect our axion parameters to be encoded in some kind of matrix structure, with a non-trivial role played by the rotations between different bases. This is due to the fact there is some physical meaning to the basis in which Dirac quantisation occurs, which in general is not the same as the diagonal basis. In general this RMT structure will give localised physical parameter distributions, where we shall suspend the exploration of coupled dark sector cosmologies as a focus of future work beyond the simple example above. The following sections detail the introduction of random matrices for the string axiverse, and the power random matrix theory can have even when considering a more complete picture of the axion landscape.

Our study consists of three models with their foundations in the universal behaviour of asymptotic RMT plus an approach to realisations of the string axiverse in G2 compactified M-theory. Below we outline our treatment of both the kinetic and mass matrix and associated parameters in these models. In the right-hand panels of Fig.~\ref{fig:massspectra} we present the eigenvalue spectra of the mass matrix in the mass eigenstate basis for each model using arbitrary prior configurations. In the left-hand panels we show an approximated theoretical density function fit for the form of the finite dimensional matrix spectra in our models. 

\subsubsection{MP RMT Model \\(Unit $\mathcal{K}_{ij}$  / White Wishart $\mathcal{M}_{ij}$)}
\label{sec:mpmodel}

This model is based on the N-flation model presented in Ref.~\cite{Easther:2005zr} (See Appendix~\ref{appendix:nflation}) whereby we encode our uncertainty using a spectrum of masses governed by the Mar\v{c}henko-Pastur density function for a population of $N$ uncoupled axions. We need only consider matrix structure for the mass matrix, $\mathcal{M}_{ij}$ where, unlike our other models in the subsequent sections, we begin in the following basis,

\beq
\label{eq:nflatbasis2}
\mathscr{L} = -\frac{1}{2}\partial_\mu {\phi}_i\partial^\mu {\phi}_j - \frac{1}{2} {\phi}_i {\mathcal{M}}_{ij} {\phi}_j \, ,
\eeq
where our mass matrix is constructed as,
\begin{align}
\mathcal{M}_{ij} &= \left(\frac{n_{\rm ax}}{\beta_{\mathcal{M}}} \right)  B^T_{ih}B_{hj}\,, \\
B_{hj} &\in \sigma_{\mathcal{M}} \times  \mathcal{N}(0,1)\,. \label{eq:sigmam}	
\end{align}
Our parameters in this model for $\mathcal{M}_{ij}$ consist of the scaling factor $\sigma_{\mathcal{M}}$ which sets the value of $\langle m^2_a \rangle$ and distribution shaping index $\beta_{\mathcal{M}}$. In this basis the role of the kinetic matrix is such that $\mathcal{K}_{ij}$ is unitary providing only trivial rotations to the fields ($U_{kl} = \mathbb{1}$) and mass matrix following the process outlined in Section~\ref{sec:effmodel}. Following the considerations in Ref.~\cite{Easther:2005zr} when setting the initial field conditions, the treatment of the kinetic terms is replaced by considering the axion vevs in the mass-eigenstate basis using an equal field condition scale parameter, $\bar{f}$ along with the initial misalignments. The initial field conditions in this model are defined as, 

\begin{equation}
\label{eq:mpvev}
{\phi_i} = V_{ij}\bar{f} \mathbb{1} \theta_l \, .
\end{equation}
   
\begin{figure*}[ht] 
\vspace*{-4.0ex}   
  \begin{subfigure}[b]{0.48\linewidth}
    \centering
    \includegraphics[height=0.27\textheight]{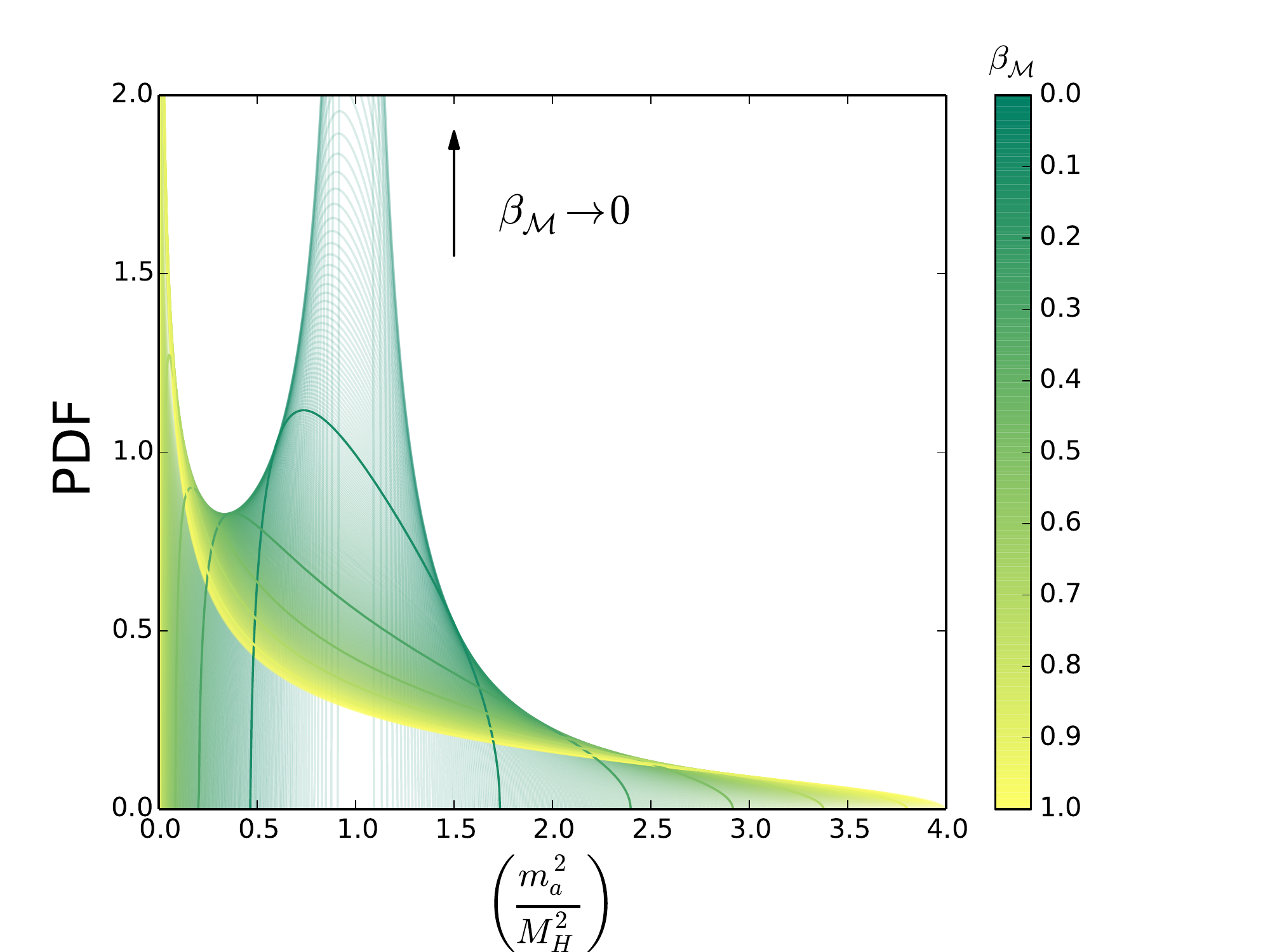} 
    \caption{{\bf MP RMT:} Mar\v{c}henko-Pastur density function for 250 values of $\beta_{\mathcal{M}} \in (0,1]$ centred about $\langle m^2_a \rangle= M_H^2$.} 
    \label{fig:1} 
  \end{subfigure} 
  \hspace{\fill}  
  \begin{subfigure}[b]{0.50\linewidth}
    \centering
    \includegraphics[height=0.27\textheight]{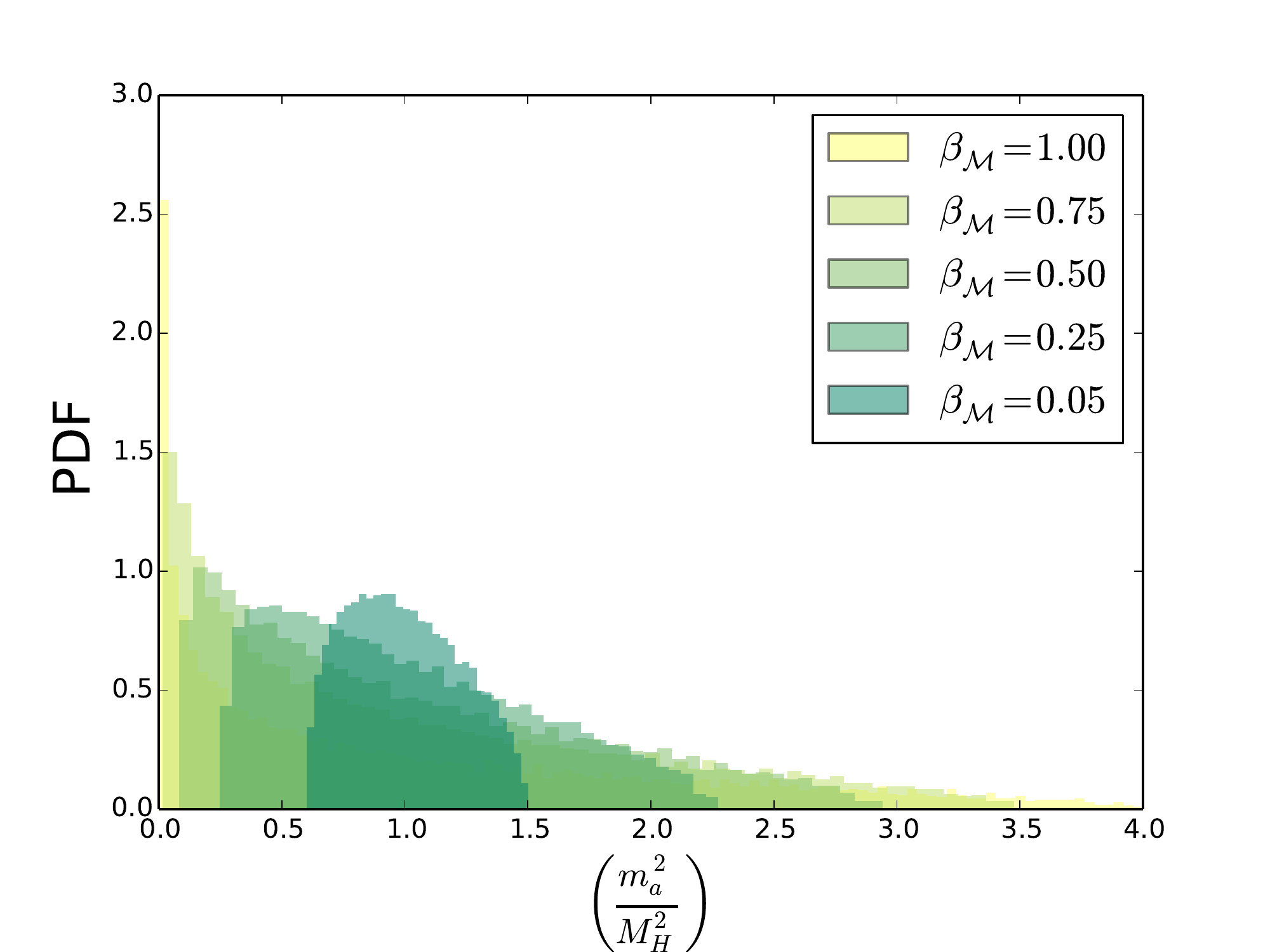} 
    \caption{{\bf MP RMT:} Probability density plots for the eigenvalues $m^2_a$ of $\mathcal{M}_{ij}$ centred about $\langle m^2_a \rangle = M_H^2$} 
    \label{fig:2} 
  \end{subfigure} 

  \vspace{0.1ex}  
  \begin{subfigure}[b]{0.48\linewidth}
    \centering
    \includegraphics[height=0.27\textheight]{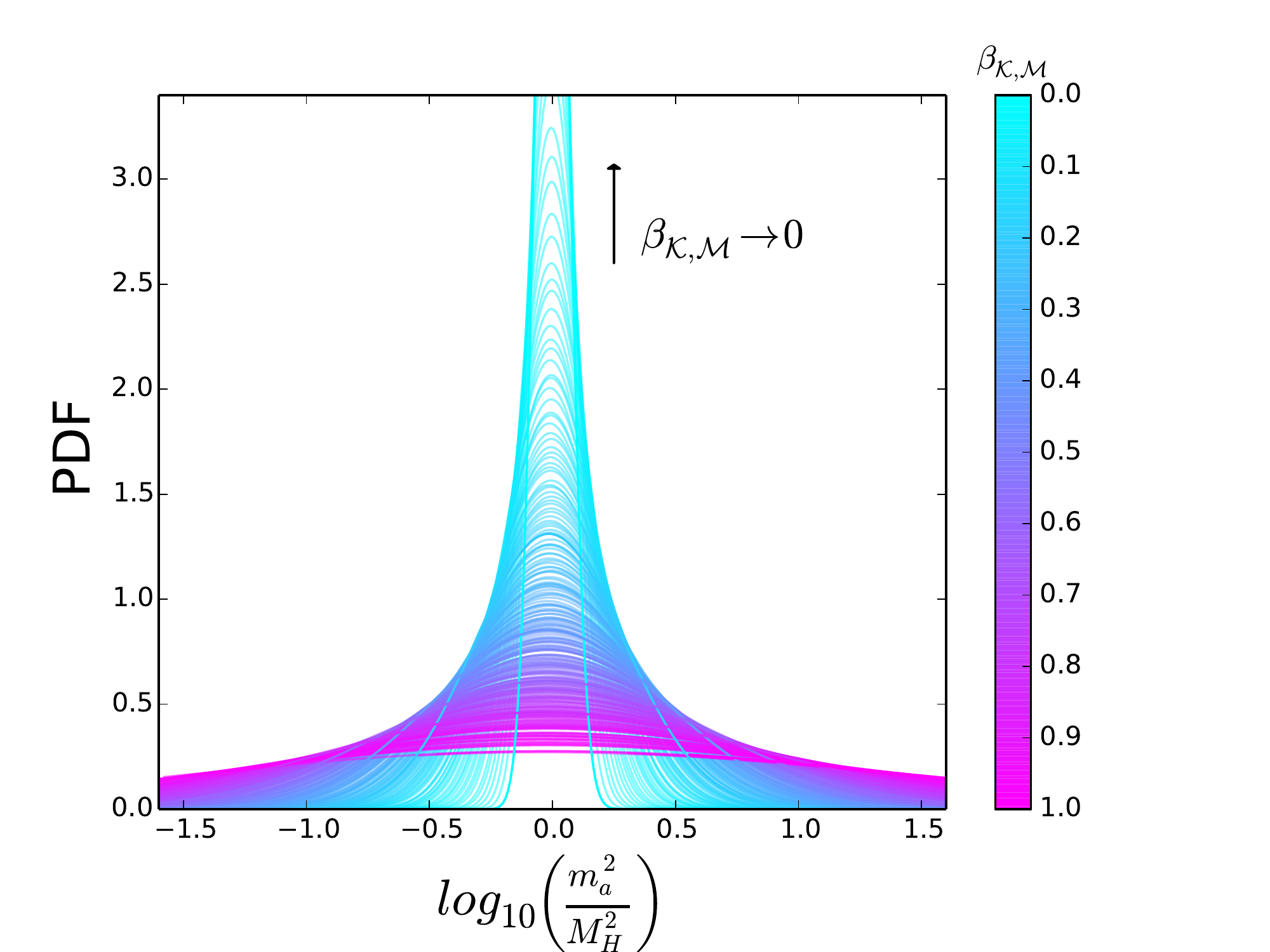} 
    \caption{{\bf WW RMT:} Log-normal density function fit using 250 values of $\beta_{\mathcal{K},\mathcal{M}}\in (0,1]$ centred about $\langle m^2_a \rangle = M_H^2$.} 
    \label{fig:3} 
  \end{subfigure} 
  \hspace{\fill}
  \begin{subfigure}[b]{0.5\linewidth}
    \centering
    \includegraphics[height=0.27\textheight]{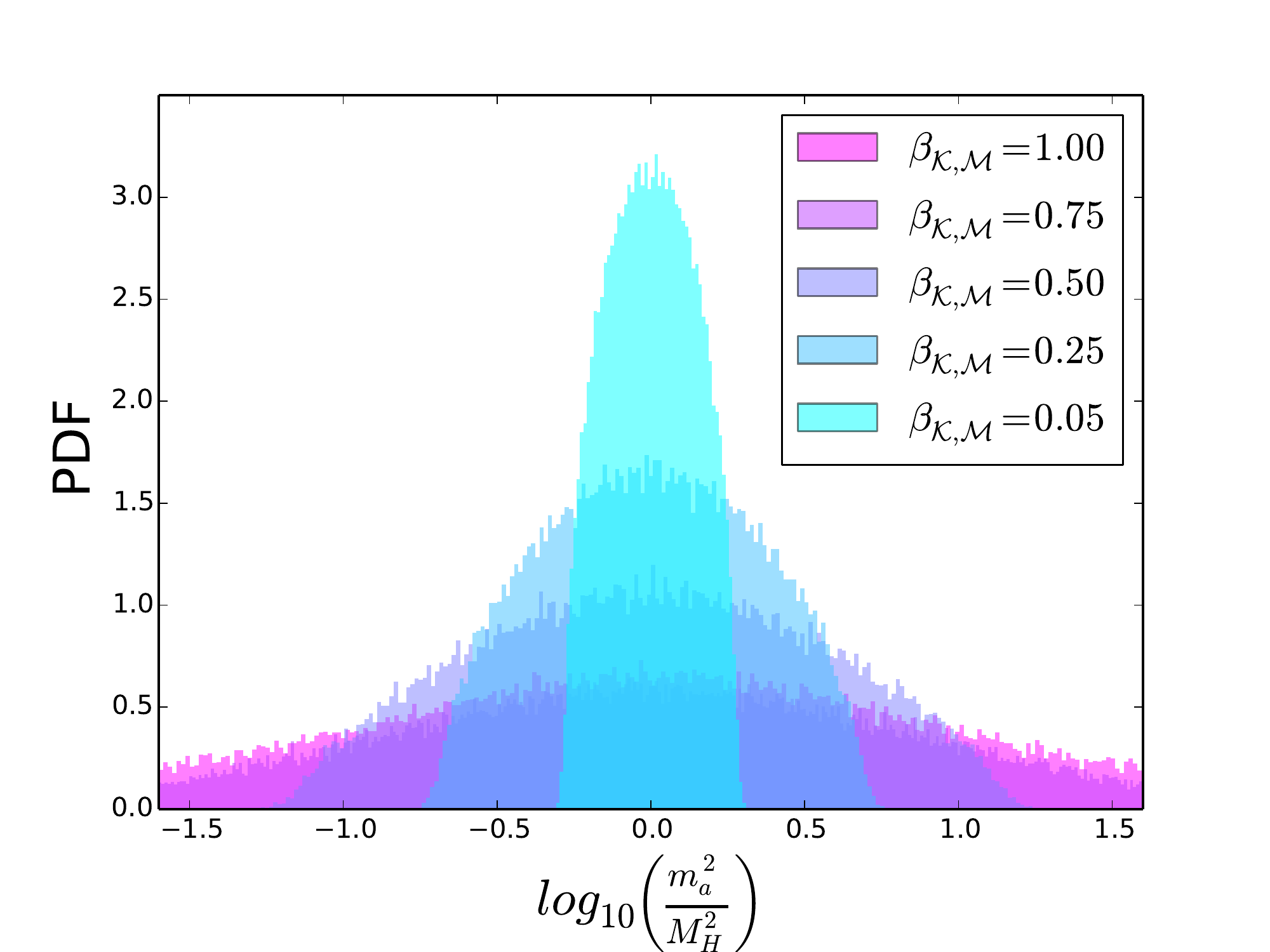} 
    \caption{{\bf WW RMT:} Probability density plots for the eigenvalues $m^2_a $ of $\mathcal{M}_{ij}$ centred about $\langle m^2_a  \rangle = M_H^2$.} 
    \label{fig:4} 
  \end{subfigure} 
\hspace*{-0.5cm}
    \vspace{0.1ex}
  \begin{subfigure}[b]{0.5\linewidth}
    \centering
    \includegraphics[height=0.27\textheight]{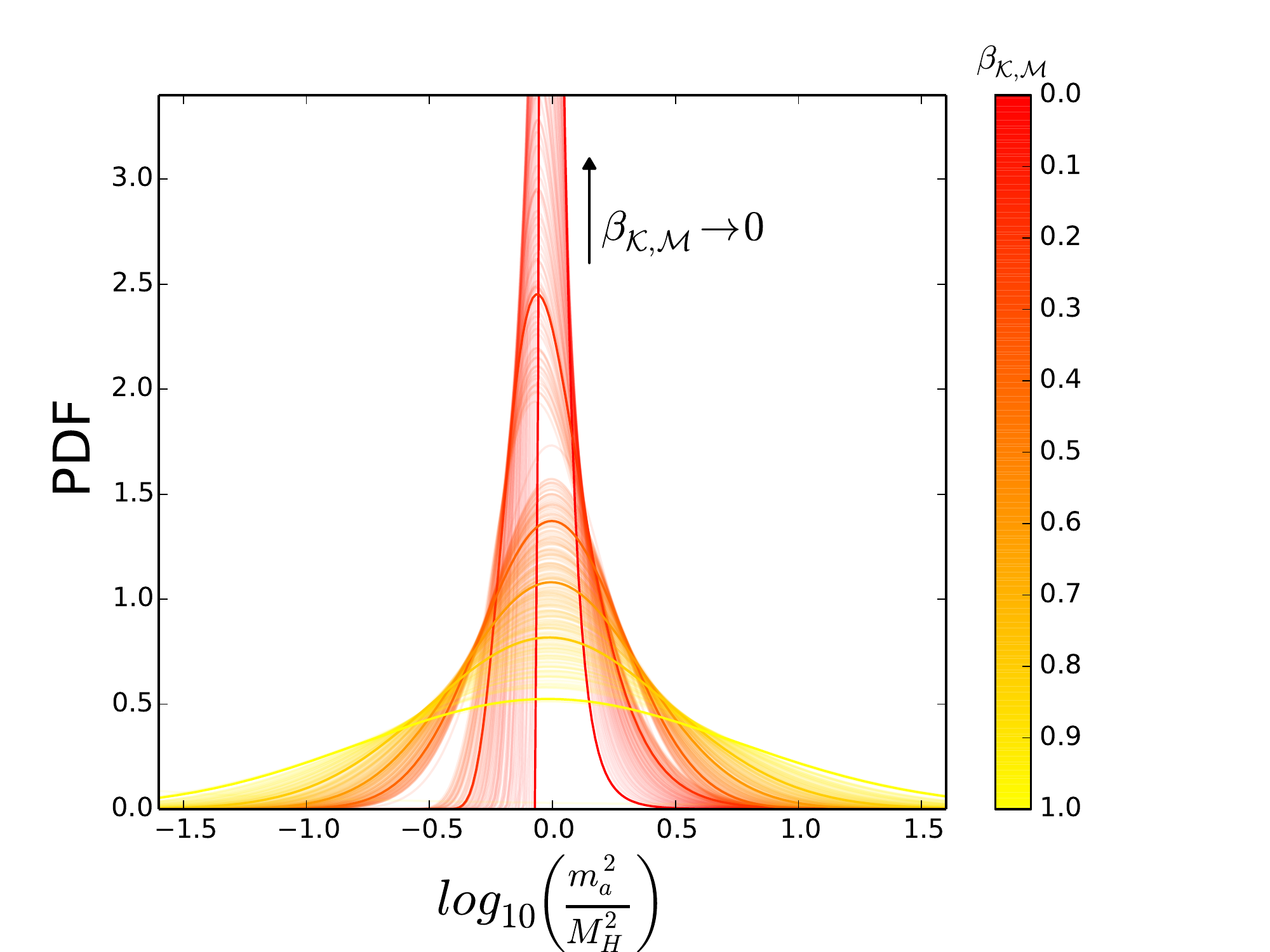} 
    \caption{{\bf LF RMT:} Log-normal density function fit using 250 values of $\beta_{\mathcal{K},\mathcal{M}}\in (0,1]$ centred about $\langle m^2_a  \rangle = M_H^2$.} 
    \label{fig:5} 
  \end{subfigure}
  \begin{subfigure}[b]{0.48\linewidth}
    \centering
    \includegraphics[height=0.27\textheight]{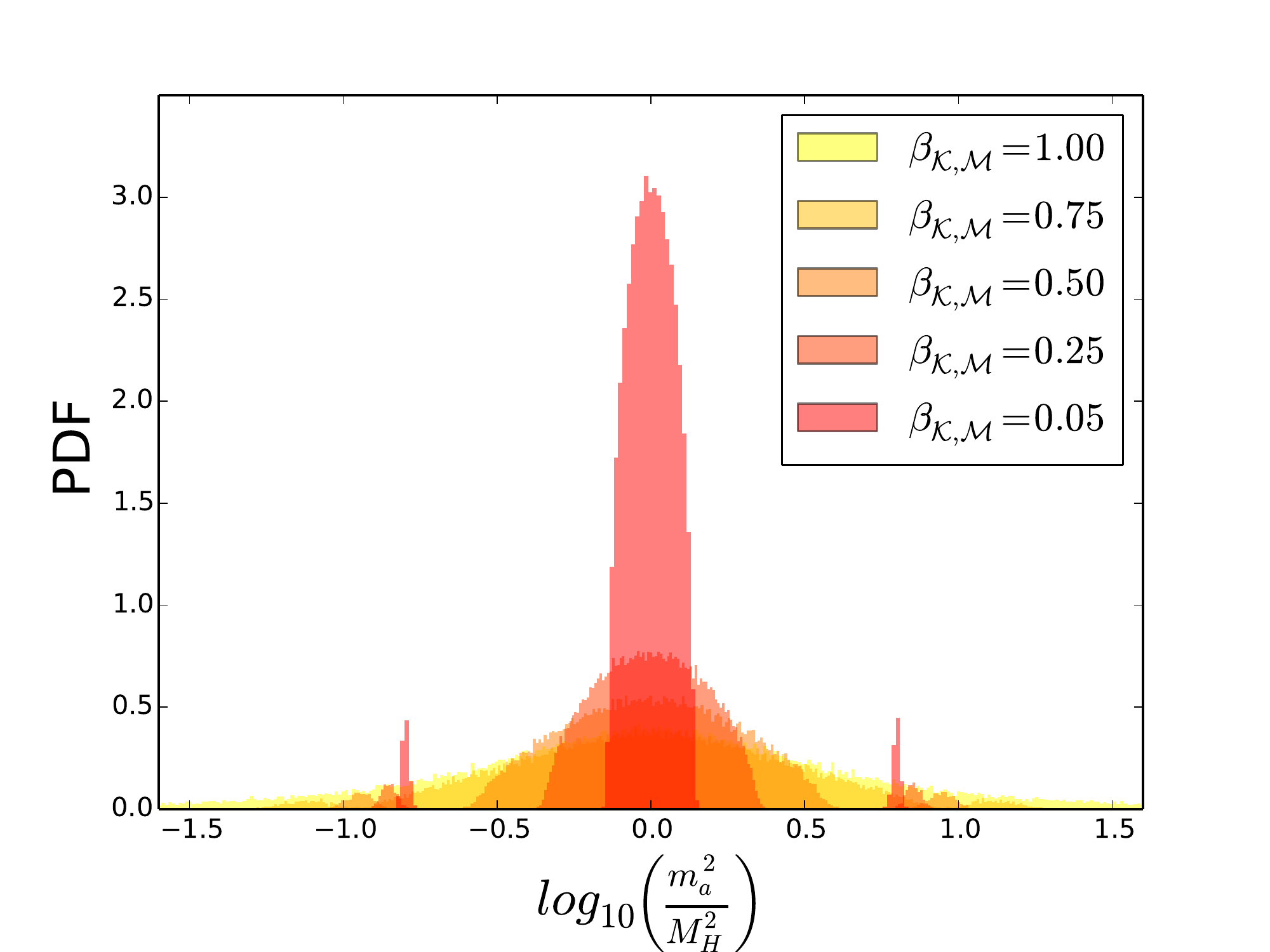} 
    \caption{{\bf LF RMT:} Probability density plots for the eigenvalues $m^2_a $ of $\mathcal{M}_{ij}$ centred about $\langle m^2_a \rangle = M_H^2$.} 
    \label{fig:6} 
  \end{subfigure} 

\caption{{\bf Theoretical mass squared value spectra density function fits and associated \boldsymbol{$\mathcal{M}_{ij}$} eigenvalue probability densities for RMT models:} \emph{Left-hand panels:} Theoretical density function fits for each of the RMT models outlined in Sections~\ref{sec:mpmodel}-\ref{sec:logflat} for 250 values of $\beta_{\mathcal{K},\mathcal{M}}\in(0,1]$. \emph{Right-hand panels:} Probability density plots for the eigenvalue spectrum of the rotated mass matrix, $\mathcal{M}_{ij}$ constructed using 1000 iterations and an a axion population size, $n_{\rm ax} =50$.}
\label{fig:massspectra} 
\end{figure*}

Fig.~\ref{fig:1} shows the theoretical eigenvalue spectrum of $\mathcal{M}_{ij}$, following the Mar\v{c}henko-Pastur density function for 250 varying values of $\beta_{\mathcal{M}}$. In Fig.~\ref{fig:2} we show the probability density convergence of the eigenvalue spectrum to the Mar\v{c}henko-Pastur law for a large number of fields ($n_{ax} = 1000$). The MP RMT model parameters are: 
\begin{equation*}
n_{\rm ax},\  \sigma_{\mathcal{M}}, \ \beta_{\mathcal{M}}, \ \bar{f}\,.	
\end{equation*}

\subsubsection{WW RMT Model \\(White Wishart $\mathcal{K}_{ij}$ / White Wishart $\mathcal{M}_{ij}$)}
\label{sec:WWmat}

It has also been suggested that the kinetic matrix, $\mathcal{K}_{ij}$ may too be well approximated by a matrix belonging to the Wishart ensemble on the basis of universality and symmetry \cite{Bachlechner:2014hsa}\cite{Bachlechner:2014gfa}\cite{Ferrari:2011is}. For the purposes of alignment the fundamental domain of such a matrix benefits from properties of eigenvector delocalisation and has well motivated features for inflationary models. In this model we include a kinetic matrix constructed with the same approach for the mass matrix in Section~\ref{sec:mpmodel} where, 

\begin{align}
\mathcal{K}_{ij} &=  \left(\frac{n_{\rm ax}}{\beta_{\mathcal{K}}} \right) A^T_{ih}A_{hj}\,, \\
A_{hj} &\in \sigma_{\mathcal{K}} \times \mathcal{N}(0,1)\,, \label{eq:sigmak}		\end{align} 
which in turn introduces the distribution shaping parameter $\beta_{\mathcal{K}}$. We begin in the basis defined in Eq.~(\ref{eq:effL}). In this basis the matrix structure for $\mathcal{K}_{ij}$ gives an axion decay constant spectrum governed by the Mar\v{c}henko-Pastur law up to canonical normalisation factors. The mass matrix, $\mathcal{M}_{ij}$ is now subject to non-trivial unitary rotations used to diagonalise $\mathcal{K}_{ij}$. In Fig.~\ref{fig:4} we show the rotated mass matrix spectrum for fixed values of $\beta_{\mathcal{K},\mathcal{M}}$. We use Fig.~\ref{fig:3} to display the approximate reduction of the spectral width for 250 different values of $\beta_{\mathcal{K},\mathcal{M}}\in (0,1]$ via a log-normal density function fit on the mass spectra. 
     \begin{figure}
    \centering
    \includegraphics[width=0.49\textwidth]{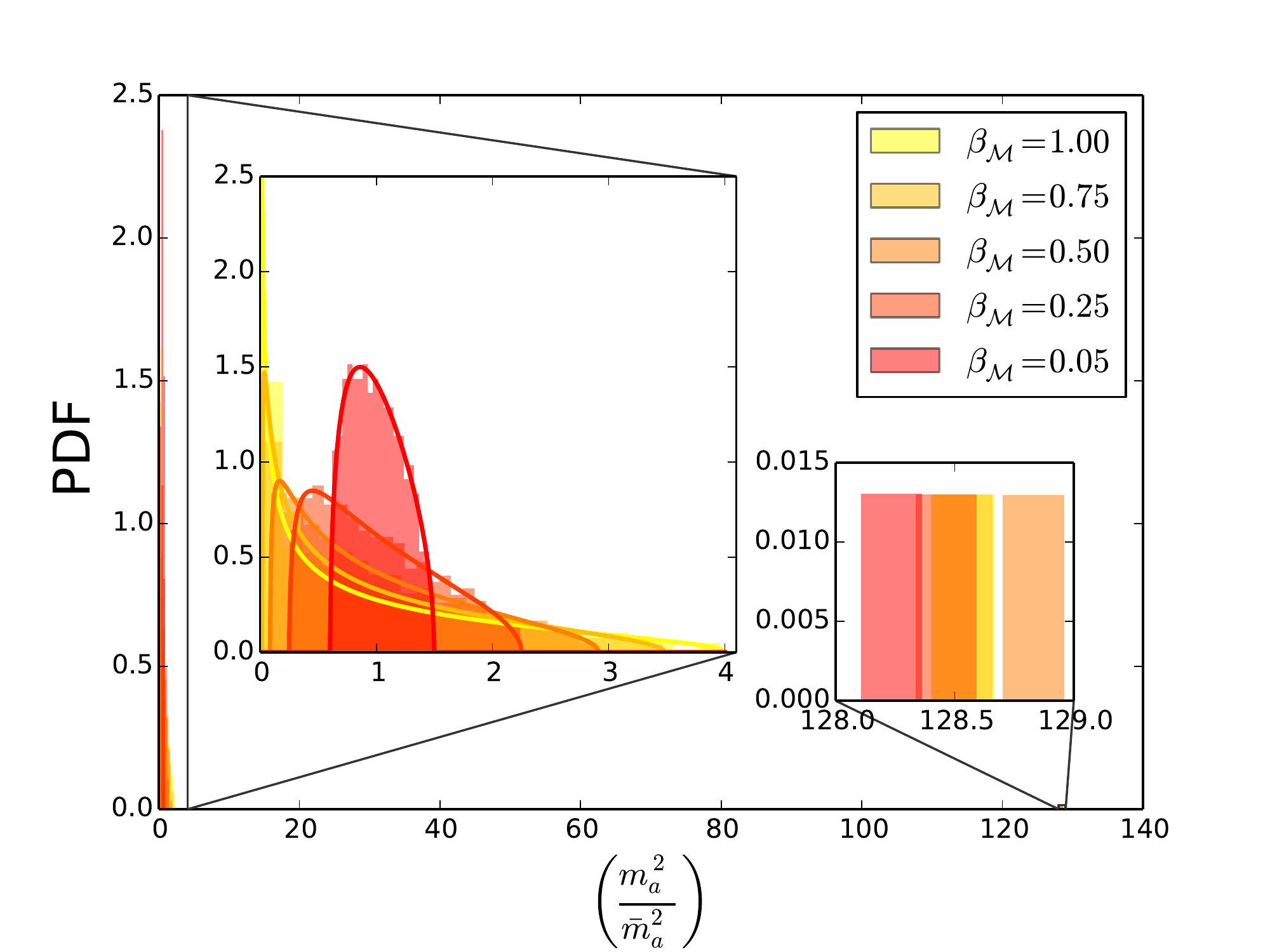}
    \caption[]%
    {{\bf LF RMT model non-rotated mass spectrum:} Eigenvalue spectrum of $m^2_a$ values for a $300 \times 300$ matrix, $\mathcal{M}_{ij}$ before basis selection rotations in the LF RMT model demonstrating the spiked Wishart spectral properties of the initial mass matrix. The bulk of the eigenvalue spectrum is governed by the Mar\v{c}enko-Pastur law (\emph{left inset}) which is partnered with one single outlying eigenvalue of $\mathcal{O}(N)$ (\emph{right inset}).}    
    \label{spikedis}     
\end{figure} 

In the limit $\beta_{\mathcal{K},\mathcal{M}} = 1$, the mass probability distributions are well modelled by a log-normal density function. When $\beta_{\mathcal{K},\mathcal{M}} \neq 1$ the mass spectrum is better approximated by truncated log-normal density functions as the edges of the distribution are hardened, simultaneously reducing the spectral width of the distribution. In the limit $\beta_{\mathcal{K},\mathcal{M}} \rightarrow 0$ we observe the convergence to a semicircular distribution within a significantly more localised mass window. The WW RMT model parameters are: 
\begin{equation*}
n_{\rm ax},\  \sigma_{\mathcal{M}}, \ \sigma_{\mathcal{M}}, \ \beta_{\mathcal{K}}, \ \beta_{\mathcal{M}}\,.
\end{equation*}

\subsubsection{LF RMT Model \\ (Spiked Wishart $\mathcal{K}_{ij}$ / Spiked Wishart $\mathcal{M}_{ij}$)}
\label{sec:logflat}

Our final RMT model will focus on the case in which we relax the condition that our sub-matrices $A_{hj}$, $B_{hj}$ are formed using statistical distributions defined with zero mean where, 
\begin{equation}
A_{hj},B_{hj} \in \sigma_{\mathcal{K},\mathcal{M}} \times \Omega(\cancel{0},\sigma)\, .
\label{eq:submlog}
\end{equation}
Our choice statistical distribution takes a log-flat prior on the elements of the sub-matrices in Eq.~(\ref{eq:submlog}), using the motivations of scale invariance highlighted in Section~\ref{sec:haarmeasure}, as displayed in Eqs.~(\ref{eq:lfel1}) and (\ref{eq:lfel2}). The random matrices $\mathcal{K}_{ij}$ and $\mathcal{M}_{ij}$ now fall under a class of matrices which exhibit the properties of a rank one spiked Wishart matrix (see Appendix~\ref{appendix:rmt}). The eigenvalue spectrum of these matrices presents a bulk distribution governed by the Mar\v{c}enko-Pastur law with one single outlier of the order $\lambda_{\rm max}\sim \mathcal{O}(n_{\rm ax})$ for a $n_{\rm ax} \times n_{\rm ax}$ dimensional matrix. Fig.~\ref{spikedis} shows the normalised mass spectrum before basis selection rotations using $n_{\rm ax}=300$, demonstrating these features in the spectrum. The axion decay constants in this model present a distribution of the form in Fig.~\ref{decayf} (log-scale) and Fig.~\ref{spikedis} (linear scale) up to canonical normalisation factors. An interesting feature of this model could be the realisation of an eigenvalue repulsion manifesting itself in the form of a single large decay constant traversing fundamental scales whilst the bulk of the distribution is contained in the sub-fundimental limit.       


 In order to construct our matrices we chose that each sub-matrix is parametrised by two upper and lower limit parameters for the elements in each matrix, denoted by $k_{\rm min}, \ m_{\rm min}$ and $k_{\rm max}, \ m_{\rm max}$. The elements of each sub-matrix $A_{hj}$ and $B_{hj}$ are drawn from:
\begin{align}
\log_{10} {A}_{hj} &\in \mathcal{U}[k_{\rm min},k_{\rm max}] \label{eq:lfel1}\, , \\ 
\log_{10} {B}_{hj} &\in \mathcal{U}[m_{\rm min},m_{\rm max}] \label{eq:lfel2} \, .
\end{align}

In accordance with the previous WW RMT model the eigenvalue spectrum of $\mathcal{K}_{ij}$ is subject to non-trivial rotations from the unitary rotations acting on $\mathcal{K}_{ij}$ where we also observe a log-normal distribution convergence of the mass spectrum in the mass eigenstate basis in the limit $\beta_{\mathcal{K},\mathcal{M}}=1$. Unlike the WW-RMT model when $\beta_{\mathcal{K},\mathcal{M}}\neq1$ the outlying eigenvalues present in the mass matrix in the initial basis, cause the formation of two outlying regions with eigenvalues separated from the bulk region of the spectrum in the mass eigenstate basis. The total spectral width of the eigenvalues is not reduced for values of $\beta_{\mathcal{K},\mathcal{M}}\neq1$ as displayed in Fig.~\ref{fig:4}, demonstrating the importance of the outlying eigenvalues in the initial basis. This model retains a non-zero probability density for fields with masses away from the bulk of the spectrum as shown in Fig.~\ref{fig:6}. 
              
Following the treatment used in Fig.~\ref{fig:3} we show the theoretical log-normal density function fit for 250 values of $\beta_{\mathcal{K},\mathcal{M}}\in (0,1]$ for the LF RMT model in Fig.~\ref{fig:5}. The separation of the distribution into three populations, a bulk and two repulsed regions when $\beta_{\mathcal{K},\mathcal{M}} \neq 1$ induces a skew in the log-normal density functions. This does not provide a very accurate theoretical fit for the total form of the mass spectrum, however we use this as an approximated measure of the effect of singular repulsed eigenvalues in the initial basis to compare to models without the properties of spiked population spectra. The skew in these distributions when $\beta_{\mathcal{K},\mathcal{M}} \neq 1$ give an indication of the potential magnitude of divergence away from the cosmologies obtained when modelling both $\mathcal{K}_{ij}$ and $\mathcal{M}_{ij}$ with standard Wishart matrices. The LF RMT model parameters are: 
\begin{equation*}
n_{\rm ax},\  k_{\rm min}, \ k_{\rm max},\ m_{\rm min}, \ m_{\rm max}, \ \beta_{\mathcal{K}}, \ \beta_{\mathcal{M}}\,.
\end{equation*}
\subsection{The M-Theory Axiverse}
\label{sec:mthethe}
In this section, we present a special type of RMT model motivated by the M-theory Axiverse \cite{Acharya:2010zx}. As we will see shortly, the matrix structure in the M-theory framework is constructed in a similar manner to the previous RMT models, guaranteeing positive definiteness in the axion masses. Since the moduli stabilisation under the framework of G2 compactified M-theory has already been extensively studied in Refs.~\cite{Acharya:2007rc,Acharya:2008zi,Acharya:2008hi}, we choose to explore the probability distribution of mass matrix eigenvalues and axion decay constants in the context of this framework. For technical details, see Appendix~\ref{sec:mthethe2}.

To formulate the structure of $\mathcal{K}_{ij}$ and $\mathcal{M}_{ij}$ we begin with a continuation of discussion in Appendix~\ref{sec:mthethe2}, starting with an expansion up to quadratic order of the superpotential given in Eq.~(\ref{eq:mtheorysuper}) which gives the mass terms with the following mass matrix,
\begin{align}
\mathcal{M}_{ij} =& \sum_{k=1}^{n_{\rm ax}} \sum_{r=1}^N \frac{4F \widetilde{\Lambda}_r^3 b_r N_r^k}{M_{S}^3} e^{- b_r \sum_m^{n_{\rm ax}} N_r^m s_m} b_r N_r^i b_r N_r^j \\
=& \sum_{r=1}^N \frac{4 F \widetilde{\Lambda}_r^3 C_r}{M_{S}^3} e^{- S_r} \widetilde{N}_r^i \widetilde{N}_r^j \,,
\end{align}
where $\widetilde{N}_i^j = b_i N_i^j$ is a rectangular matrix of size $(n_{\rm ax}, N)$, $C_r = \sum_k ^{n_{\rm ax}}\widetilde{N}^k_r$ and $S_r = \sum_m^{n_{\rm ax}} \widetilde{N}_r^m s_m$. The dimensions of the $\widetilde{N}_i^j$ are controlled by the axion population size, $n_{\rm ax}$ and the number of instantons, N. This expression allows us to parametrise the mass matrix term as the product of two rectangular matrices,

\beq
\mathcal{M}_{ij} = {1\over N} A_{ir}A_{jr} \label{eq:AxiverseM1}\,.
\eeq
This leaves us with the following form for the sub-matrix, 
\beq
A_{ir} = \left(2 \sqrt{\frac{F \widetilde{\Lambda}_r^3 C_r}{M_{S}^3}} \right) e^{- S_r/2} \widetilde{N}_r^i \, , \label{eq:AxiverseM2}
\eeq
where $i,j = 1,\ldots,n_{\rm ax}$ and $r = 1,\ldots,N$.
Note that $A_{ir}$ is a rectangular matrix of size $(n_{\rm ax}, N)$ where the normalisation factor ${\sfrac{1}{N}}$ is introduced to provide a consistent construction structure compared to the generalised form of the matrices we consider in our RMT models. Since $N>n_{\rm ax}$, this implies that the shape parameter, $\beta_\mathcal{M}$ should take values of $\beta_\mathcal{M} < 1$.

An analysis of the kinetic terms allows us to find the axion decay constants, $f_a$. In the moduli sector the K\"{a}hler potential takes the form,
\beq
K = -\ln (\mathcal{V}_X) \,,
\eeq
where $\mathcal{V}_X$ is a homogeneous function of the moduli $s_i$, of degree $\alpha$ depicting the volume of the hidden manifold in 11D Planck length. One important feature of the K\"{a}hler potential is that it leads to a non-trivial K\"{a}hler metric (which in this case is also the axion kinetic matrix) $\mathcal{K}_{ij} \equiv \frac{\partial^2 K}{\partial z_i \partial z_j}$ which is a homogeneous function of degree minus 2. We can assume the simplest form parametrising the non-trivial kinetic matrix is,
\beq
\mathcal{K}_{ij} = \frac{a_i a_j}{s_i s_j} \,, \label{eq:AxiverseK}
\eeq
where $a_i$ are constants and $s_i$ represent the moduli fields.
However, a generic matrix usually contains negative eigenvalues.
To avoid such an issue, we will allow for the further simplification the kinetic matrix such that $\mathcal{K}_{ij}$ is diagonal,
\beq
\mathcal{K}={\rm diag}[(a/s)] \, .
\label{eq:mkmatrix}
\eeq
It has been shown that such a form for the kinetic matrix can relieve tensions arising from dark radiation constraints in string axiverse models~\cite{Acharya:2015zfk}. 
\begin{figure}
\includegraphics[width=0.48\textwidth]{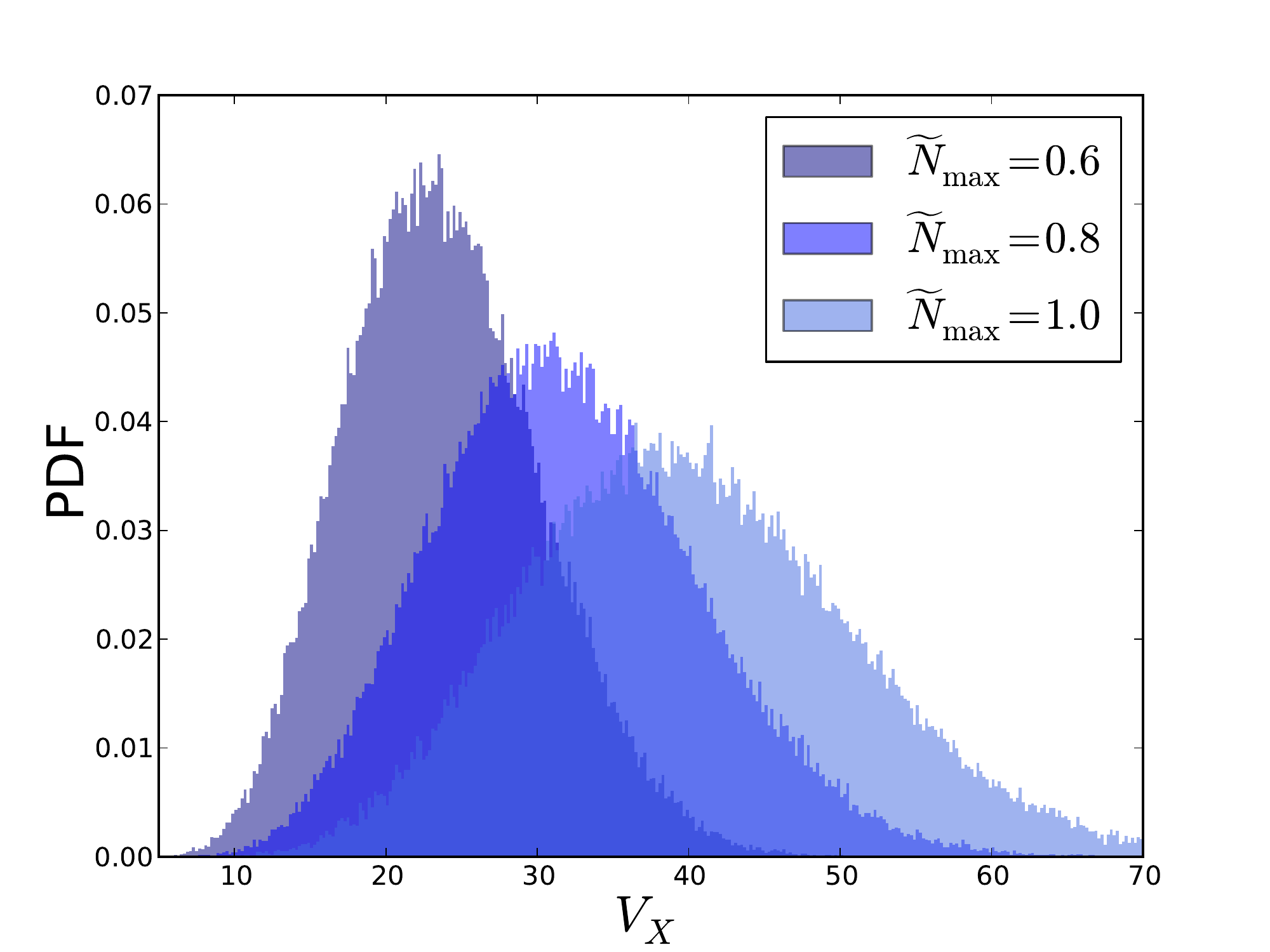}
 \caption[]
{{\bf M-theory RMT model 3-cycle volume distribution spectra:} Probability density plots for the 3-cycle volume using $\widetilde{N}_{\rm max} = 0.6, 0.8, 1.0$ with an axion population size, $n_{\rm ax}=10$. The moduli vev is uniformly distributed between 10 to 100 in units of the string scale, $P(s_i) = \mathcal{U}(10,100)$. The probability density of retrieving the GUT-value, $V_X = 25$, is found to be enhanced for values of $\widetilde{N}_{\rm max} \approx 0.6$.}\label{Mth-vol}
\end{figure}
For convenience we introduce a rescaling of the parameters so that all the physical parameters we consider are dimensionless:
\begin{align}
F \rightarrow& \ F/M_H^2 \label{eq:fdim}\,,\\
\widetilde{\Lambda}_i \rightarrow& \ \widetilde{\Lambda}_i/M_{S} \label{eq:lamdim}\,, \\
\mathcal{M}_{ij} \rightarrow& \ \mathcal{M}_{ij}/M_H^2\,, \\
\end{align}
such that:
\beq
\mathcal{M}_{ij} = \sum_{r=1}^N 4 F \widetilde{\Lambda}_r^3 C_r e^{- S_r} \tilde{N}_r^i \tilde{N}_r^j\,, \label{eq:mmassmatrix}
\eeq
where we note that the moduli and axion fields are expressed with respect to the string scale.

\begin{figure*}
\centering
\begin{tabular}{cc}
    \includegraphics[width=0.49\linewidth]{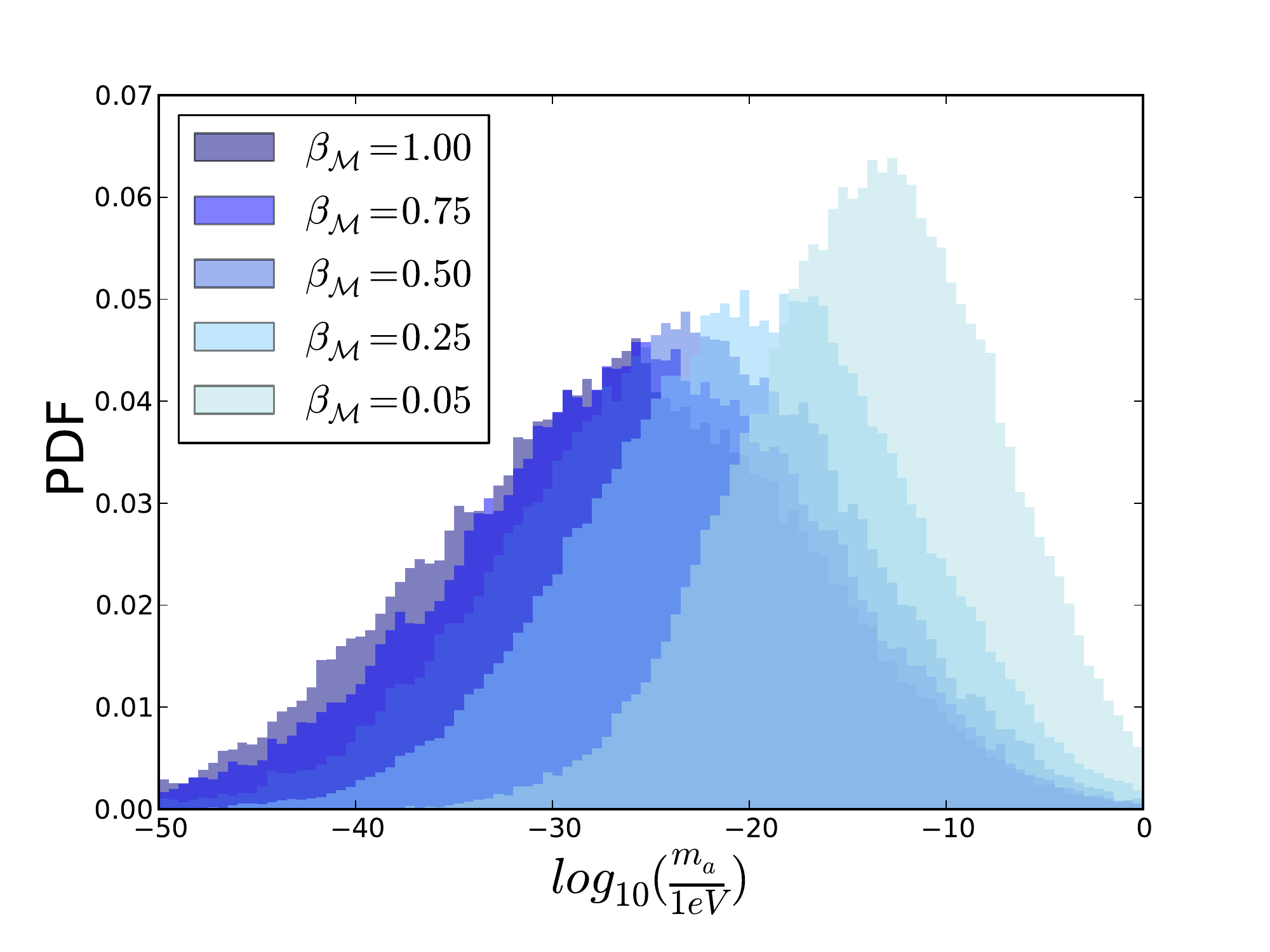}&
    \includegraphics[width=0.49\linewidth]{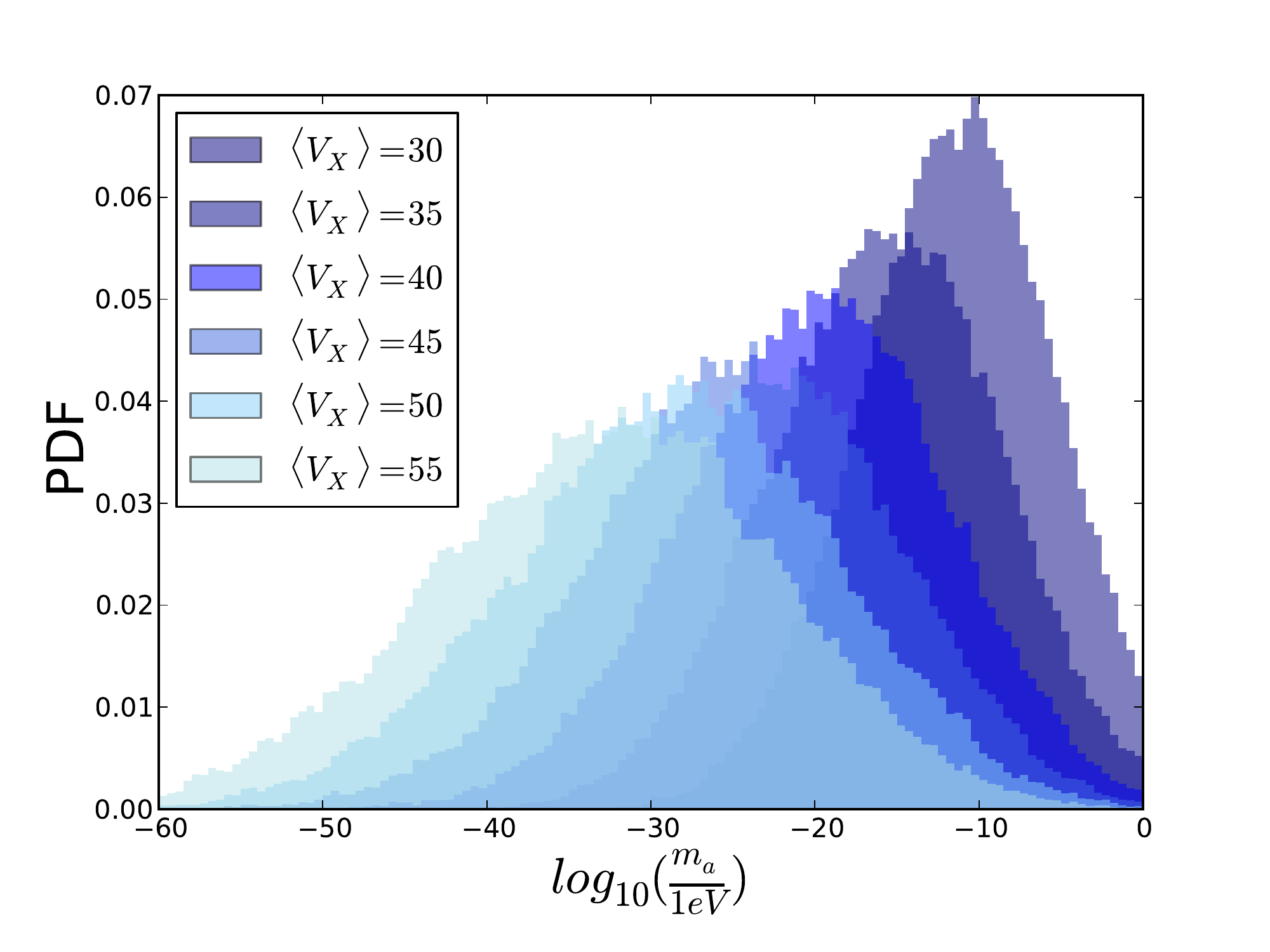}\\
\end{tabular}
\caption{{\bf M-theory RMT model mass spectra:} \emph{Left panel}: Probability density plots for axion masses using the fixed value $\langle V_X \rangle = 25$ for $\beta_\mathcal{M} = 1.00, 0.75, 0.50, 0.25, 0.05$. \emph{Right panel}: Probability density plots for axion masses with fixed $\beta_M = 0.5$ for $\langle V_X \rangle = 30, 35, 40, 45, 50, 55$. Both panels are constructed using 10000 iterations in the case of an axion population size, $n_{\rm ax}=10$. }
\label{fig:Mtmass}
\end{figure*}

The results of moduli stabilisation in M-theory show that the moduli vacuum expectation value should range between values of $\sim(10 \rightarrow 100)$ in units of the string scale \cite{Acharya:2007rc,Acharya:2008zi,Acharya:2008hi}. It is then natural to assume that our choice of prior should be a uniform distribution where,
\beq
\label{eq:modvev}
P(s_i) = \mathcal{U}(s_{\rm min},s_{\rm max})\,,
\eeq
with $s_{\rm min}\approx 10$, $s_{\rm max}\approx 100$. We also explore the values of the moduli vevs using a Gaussian distribution in some of our example cosmologies in Section~\ref{sec:mt}:
\beq
\label{eq:modvev2}
P(s_i) = \mathcal{N}(\bar{s},\sigma_s)\,.
\eeq

There is no assumption made on the topology of the manifold such that the K\"{a}hler metric parameters are fixed to,
\beq
\label{eq:mthea}
a_i = 1\,.
\eeq
The axion decay constants are then distributed between, 
\begin{equation}
f_{a,i} = \sqrt{2}a_i/s_i \sim (10^{-2} - 10^{-1})\,.	
\end{equation}
The shape of the M-theory axion decay constant spectrum using arbitrary limits of the moduli vev distribution in Eq.~(\ref{eq:modvev}) is shown in Fig.~\ref{decayf}.
The volume of the corresponding 3-cycles is calculated from,
\beq
V_{X}^i = \text{Im}(F_i) = \sum_{k=1}^{n_{\rm ax}} N_{i}^k s_k = \frac{1}{2\pi} \sum_{k=1}^{n_{\rm ax}} \widetilde{N}_{i}^k s_k \,,  
\eeq
where $F_i$ are the gauge kinetic functions (see Appendix~\ref{sec:mthethe2}) and in the final step we make the assumption that the membrane instanton integers are equal to unity ($b_i = 2\pi$). Since we are considering M-theory models which are Grand Unified Theories (GUTs) in their low energy limits, at least one of the gauge kinetic functions must give rise to the expected value of the Grand Unified coupling constant $\alpha_{\rm GUT} = 1/V_X \approx 1/25$. 

The distribution of $\widetilde{N}_i^k$ is uniform from 0 to $\widetilde{N}_{\rm max}$ such that,
\beq
\label{eq:nmax}
P(\widetilde{N}_i^k) = \mathcal{U}(0,\widetilde{N}_{\rm max})\,.
\eeq
For some of our example cosmologies in Section~\ref{sec:mt} the values of $\widetilde{N}_i^k$ are sampled using a Gaussian distribution:
\beq
P(\widetilde{N}_i^k) = \mathcal{N}(\bar{N},\sigma_N)\,.
\eeq
In Fig.~\ref{Mth-vol} we show the enhanced probability density for retrieving values of $V_{X}\approx 25$ when using $\widetilde{N}_{\rm max} \approx 0.6$. Increasing the value of $\widetilde{N}_{\rm max}$ serves to increase the spread of the distributions for $V_{X}$ at values centred around $V_{X} > 25$ which are too high for GUT coupling constant unification. Due to the uniform nature of the distributions, we can chose to parametrise the axion mass distribution using the average value of 3-cycle volume distribution $\langle V_X \rangle$ instead of $\widetilde{N}_{\rm max}$ as they are related by,
\beq
 \langle V_X \rangle = \frac{n_{\rm ax} \widetilde{N}_{\rm max} \langle s \rangle}{4\pi}\,.
\eeq 
The values of the other mass scales and coefficients coming from the form of the mass matrix defined in Eq.~(\ref{eq:mmassmatrix}) are taken as the following values:
\begin{align}
\label{eq:mtheoryscale1}
\widetilde{\Lambda}_i &= \Lambda =\mathcal{O}(1) \, ,\\
F &= 5.4\times 10^{104} \left(\frac{m_{3/2}}{1\text{ TeV}}\right)\,,
\label{eq:mtheoryscale2}
\end{align}
where the large value of $F$ is imposed by our choice of units. The mass scales in the mass matrix, $\mathcal{M}_{ij}$ are measured in units of $M_H$ and the scale of the quantities which give the value of $F$ come naturally from a SUSY/high energy physics/string theory perspective. These choices are made to account for the fact that non-perturbative scales are expected to show up around the string scale. The SUSY breaking order parameter is approximated using $m_{3/2}M_{Pl}/M_H^2$ where the gravitino mass is assumed to be of order 1 TeV from naturalness arguments. In practice, we will use a single scale parameter, $F\Lambda^3\sim\mathcal{O}(10^{105})$, which we allow to vary in our MCMC analysis.

In each panel in Fig.~\ref{fig:Mtmass} we construct the probability density plots for the axion mass spectrum using $10000$ points in the parameter space. The left-hand panel of Fig.~\ref{fig:Mtmass} shows the effect of varying $\beta_{\mathcal{M}}$ for fixed values of $\langle V_{X} \rangle$. As $\beta_{\mathcal{M}} \rightarrow 0$ it shifts to mass spectrum to be centred around higher mass scales whilst also decreasing the spread of the masses. In the right-hand panel of Fig.~\ref{fig:Mtmass} we show the expected result that larger values for the average volume lead to the axion masses centred about smaller values with a wider spread. For both of these configurations we see axion masses covering many orders of magnitude, which is a key result common to many string axiverse models. 



\section{Results I : Dark Sector Cosmologies} \label{sec:results1}

We define two example cosmologies via contributions to the total energy density at the present time: 
\begin{itemize}
\item \textbf{\textit {Dark matter cosmology}} - We will refer to the effective dark matter density as $\Omega_{\rm DM}$ coming from a population of axions. The total matter density parameter is therefore $\Omega_{m} = \Omega_{b} + \Omega_{\rm DM}$ where we decompose the total density into four components $\Omega = \Omega_{b}+\Omega_{\rm DM} + \Omega_{\Lambda}+\Omega_{r}$.  We initially look for values of $\Omega_{\rm DM}$ falling in the the very rough bounds, $0.2\leq\Omega_{\rm DM}\leq0.4$ in our example cosmologies with proper constraints addressed later. 
 \item \textbf{\textit  {Dark energy cosmology}} - We will refer to the effective dark energy density as $\Omega_{\rm DE}$ coming from a population of axions. We set $\Omega_{\Lambda}=0$ where we decompose the total density into three components $\Omega = \Omega_{\rm DE} + \Omega_{m}+\Omega_{r}$. We initially look for values of $\Omega_{\rm DE}$ falling in the the very rough bounds $0.6\leq \Omega_{\rm DE}\leq0.8$.       
\end{itemize}
  
We define the rough limits of the axion masses we require for each cosmology as the following. If axions are to account for the total dark matter, axion field oscillations should roughly begin in the radiation dominated era. This requires at least one axion with a mass larger than the Hubble rate at matter-radiation equality which defines the mass limit,
\begin{equation}
 m_a \gtrsim 10^{-27} {\rm eV}\,.
 \label{Eq:dmlimit}
\end{equation}
The energy-density of fields above this limit scales just as non-relativistic matter throughout the matter dominated era, fixing them as plausible dark matter candidates. Axions behaving as dark energy are limited to masses defined by the upper mass bound, 
\begin{equation}
 m_a \lesssim 10^{-32} {\rm eV}\,,
 \label{Eq:delimit}
\end{equation}
 as motivated by Ref.~\cite{Hlozek:2014lca}. 
 
 Our example figures in Sections.~\ref{sec:mpex} to \ref{sec:logflatex} contain data for 2500 example cosmologies. Our contour density plots are constructed using $50\times50$ gridded scans in multidimensional parameter space with gaussian filtering and cubic spline interpolation. The M-theory examples in Section~\ref{sec:mt} use $10\times10$ (Fig.~\ref{Mth-omegadm1} and Fig.~\ref{Mthcontourde}) and $20\times20$ (Fig.~\ref{Mth-omegadm2}) gridded scans with cubic spline interpolation, consisting of 10 samples at each point giving a total of 1000 and 4000 cosmologies respectively.    

 \begin{figure*}[t]
 \centering
\begin{tabular}{cc}
    \includegraphics[width=0.49\linewidth]{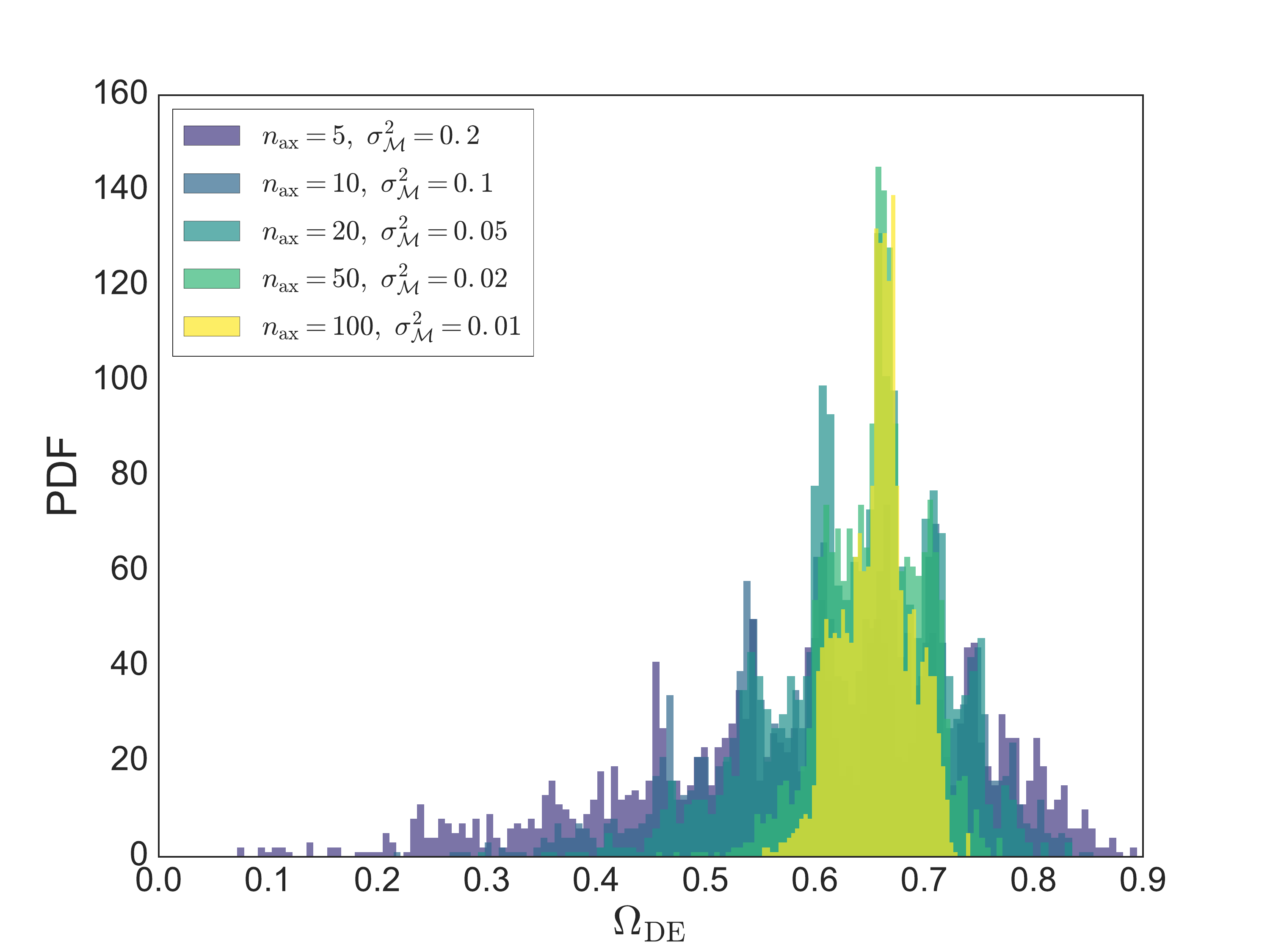}&
    \includegraphics[width=0.49\linewidth]{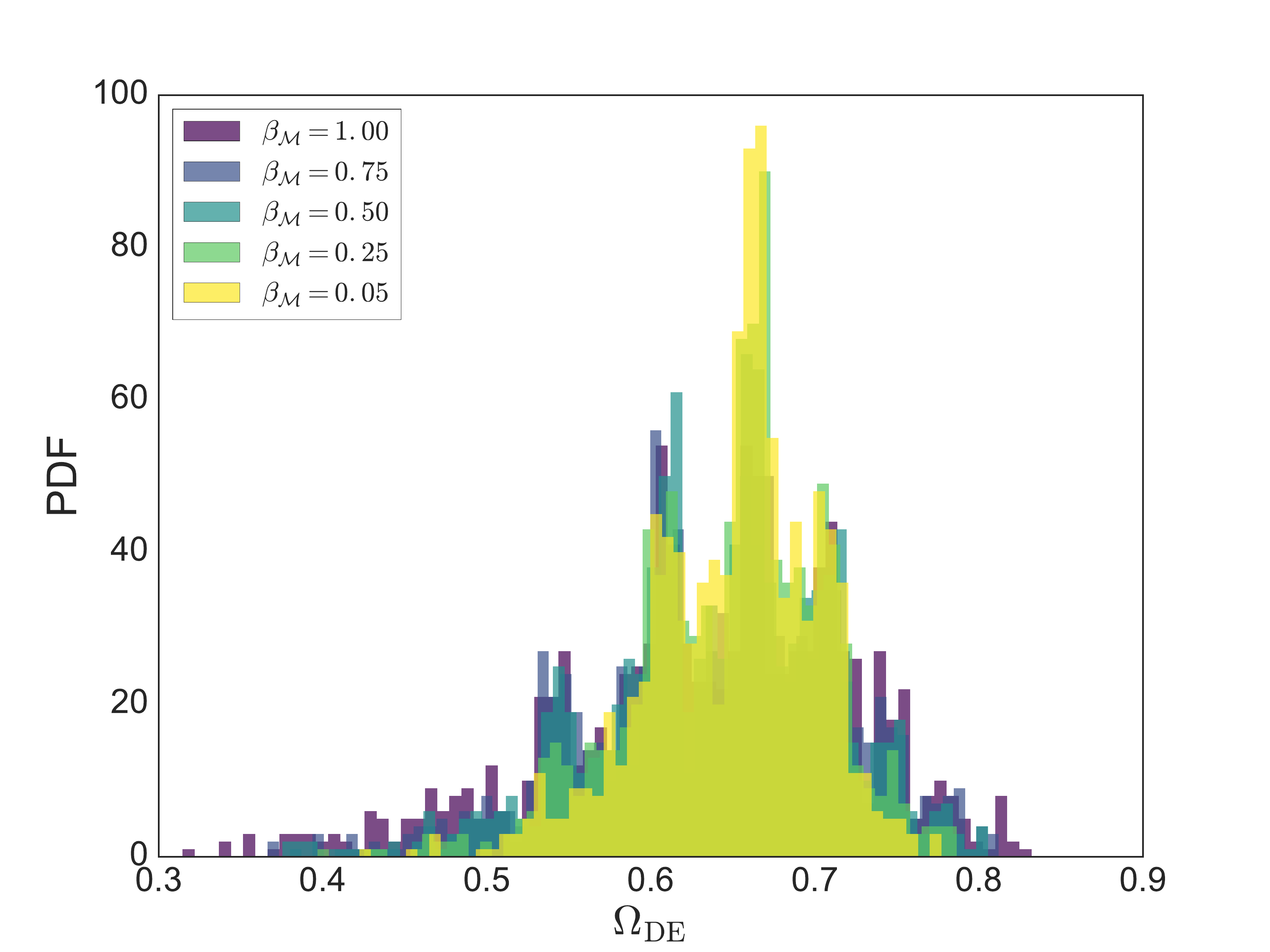}\\
\end{tabular}
\caption{{\bf Mar\v{c}enko-Pastur RMT model DE cosmology examples:}  \emph{Left panel}: Probability density plots for $n_{\rm ax}=\mathcal{O}(1) \rightarrow \mathcal{O}(100)$ with fixed values of $\sigma^2_{\mathcal{M}}$ according to the approximation in Eq.~(\ref{eq:nflatdewin}) with further fixed parameter values $\beta_{\mathcal{M}} = 0.5$ and $\bar{f}=1$. \emph{Right panel}: Approximate degeneracy for values of $\beta_{\mathcal{M}}\in (0,1]$ for the axion dark energy density parameter $\Omega_{\rm DE}$ using $n_{\rm ax}=20$ axions with fixed parameter values $\sigma^2_{\mathcal{M}} = 0.05$ and $\bar{f}=1$.}
\label{fig:nflatdeom}
\end{figure*}

\subsection{MP RMT Model}
\label{sec:mpex}

In Fig.~\ref{fig:nflatdeom} and Fig.~\ref{fig:mpdm1} we present our first example cosmologies in the simplest RMT model containing the smallest number of parameters to consider. The matrix eigenvalues have a bounded spectral width governed by the Mar\v{c}enko-Pastur distribution law. When fixing our mass spectrum shape with $\beta_{\mathcal{K},\mathcal{M}}=0.5$, this sets a configuration where each field provides approximately degenerate contributions to the total energy density up to variations in both the initial fields misalignment and random rotations from our choice of basis due to the absence of any treatment of $\mathcal{K}_{ij}$. The scale of the mass distribution defining the nature of the fields, fixed by $\sigma^2_{\mathcal{M}}$ acts as a free scaling parameter to switch between each type of cosmology. 
 
\subsubsection{MP-DM}
 \label{sec:mpexdm}

  \begin{figure}
  \vspace{0.5cm}
    \centering
    \includegraphics[width=0.49\textwidth]{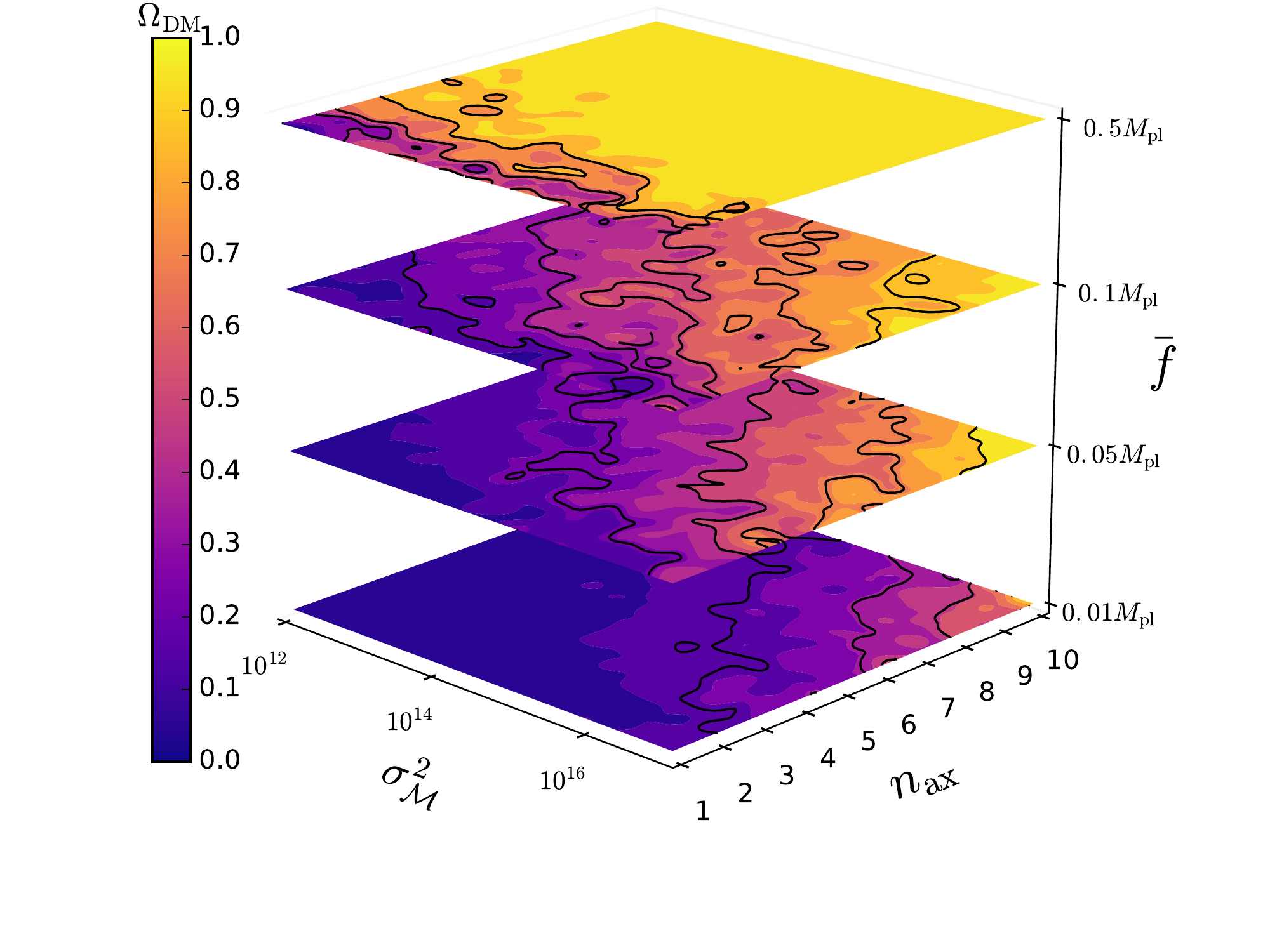}
    \caption[]{{\bf Mar\v{c}enko-Pastur RMT model DM cosmology example:} Contour density plots for the axion dark matter density parameter, $\Omega_{\rm DM}$ for $\sigma^2_{\mathcal{M}}$ and $n_{\rm ax} = \mathcal{O}(1\rightarrow 10)$ using different fixed values of the initial field displacement scaling, $\bar{f}$.}    
    \label{fig:mpdm1}     
\end{figure}

In Fig.~\ref{fig:mpdm1} we display contour density plots for different mass distribution scales against the axion population size at fixed values for the initial field condition scaling. We demonstrate the emergence of axion dark matter density domination at the present time with large initial field displacement scalings, $\bar{f}\approx M_{pl}$ for $n_{\rm ax} \gtrsim 1$. See Appendix~\ref{sec:rejec} for a visual example of the evolution of the cosmological densities in these configurations. 

In each of our RMT models the form of the mass matrix is such that a population of axions behaving as the total dark matter requires initial field oscillations onset at a scale where the requirements on the heaviest axion mass in the population set the order of the total mass scale, $\sigma^2_{\mathcal{M}} \gg M_H$. The equal field conditions, $\bar{f}$ along with the uniform sampling of $\theta$ restrict the total number of axions, $n_{\rm ax}$ allowed in the population at any given mass scale. Only when $n_{\rm ax} \approx 1$ do we recover the potential for values of $\Omega_{\rm DM}$ consistent with expectations presenting an approximate degeneracy along the total mass scale interval we consider. Larger population numbers feel both the linear sum of field density contributions along with the convergence of the initial misalignments in our prior sampling to their averaged value, $\langle \theta \rangle \approx \sfrac{\pi}{2}$, giving the large region of parameter space returning values of $\Omega_{\rm DM} \gtrsim 0.8$. 

A significant increase the potential for larger population sizes returning values of $0.2 \leq \Omega_{\rm DM} \leq 0.4$ is seen by relaxing the scaling of the initial field displacements to $\bar{f} = \mathcal{O}(0.1M_{pl})$ as demonstrated in the lower panels. The degeneracy relationship between the number of fields allowed in the population and the mass distribution scale becomes more apparent in the second and third panels. As expected larger values of $n_{\rm ax}$ quickly return values of $\Omega_{DM}$ far in excess of what is required as the mass distribution scale is increased. Our simple example highlights this when $\bar{f}=0.1M_{pl}$, mass distributions with $\sigma^2_{\mathcal{M}}  \approx 10^{12}$ require a population size, $n_{\rm ax} \approx 10$. Distributions with $\sigma^2_{\mathcal{M}} \approx 10^{17}$ require $n_{\rm ax} \approx 1$. The lower panels shift the preferred values of $\sigma^2_{\mathcal{M}}$ as we reduce the scaling for the initial field displacements.   

\begin{figure*}[t]
    \centering
    \includegraphics[width=1\textwidth]{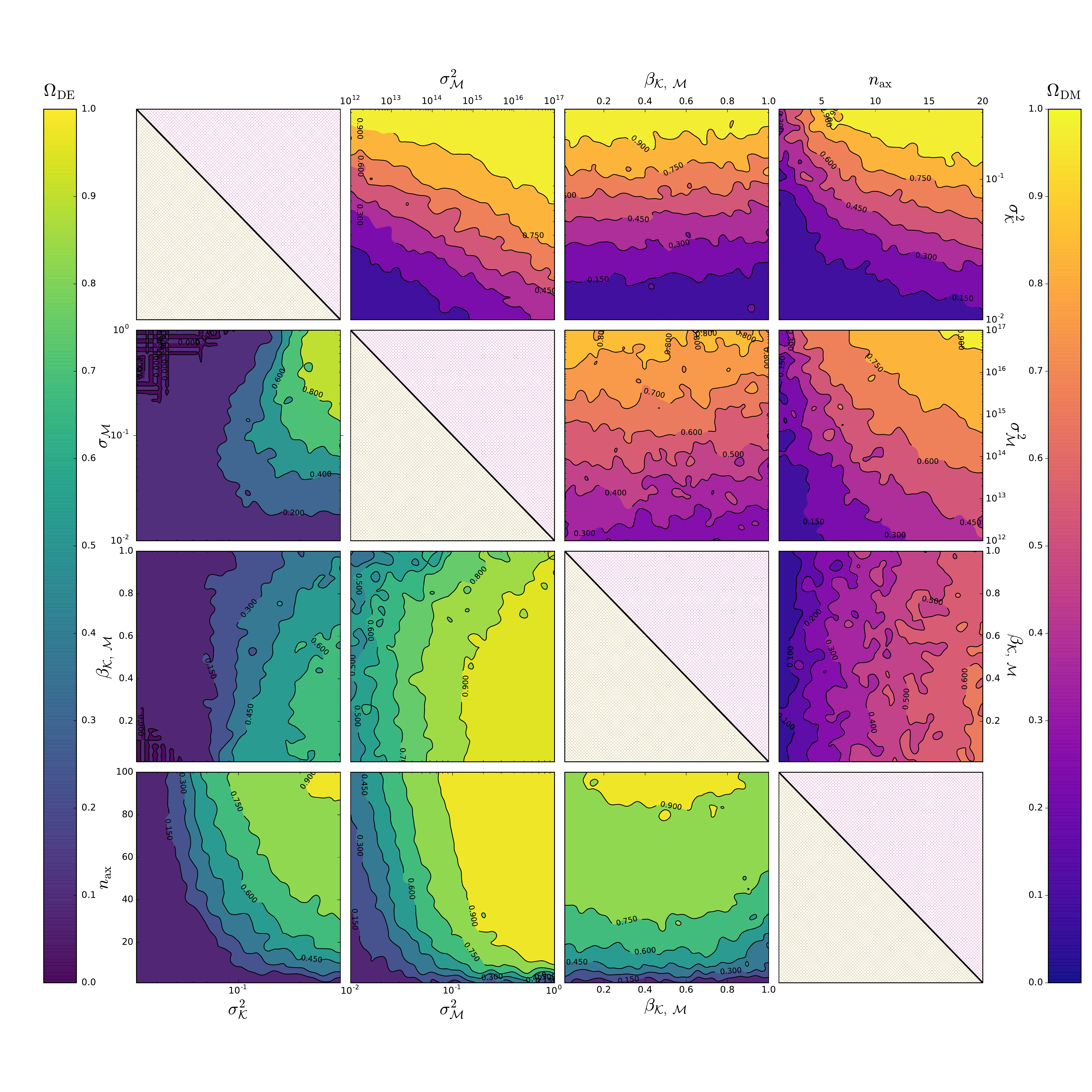}
    \caption[]{{\bf Wishart/Wishart RMT model DM and DE cosmology examples:} Contour density plots for two dimensional slices of the model parameter space for each parameter in the WW RMT model. \emph{Upper-triangle panels}: Example contours for excluded regions of parameter space for the axion dark matter density parameter, $\Omega_{\rm DM}$ using the intervals outlined in Eqs.~(\ref{eq:wwrun1})-(\ref{eq:wwrun2}) along with fixed values in Eqs.~(\ref{eq:wwfix1})-(\ref{eq:wwfix2}). \emph{Lower-triangle panels}: Example contours for excluded regions of parameter space for the axion dark energy density parameter, $\Omega_{\rm DE}$ using the intervals outlined in Eqs.~(\ref{eq:wwrun3})-(\ref{eq:wwrun4}) along with fixed values in Eqs.~(\ref{eq:wwfix3})-(\ref{eq:wwfix4}).}    
    \label{fig:wwexamples}     
\end{figure*}

\subsubsection{MP-DE}
\label{sec:mpexde}

It is easy to find parameters of the MP model that give rise to DE as the requirements are simple. Our MP-DE cosmologies begin with the approximation that the mass scale at which axion field oscillation begins follow the simple limiting constraint, $\langle m^2_a \rangle \lesssim M_H^2$. We maximise the range of the initial field conditions by fixing $\bar{f} = M_{pl}$ as well as fixing the shape of the distribution with $\beta_{\mathcal{M}}=0.5$. When searching for a population of non-oscillating fields we approximate the value of $\sigma^2_{\mathcal{M}}$ for a significant number of low mass axions driving a phase of acceleration using, 
\begin{equation}
\sigma^2_{\mathcal{M}} \approx \frac{\sigma^2_{M_H}}{n_{\rm ax}} \approx \frac{1}{(5\rightarrow 
100)} \approx 0.2 \rightarrow 0.01\, ,
\label{eq:nflatdewin}
\end{equation}
In the left hand panel of Fig.~\ref{fig:nflatdeom} we display the probability densities for, $n_{\rm ax} = \mathcal{O}(1 \rightarrow 100)$ for corresponding values of $\sigma^2_\mathcal{M}$ determined by Eq.~(\ref{eq:nflatdewin}). Seemingly larger values of $n_{\rm ax}$ tailor the potential for desirable values of $\Omega_{\rm DE}$ by reducing the spread. A population size of $n_{\rm am} = 100$ returns a high probability density of cosmologies with values of $\Omega_{\rm DE}$ contained in the window of interest. As $n_{\rm ax} \rightarrow \mathcal{O}(100)$ the initial field misalignments in the population will converge to their averaged value $\langle \theta \rangle$ where the linear combination of the field density contributions cause the probability density of the dark energy density parameter to converge towards the modal value. Decreasing the value of $n_{\rm ax}$ increases the chance of returning cosmologies failing the acceleration criterion, $\ddot{a}>0$ at $z=0$ used in Section~\ref{sec:mcmcresults}.   
 
Using the relationship in Eq.~(\ref{eq:nflatdewin}) we address the role of the final parameter in this model, $\beta_{\mathcal{M}}$. The right-hand panel of Fig.~\ref{fig:nflatdeom} shows the spread of $\Omega_{\rm DE}$ values for fixed values of $\beta_{\mathcal{M}}$, distributed about  $\sfrac{\sigma^2_{M_H}}{n_{\rm ax}}=\sfrac{1}{20} $. We highlight the approximate degeneracy across our five fixed values of $\beta_{\mathcal{M}}$. Given the statistical sampling of $\beta_{\mathcal{M}}$ with either a uniform distribution or Gaussian sampling as shown in Appendix~\ref{sec:axmod}, only extremal values will induce limited variations to the spread of $\Omega_{\rm DE}$ as compared to $\beta_{\mathcal{M}} =0.5$ with each value retaining a a mean value of $\Omega_{\rm DE} \approx 0.65$.

   \begin{figure*}[t]
   \hspace*{-1.5cm}
 \centering
    \includegraphics[width=1.05\linewidth]{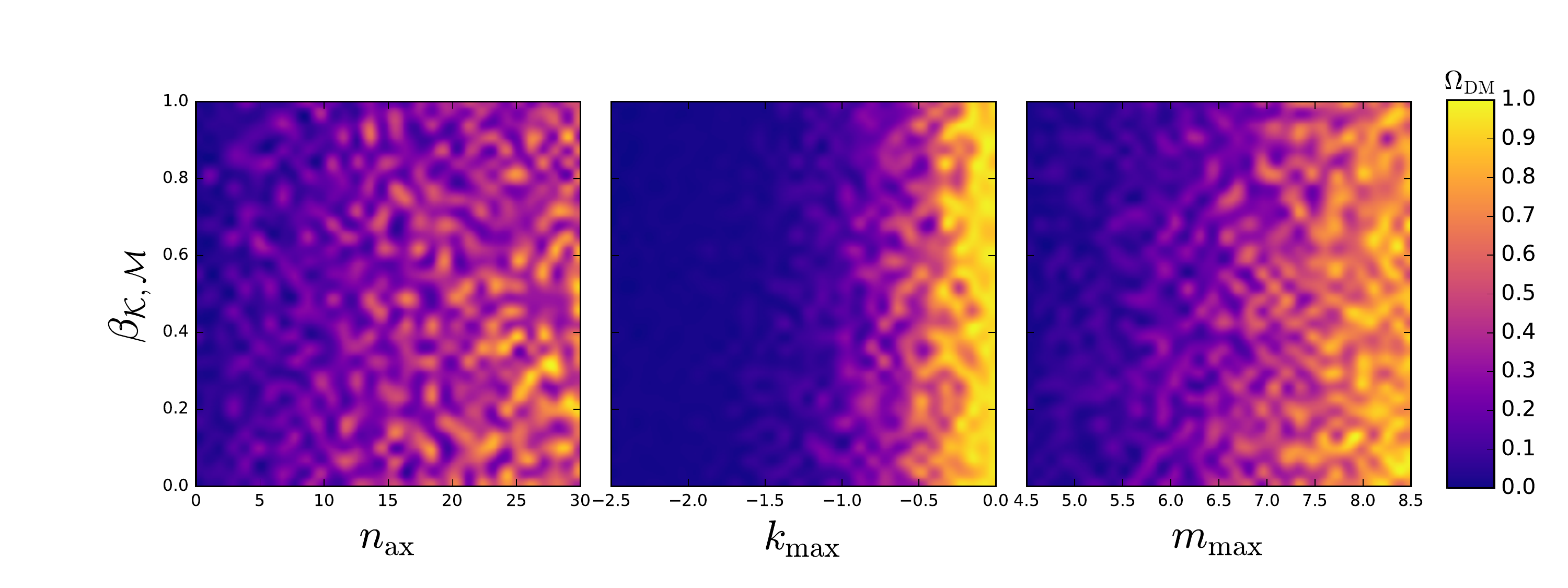}
\caption{{\bf Log-Flat/Log-Flat RMT model DM cosmology example:} Density heat maps for the axion dark matter density parameter, $\Omega_{\rm DM}$ for values of $\beta_{\mathcal{K},\mathcal{M}} \in (0,1]$ along with the remaining model parameters. We use $n_{\rm ax}= [1-30]$ axions and varied limits on both the decay constant spectra parameterised by $k_{\rm max}$ and the mass spectra parameterised by $m_{\rm max}$.}
\label{fig:lfdm}
\end{figure*}

\subsection{WW RMT Model}
\label{sec:wwex}

In Fig.~\ref{fig:wwexamples} we display contour density plots for intervals of two dimensional parameter space for each parameter in the WW RMT model. 

\subsubsection{WW-DM}
\label{sec:wwexdm}

The parameters in this model which we allow to run are scanned over the following intervals, 
\begin{align}
\log_{10}(\sigma^2_\mathcal{K}) &\in [-4.0,-1.0]\, , \label{eq:wwrun1} \\
\log_{10}(\sigma^2_\mathcal{M}) &\in [12.00,17.0]\, , \\ 
\beta_{\mathcal{K},\mathcal{M}} &\in [0.01,1.0]\, , \label{eq:wwrun2}\\
n_{\rm ax} &\in [1,20] \, ,
\end{align}
where we use the following values if parameters remain fixed,
\begin{align}
\log_{10}(\sigma^2_\mathcal{K}) &= -2.60 \, , \label{eq:wwfix1}\\
\log_{10}(\sigma^2_\mathcal{M}) &= 5.70\, , \\ 
\beta_{\mathcal{K},\mathcal{M}} &=0.5\, ,\\
n_{\rm ax} &= 20 \,\label{eq:wwfix2} .	
\end{align}

In the top row of panels we show the banding of dark matter density whilst increasing the distribution scale of our kinetic matrix, $\sigma^2_{\mathcal{K}}$. As seen in the upper left panel the probability density for axion dark matter domination widens as the distribution scale of the initial mass matrix, $\sigma^2_{\mathcal{M}}$ leaves the lower dark matter mass limit. Indeed it is expected that the limited spectral width of the matrix spectra in these models is such that we should not expect large amounts of freedom to reposition ourselves in the parameter space before traversing into the bounds of the contours with non-desirable quantities of dark matter. The limited width of the \emph{purple} and \emph{mauve} bands indicate the freedom we have to centre the decay constant spectra at fixed mass scales. The gradient of the bands corresponds to the notion that in general one would expect far-in excess the quantities of dark matter required when considering axion populations at the mass scale limit detailed in Eq.~(\ref{eqn:strict_mass}), unless we compensate the distribution scales for $\mathcal{K}_{ij}$. Indeed we would expect, sub-GUT scales for our kinetic matrix distributions in this model when addressing a significant population size, $n_{\rm ax}$.
  
The convergence of the contour bands to values of $\Omega_{\rm DM}\lesssim 0.9$ is shown in the upper right panel for a spectrum of high scale decay constants when $n_{\rm ax} \lesssim 5$. Correspondingly the panel below details the convergence in the same regard as the mass matrix scale increases. The bands widen when considering a larger number of fields $n_{\rm ax} \approx \mathcal{O}(10)$, at lower mass scales in the approximate regions (\emph{purple} and \emph{mauve}) for fixed $\sigma^2_{\mathcal{K}}$. Likewise at lower values of $\sigma^2_{\mathcal{K}}$ we see a widening when $n_{\rm ax} \approx \mathcal{O}(10)$. The simplicity of the matrix structure we use will provide very comparable results between the WW RMT and MP RMT models, with approximate comparisons to be drawn from the middle right hand panels of Fig.~\ref{fig:wwexamples} and the panel second from top in Fig.~\ref{fig:mpdm1}. Indeed it is expected the averaging of the field contributions with $n_{\rm ax} \gtrsim \mathcal{O}(10)$ will give comparable results when using the equal initial field conditions for the field vevs in Eq.(\ref{eq:mpvev}), given the bounded spectra for $f_a$ when partnered with the random rotations and sampling on the misalignments.

\subsubsection{WW-DE}
\label{sec:wwexde}

Our WW-DE examples reside in the lower triangle of panels in Fig.~\ref{fig:wwexamples}. Unlike this models dark matter counterpart the requirement for non-oscillating fields with the limiting upper mass bound in Eq.~(\ref{Eq:delimit}) at the approximate scale $\sigma^2_{\mathcal{M}} \approx M_H$ will be more susceptible to both the freedom in the distribution for $f_a$ and the shape of the rotated mass spectra. Our parameters allowed to run are scanned over the following intervals, 

\begin{align}
\log_{10}(\sigma^2_\mathcal{K}) &\in [-2.0,0.0] \,,\label{eq:wwrun3}\\
\log_{10}(\sigma^2_\mathcal{M}) &\in [-2.0,1.0]\,, \\ 
\beta_{\mathcal{K},\mathcal{M}} &\in [0.01,1.0]\,,\label{eq:wwrun4} \\
n_{\rm ax} &\in [1,100]\,, 
\end{align}
where if parameters remain fixed we use the following values,
\begin{align}
\log_{10}(\sigma^2_\mathcal{K}) &= -0.60\,,\label{eq:wwfix3} \\
\log_{10}(\sigma^2_\mathcal{M}) &= -1.65\,, \\ 
\beta_{\mathcal{K},\mathcal{M}} &=0.5\,, \\
n_{\rm ax} &= 20 \,\label{eq:wwfix4}.	
\end{align}

In the upper left, lower left and lower central panels we show the relationship between the population size and the scale of each of the distributions for the physical parameters. In general we do require scaling parameters of the order, $\sigma^2_{\mathcal{K}} \approx M_{pl}$ and $\sigma^2_{\mathcal{M}} \approx M_{H}$ (upper left panel) with the regions of parameter space with either $\sigma^2_{\mathcal{K}} \lesssim 0.1M_{pl}$ or $\sigma^2_{\mathcal{M}} \lesssim 0.1M_{H}$ quickly providing insufficient dark energy density unless the population size is increased to $n_{\rm ax} \rightarrow \mathcal{O}(100)$ (lower left and central panels).  

In the upper and left central panels show the preference for the incorporation of the full tail of the distributions corresponding to values of $\beta_{\mathcal{K},\mathcal{M}}\rightarrow 1$ as the defining scales of the distributions are increased. The reduction of the spectral width gives a degeneracy in the contours for values of $\beta_{\mathcal{K},\mathcal{M}} \lesssim 0.5$ which can be seen more prominently in the upper central panel following fixed values for $\sigma^2_{\mathcal{M}}$. The preferential defining shape of the sub-matrices is dependant on the distribution scales. In the left central panel we see the recovery of a full degeneracy across all values of $\beta_{\mathcal{K},\mathcal{M}}$ when the initial conditions for the fields are are at insufficient scales required for any form of axion dark energy presence at the current time. 

Finally in the lower right panel we show relationship between the shape of the distribution and the axion population size. Fixed population sizes give a degeneracy for values $\beta_{\mathcal{K},\mathcal{M}} \lesssim 0.5$. The contour curvature as $n_{\rm ax} \rightarrow \mathcal{O}(100)$ potentially corresponds to a spreading of the mass spectrum, increasing the probability density of lighter fields. It could also potentially correspond to the inclusion of heavier oscillating late time dark matter like fields at $z=0$ as the tails of the distributions are sampled for large $n_{\rm ax}$. 

\subsection{LF RMT Model}
\label{sec:logflatex}
 \begin{figure}
    \centering
    \includegraphics[width=0.49\textwidth]{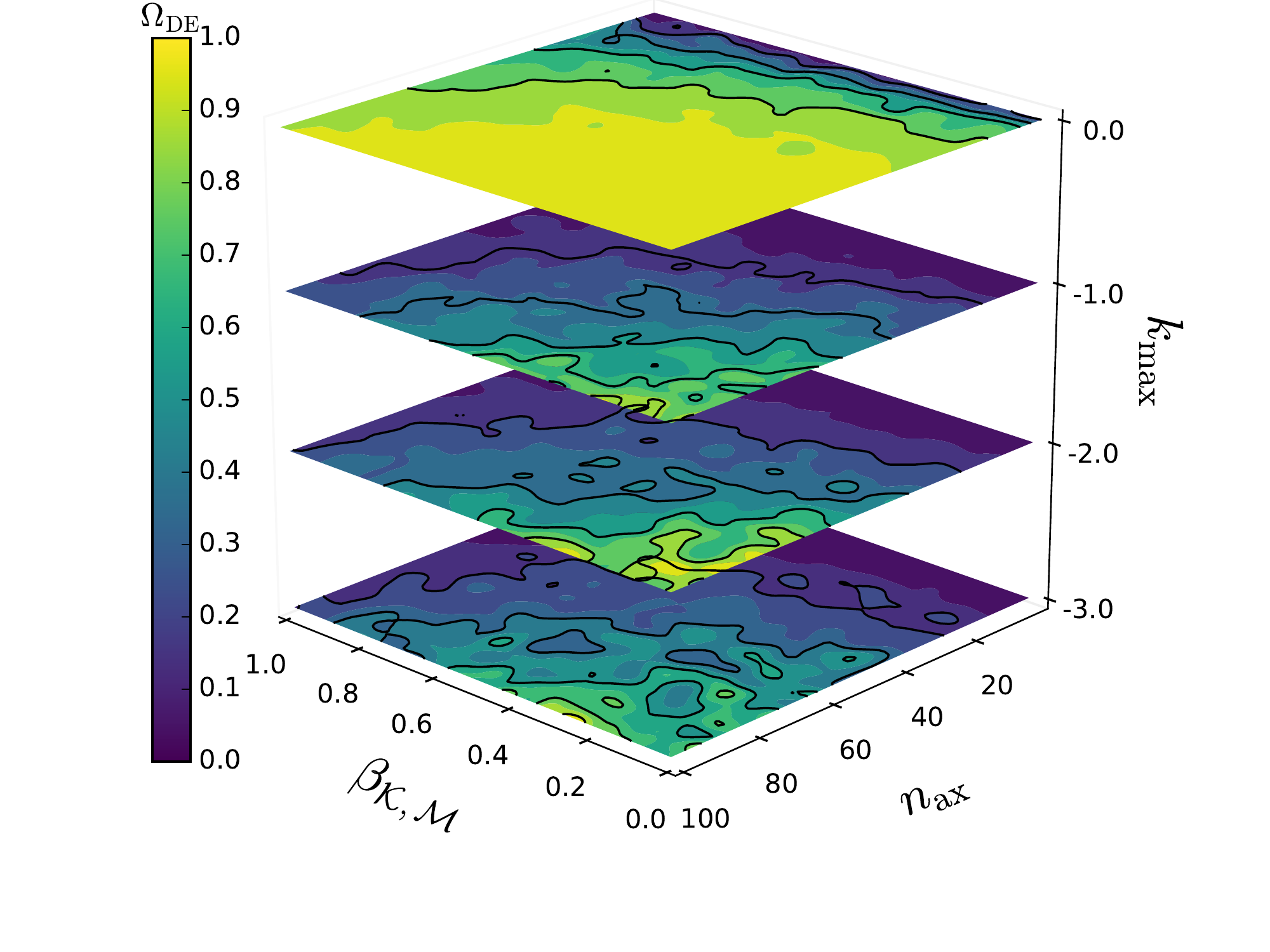}
    \caption[]{{\bf Log-Flat/Log-Flat RMT model DE cosmology example for \boldmath{$k_{\rm max}$} limits:} Contour density plots for the axion dark energy density parameter, $\Omega_{\rm DE}$ for $\beta_{\mathcal{K},\mathcal{M}}$ and $n_{\rm ax}=\mathcal{O}(1\rightarrow 100)$ for different fixed values of $k_{\rm max}$.}    
    \label{fig:lfdeex2}     
\end{figure}
The LF RMT model examples in this section investigate potential differences in our example cosmology outputs from the $\mathcal{O}(n)$ enhanced eigenvalues present in each of the spectra for our physical quantities when compared to the limited bulk spectra in the previous models. It is worth noting by construction our examples should see very little variation compared to the WW RMT models output due to the order of magnitude of our axion population number we select. We will leave the study of large population numbers where our largest eigenvalues could obtain significant enhancements in the form of both large singular decay constants and a widening of the spectral width of the mass distribution for future study. We choose to limit the number of parameters we consider in our examples in this model by fixing the values of our lower bounds on our distributions controlled by $k_{\rm min}$ and $m_{\rm min}$ throughout. The values of $\langle f_a \rangle$ and $\langle m_a \rangle$ are scaled by changing the values of $k_{\rm max}$ and $m_{\rm max}$ accordingly.

\subsubsection{LF-DM} 
\label{sec:lfdmex}

We are interested in the role of a spectrum of high scale decay constants in the low mass axion window for LF-DM, to explore the possible effects of the largest eigenvalues in both spectra. Our LF-DM parameter intervals are defined as, 
\begin{align}
k_{\rm max}  &\in [-2.5,0.0] \,,\\
m_{\rm max} &\in [4.5,8.5] \,,\\ 
\beta_{\mathcal{K},\mathcal{M}} &\in [0.01,1.0] \,,\label{eq:lfbeta} \\
n_{\rm ax} &\in [1,30] \,,
\end{align}
with the defined fixed values,
\begin{align}
k_{\rm min}  &= -5.0 \label{eq:lfde111}\,, \\ 
k_{\rm max} &= -1.0 \,,\\
m_{\rm min} &= 4.0 \label{eq:lfde3}\,,\\
m_{\rm max} &= 6.0\,, \\
n_{\rm ax} &= 20 \,,\\
\beta_{\mathcal{K},\mathcal{M}} &= 0.5\,. 	
\end{align}

The values in Eq.~(\ref{eq:lfde111}) in chosen to fix the lowest scale for $\langle f_a \rangle$ for the bulk of the spectrum to sub-GUT values when $k_{\rm max}$ is at its lowest value. The upper limit of $k_{\rm max}$ corresponds to the decay constant scale, $\langle f_a \rangle = \mathcal{O}(0.1M_{pl})$. Our lower limit on $m_{\rm min}$ in Eq.~(\ref{eq:lfde3}) is to ensure we have fields oscillating with masses $m_a>10^6 M_H$. The maximum fixed value of $m_{\rm max}$ corresponds to fields drawn about mass distribution centred around, $\langle m_a \rangle \approx \mathcal{O}(10^7M_H)$ with the upper limit $m_{\rm max}$ giving a mass distribution scale, $\langle m_a \rangle \approx \mathcal{O}(10^9M_H)$.   

Fig.~\ref{fig:lfdm} details regions of two-dimensional parameter space for each model parameter against values of $\beta_{\mathcal{K},\mathcal{M}}$ defined in the interval in Eq.~(\ref{eq:lfbeta}). In each of the panels we see reproduce the approximate degeneracy across all values of $\beta_{\mathcal{K},\mathcal{M}}$ reflecting the corresponding panels in Fig.~\ref{fig:wwexamples}. It is clear that the this model will offer little deviation from the previous model considerations for dark matter cosmologies given the low number of fields we are considering and the mass scales we are considering. In the middle panel we see the clustered heat density for large values of $\Omega_{\rm DM}$ as we scale the distribution for $f_a$ towards $M_{pl}$ once again indicating a preference away from values of $f_a \approx M_{pl}$. The left hand panel shows a measure of the potential, at fixed physical parameter scales, to find acceptable quantities of dark matter as the population size increase via the ``speckled'' nature of the probability densities.

\subsubsection{LF-DE}
\label{sec:lfdeex}

\begin{figure}
    \centering
    \includegraphics[width=0.49\textwidth]{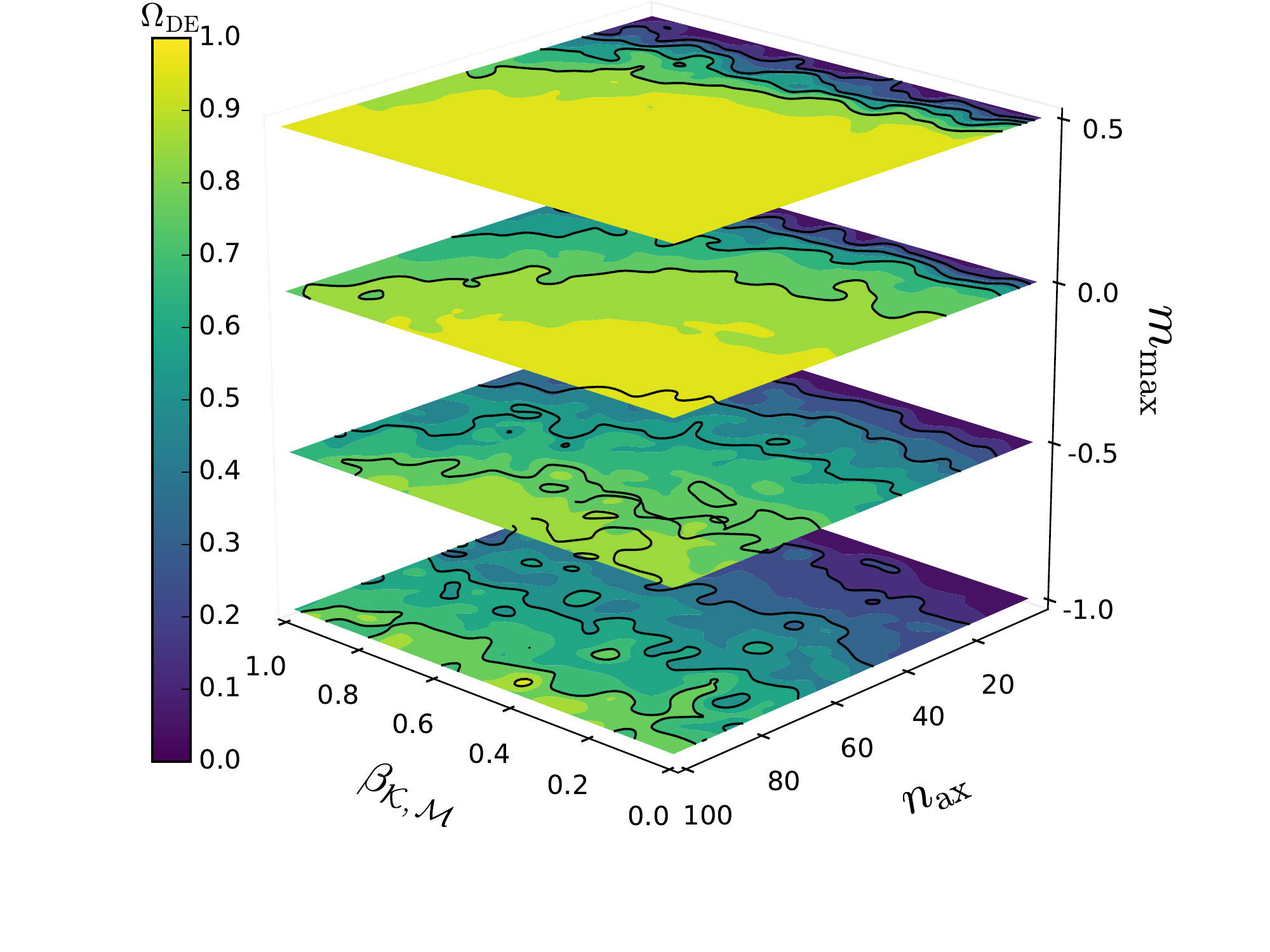}
    \caption[]{{\bf Log-Flat/Log-Flat RMT model DE cosmology example for \boldmath{$m_{\rm max}$} limits:} Contour density plots for the axion dark energy density parameter, $\Omega_{\rm DE}$ for $\beta_{\mathcal{K},\mathcal{M}}$ and $n_{\rm ax}=\mathcal{O}(1\rightarrow 100)$ for different fixed values of $m_{\rm max}$.}    
    \label{fig:lfdeex3}     
\end{figure}

In both Fig.~\ref{fig:lfdeex2} and Fig.~\ref{fig:lfdeex3} we introduce a small step into the three dimensional parameter space for $\Omega_{\rm DE}$ contour densities. We initially focus on the configuration where the scales of our dimensional quantities are determined by $m_{\rm min} = k_{\rm min}$ and $m_{\rm max} = k_{\rm max}$. This ensures that our rotated mass spectrum is centred about, $\langle m_a \rangle = M_H$ with a spectral width determined by the value we fix for $k_{\rm max}$. Our LF-DE parameters which we allow to run are scanned over the following intervals, 
\begin{align}
k_{\rm max}  &\in [-3.0,0.0]\,, \\
m_{\rm max} &\in [-1.0,0.5] \,,\\ 
\beta_{\mathcal{K},\mathcal{M}} &\in [0.01,1.0]\,, \\
n_{\rm ax} &\in [1,100]\,, 
\end{align}
with the following values of the fixed model parameters,
\begin{align}
k_{\rm min} &= m_{\rm min} = -5.0\,, \\
k_{\rm max} &= m_{\rm max} = 0.0 \label{eq:lfkmax}\,,\\
n_{\rm ax} &= 20 \,,\\
\beta_{\mathcal{K},\mathcal{M}} &= 0.5\,. 
\end{align}

Fig.~\ref{fig:lfdeex2} shows the contour densities for $\beta_{\mathcal{K},\mathcal{M}}$ against $n_{\rm ax}$ for stacked decay constant distribution scales, emphasising the previously determined preference for high scale decay constants for sufficient $\Omega_{\rm DE}$ when using mass centred distributions about $M_H$. Lower vales of $k_{\rm max}$ slowly recover the degeneracy across all values of $\beta_{\mathcal{K},\mathcal{M}}$ providing little dark energy density. For $k_{\rm max} = 0.0 $, as the population number $n_{\rm ax}$ increases significantly, a preference is made for the inclusion of the full tail of the mass spectrum as $\beta_{\mathcal{K},\mathcal{M}} \rightarrow 1$ maximising the spread of mass values fields can take. Values of $k_{\rm max}$ minimally offset from this value require $\beta_{\mathcal{K},\mathcal{M}} \rightarrow 0$ to ensure a large population of fields have approximately degenerate and sufficient mass values ($\approx M_H$), in order to furnish our cosmologies with a sufficient quantity of dark energy density at the current time.     
 
 Correspondingly Fig.~\ref{fig:lfdeex3} presents contour density plots for $\beta_{\mathcal{K},\mathcal{M}}$ against $n_{\rm ax}$ for stacked mass distribution scales offset with respect to the scale $\langle m_a \rangle \approx M_H$ fixed by $m_{\rm max}$. Each configuration uses a fixed distribution of high scale decay constants (Eq.~\ref{eq:lfkmax}). It is clear in the upper panel that distributions offset towards the upper mass limit in Eq.~(\ref{Eq:delimit}) quickly produce high probability densities for cosmologies with axion dark energy domination. Scales centred about $\langle m_a \rangle \approx M_H$ increase the width of the contour bands with acceptable values of $\Omega_{\rm DE}$ (\emph{green} and \emph{light green}). Large population sizes ($n_{\rm ax}\approx \mathcal{O}(100)$) at this scale make a preference a wider bulk in the mass distribution for values of $\beta_{\mathcal{K},\mathcal{M}} \rightarrow 1$, a feature consistent with the previous models behaviour. Mass scales offset below $M_H$ ($m_{\rm max}=-0.5$) give a preference for $\beta_{\mathcal{K},\mathcal{M}} \rightarrow 0$ whilst also requiring large population sizes. A further increase in the offset below the mass scale of $M_H$ recovers approximate degeneracies across all values of $\beta_{\mathcal{K},\mathcal{M}}$ with significantly reduced probability densities for the required values of $\Omega_{\rm DE}$.

\subsection{M-Theory RMT Model}
\label{sec:mt}

\begin{figure}
\includegraphics[width=0.5\textwidth]{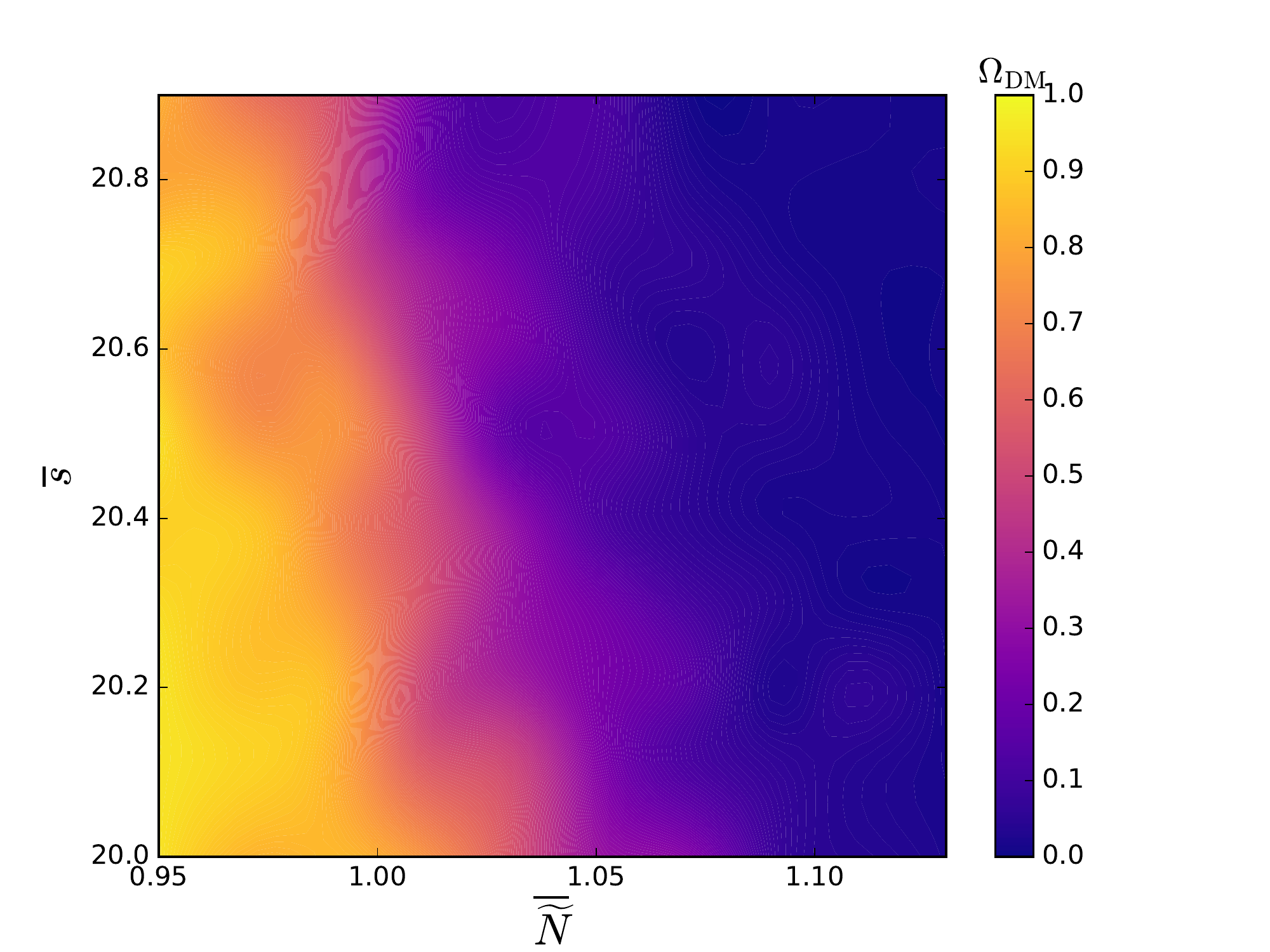}
\caption{{\bf M-theory RMT model DM cosmology example with narrow priors:} Contour density plot for the axion dark matter density parameter, $\Omega_{\rm DM}$ with narrow priors for the moduli vev, $s$ and instanton index parameter $\bar{\tilde{N}}$.}\label{Mth-omegadm1}
\end{figure}

\begin{figure}
\includegraphics[width=0.5\textwidth]{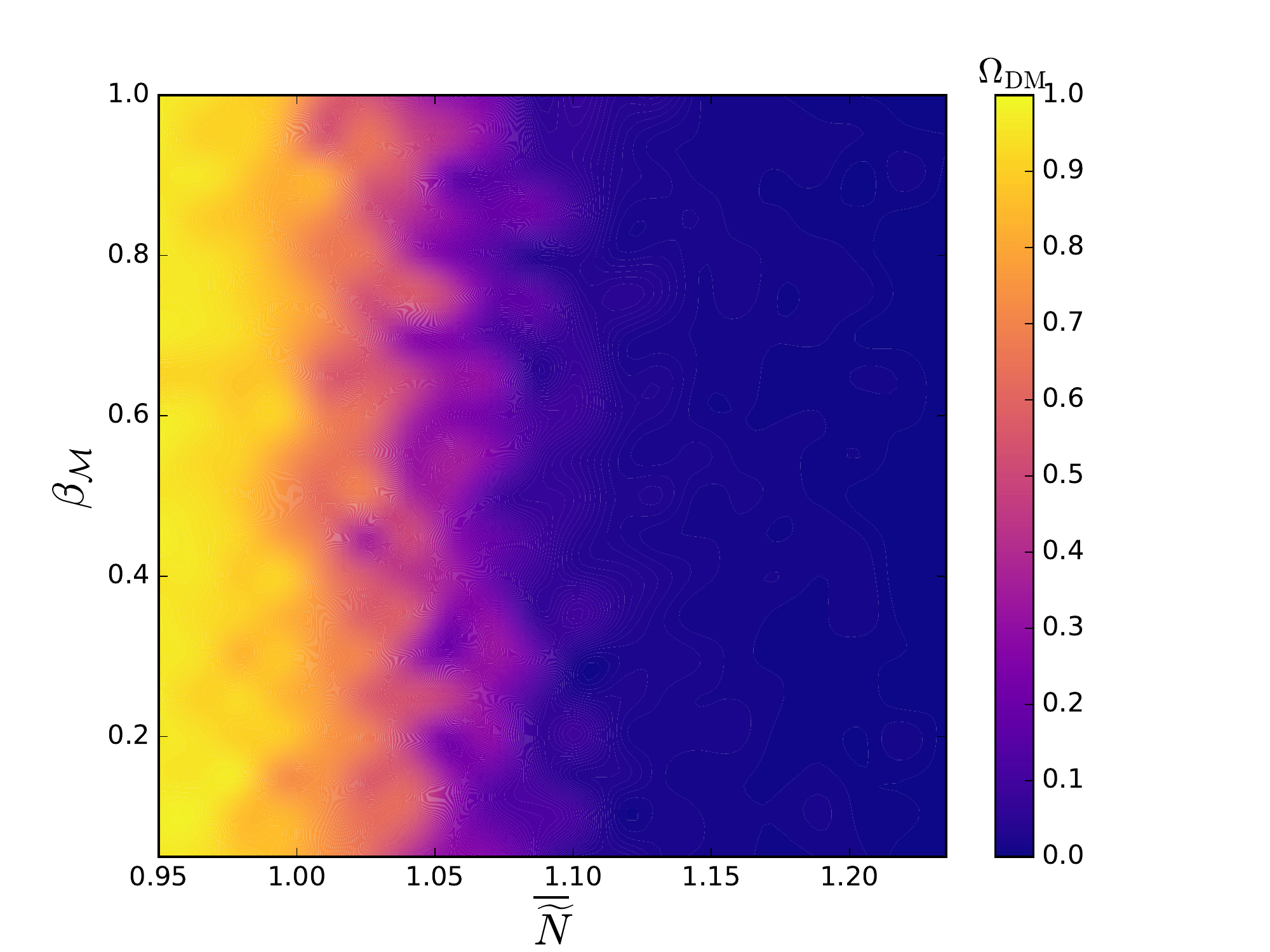}
\caption{{\bf M-theory RMT model DM cosmology example with narrow priors:} Contour density plots for the axion dark matter parameter, $\Omega_{\rm DM}$ for $\beta_{\mathcal{M}}$ and using narrower priors on the instanton index parameter $\bar{\tilde{N}}$.}\label{Mth-omegadm2}
\end{figure}

In this section we look at cosmologies returning the required values of $\Omega_{\rm DM}$ and $\Omega_{\rm DE}$ drawn from the M-theory landscape where we fix the number of fields in our examples to $n_{\rm ax} = 10$ throughout. Our choice of initial scales we use consist of the values given in Eq.~(\ref{eq:mtheoryscale1}) and Eq.~(\ref{eq:mtheoryscale2}).  In order to account for gauge couplings consistent with the known elementary particles we chose to sample the average values for the 3-cycle volume in the interval $\langle V_X \rangle = [25,60]$. 
\begin{figure}
\includegraphics[width=0.5\textwidth]{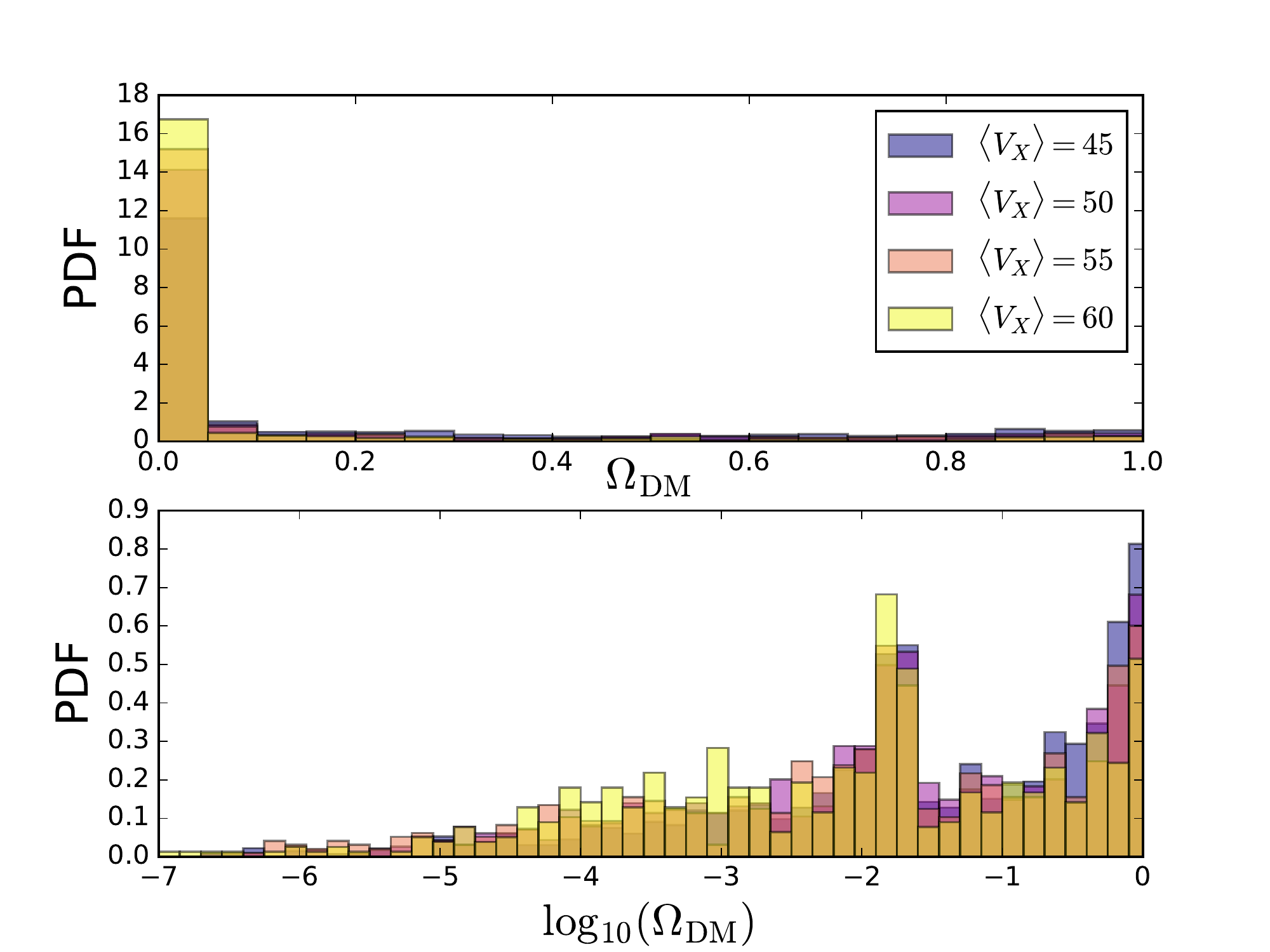}
\caption{{\bf M-theory RMT model DM cosmology example:} Probability density plots for the axion dark matter density parameter, $\Omega_{\rm DM}$ for $\langle V_X \rangle = 45,50,55,60$ presented in both linear and logarithmic scales.}\label{Mth-omegadm}
\end{figure}
In Figs.~\ref{Mth-omegadm1}-\ref{Mthcontourde} we make use of narrow prior windows incorporating Gaussian distributions in our sampling (see Section~\ref{sec:toy}).

\subsubsection{MT-DM}
\label{sec:mtdm}

For our initial look at how axions in the M-theory axiverse model could give rise to dark matter we begin by fixing the average value of the 3-cycle volume distribution, $\langle V_{X} \rangle$ to maximise the probability density of retrieving axions with masses in the window, 
\begin{equation}
10^{-32} {\rm eV} \leq m_a \leq 10^{-25} {\rm eV}\,,	
\end{equation}
 which is done by selecting the following values, 
\begin{equation} 
\label{eq:mtvchoice}
 \langle V_{X} \rangle = 45,50,55,60\,.   
\end{equation}
Fig.~\ref{Mth-omegadm} gives the probability density plots for the axion dark matter density parameter, $\Omega_{\rm DM}$ for each of our slected values of $\langle V_{X} \rangle$ in Eq.~(\ref{eq:mtvchoice}). 

In the upper panel of Fig.~\ref{Mth-omegadm} we show the high probability density to return values of $\Omega_{\rm DM} \lesssim 0.05$. The lower panel of Fig.~\ref{Mth-omegadm} details the spread of these values on a logarithmic scale with an enhanced probability density of returning values of $\Omega_{\rm DM} = \mathcal{O}(10^{-2})$. The low quantities of dark matter arise from the M-theory mass spectrum consistent with many axiverse model spanning many decades giving a significantly lower percentages of cosmologies with values of $\Omega_{\rm DM}$ falling in the window $0.2 \leq \Omega_{\rm DM} \leq 0.4$ as compared to the localised scale RMT models of the string axiverse with far more localised spectra.     
The spread of the axion masses is such that for the average 3-cycle volume distribution values, $\langle V_{X} \rangle = 45$ and $\langle V_{X} \rangle = 60$ we only see a increase in the number of cosmologies with values of $\Omega_{\rm DM}$ falling in the window above go from $\approx 3\%$ to $\approx 8.5\%$. 

\subsubsection{MT-DE}
\label{sec:mtde}

Initial searches for axions with the properties of dark energy in the M-theory model show that there is no mass distribution which gives any form of satisfactory probability density for values of the dark energy density parameter, $\Omega_{\rm DE}$ falling in the bounds $0.6 \leq \Omega_{\rm DE} \leq 0.8$. This feature arises due to the nature of the axion decay constants in the model which are typically too small, $f_a \sim a/s_i \sim 10^{-2}M_{pl}$. The dark energy density can be increased using a significantly larger number of axions or utilising the alignment mechanism which could potentially sufficiently enhance the decay constants, however our assumption on the diagonal form of the kinetic matrix in Eq.~(\ref{eq:mkmatrix}) does not allow for the inclusion of any such an alignment mechanism. Therefore, we postpone an initial look into the possibility of sampling the M-theory axiverse models for dark energy to a topic of interest for future work.

\subsubsection{M-theory Toy Model}
\label{sec:toy}
In order to paint a better picture of the potential of the dark sector in the M-theory model, we consider a toy model with narrow prior probability densities of the associated hyperparameters in order to address some of the issues highlighted in previous sections. In particular, if the priors on the moduli vev $s$, and the instanton index parameter $\widetilde{N}_i^j$, which control the volume function are narrow, our M-theory mass distributions will generically only spread over a few orders of magnitude instead of the many decades we would typically expect. As a result, the axion dark sector density parameters, $\Omega_{\rm DM}$ and $\Omega_{\rm DE}$ will also be concentrated around particular values. This configuration allows us to study correlations between mean values of the M-theory model parameters in a relatively simple manner.

\begin{figure}
\includegraphics[width=0.5\textwidth]{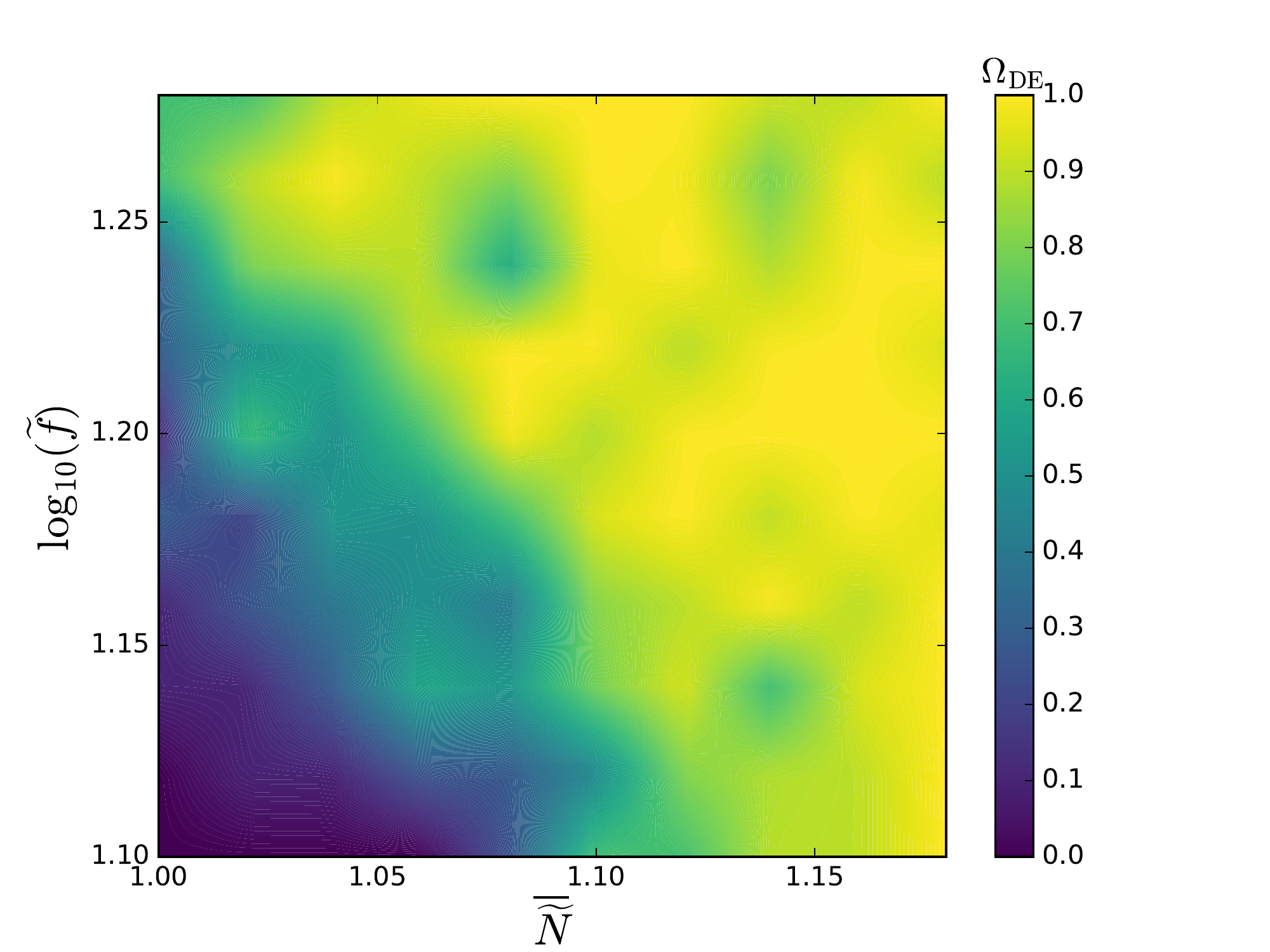}
\caption{{\bf M-theory RMT model DE cosmology example with narrow priors:} Contour density plot for the axion dark energy parameter, $\Omega_{\rm DE}$ using narrower priors for $\widetilde{N}$ along with an enhancement factor on decay constant, $f_a$.}
\label{Mthcontourde}
\end{figure}

We restrict the sampling of the parameters by fixing the priors distributions for $s$ and $\widetilde{N}_i^j$ to be drawn from a Gaussian distribution, $\mathcal{N}(\mu,\sigma)$. We limit the width of the prior sampling by fixing to the distribution standard deviation for $s$ and $\widetilde{N}$ to $\sigma_s = 1$ and $\sigma_{\widetilde{N}} = 0.01$ respectively. For our dark matter examples, we simulate cosmologies for a range of mean values of $s$ and $\widetilde{N}_i^j$ as shown in Fig.~\ref{Mth-omegadm1}. The contour density plot shows a trend of hyperbolic constraint as expected from the relation $\overline{V_X} \sim \overline{s} \times \overline{N}$. Our example cosmologies where we allow for variations in $\beta_\mathcal{M}$ are given in Fig.~\ref{Mth-omegadm2}. 

When considering dark energy, this toy model gives us a quick insight on how much enhancement the decay constants could require in the M-theory model. We study this effect by parametrising the enhancement by the factor $\widetilde{f} = \sfrac{f'_a}{f_a}$. Fig.~\ref{Mthcontourde} shows that the enhancement factor necessary to accomplish the observed dark energy is of the order $f_a \sim [10-100$].  


\section{Results II : The String Axiverse as A Bayesian Network} \label{sec:results2}
\begin{figure*}
    \centering
    \includegraphics[width=1\textwidth]{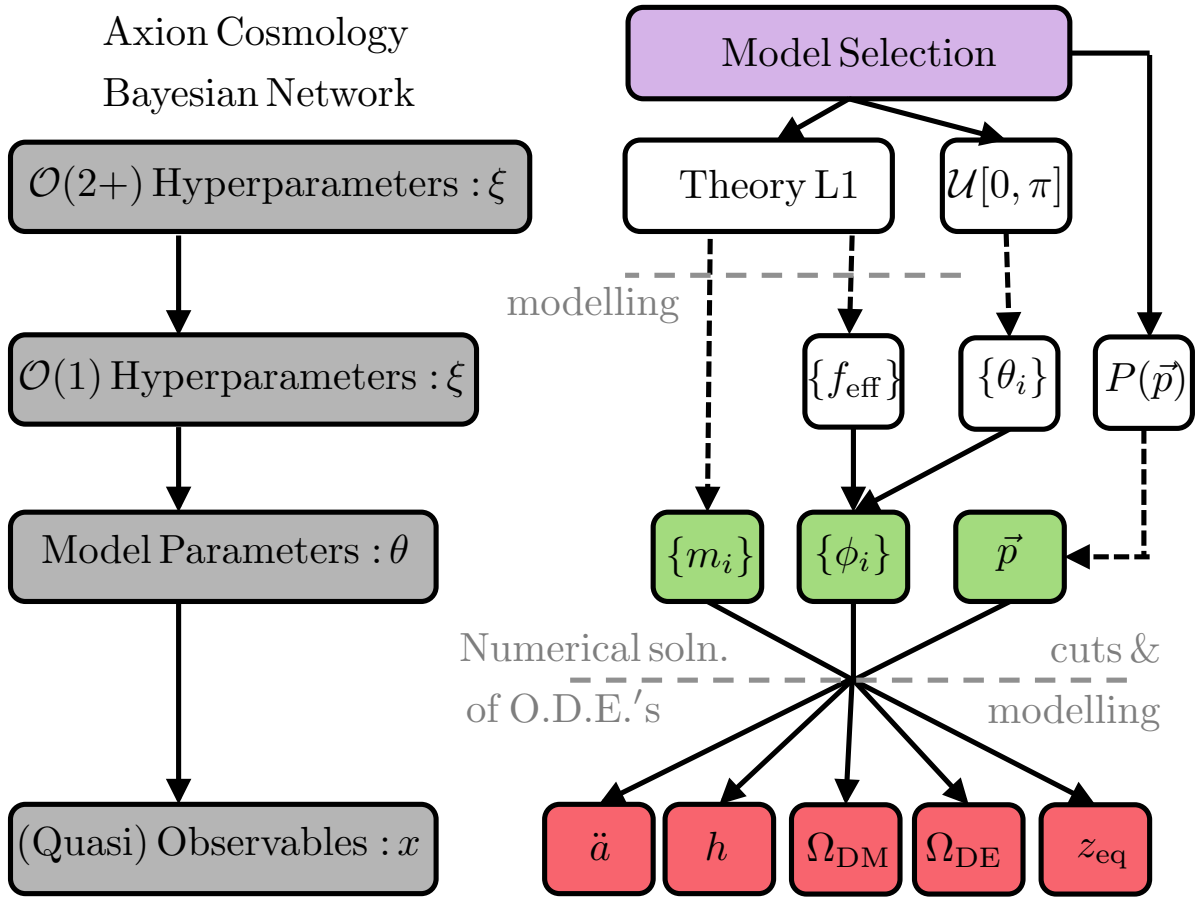}
    \caption[]%
    {{\bf A Generic Bayesian Network for Axion Cosmology:} Arrows indicate the direction of dependence, with dashed arrows indicating stochastic dependence, and solid arrows indicating deterministic dependence.}    
    \label{fig:BayesNetAxion}     
\end{figure*}

\subsection{Bayesian Networks}
\label{sec:baye}
We present here some brief examples treating the String axiverse as a Bayesian network, following the Bayesian networks approach to inflation in Ref.~\cite{2016JCAP...02..049P}. A complete treatment will be presented in a forthcoming paper. A generic Bayesian network for axion cosmology is shown in Fig.~\ref{fig:BayesNetAxion}. We apply the Bayesian network using Markov Chain Monte Carlo (MCMC) techniques. For this purpose we use the affine-invariant ensemble sampler~\cite{goodman-weare} implemented in \textsc{emcee}~\cite{emcee}. Plots detailing the constraints for model hyperparameters are made using \textsc{corner}~\cite{corner}. 

The cosmological parameters are $\vec{p}=(\Omega_r h^2,\Omega_{\rm mat} h^2,\Omega_\Lambda h^2)$. In principle the cosmological parameters are determined stochastically from the hyperparameters of a higher level distribution, though in practice here we take these as fixed Dirac delta distributions determined by the model under consideration. The matter density $\Omega_{\rm mat}=\Omega_b+\Omega_c$ contains ordinary CDM and baryons, and the total matter density includes in addition the contribution from axions that have begun oscillations: $\Omega_m=\Omega_{\rm mat}+\Omega_a^{\rm osc.}$. The axion model parameters fixed by the theory are $\{m_i\}$ and $\{\phi_i\}$. Given the complete set of model parameters the quasi observables are found deterministically by solving the equations of motion. For more details on the numerics, see Appendix~\ref{appendix:numsim}.

The level 1 (L1) theory hyper parameters stochastically determine the model parameters $\{\phi_i\}$ and $\{m_i\}$. Model selection (theory L2) sets the model, the number of axions, and the prior distributions for the L1 hyper parameters. The theoretical modelling from L1 to the model parameters accounts for treating the axion potential as a pure mass matrix, and in fixing the moduli. Theoretical modelling and cuts going from L1 to the quasi-observables includes a cut on the maximum $m_a$, and the choice of cosmological model.

The quasi-observables are the fractional densities in each part of the dark sector, the Hubble parameter, the redshift of matter-radiation equality, and the acceleration of the scale factor. In principle we could consider also the evolution of the background quantities with redshift. For simplicity in the examples shown we simply apply a Gaussian likelihood to $\Omega_m$, $h$, and $z_{\rm eq}$,  assuming the \emph{Planck} (2015) TT+lowP results~\cite{Ade:2015xua} presented in Table~\ref{tab:cosmopar}. We assign axions to the matter or DE density according to whether the equation of state has begun oscillating. We also apply a cut demanding an accelerating expansion of the Universe, $\ddot{a}>0$. 

Our treatment of the quasi-observables should be considered only as giving approximate constraints on the models. Our models can have non trivial effects on the equation of state for dark energy, $w(z)$, and for light DM axions also on structure formation and the CMB power spectrum~\cite{Hlozek:2014lca}, which are not accounted for in the simplified quasi-observables with Gaussian likelihood. 

In ordinary $\Lambda$CDM, $\Omega_m$, $z_{\rm eq}$, and $h$ are not independent. However, in axion models the change in the equation of state at late times can alter these relationships by the creation of additional matter-like axion density after $z_{\rm eq}$. Our use of $z_{\rm eq}$ as an independent quasi-observable from the matter density and $h$ serves as an approximation of the constraints of Ref.~\cite{Hlozek:2014lca}, which disfavour large energy densities of ultralight axions that begin oscillating after equality. We ignore covariance between the quasi-observales for the same reason that dependences are not the same in axion models as in $\Lambda$CDM.

Our quasi-observables are particularly simple. A more advances compression of the CMB, baryonic acoustic oscillation and growth data appropriate for DE models is given by the treatment of Refs.~\cite{2015PhRvD..92l3516A, 2017arXiv170510031Z}. In this treatment, the CMB data are compressed into a vector of measurements for the matter densities, matter power spectrum amplitude, and the angular size of the sound horizon, including covariance. Use of a wide variety of datasets will be possible by integrating our random axion models into \textsc{cosmosis}~\cite{2015A&C....12...45Z}.

\subsection{Constraints on the String Axiverse} \label{sec:mcmcresults}

All the constraints shown hold the number of axions fixed at $n_{\rm ax}=20$. Numerical accuracy settings are defined in Appendix~\ref{appendix:numsim}. All \textsc{emcee} walkers are initialised from the priors, and chains are ran to convergence as evaluated according to the spectral method of Ref.~\cite{2005MNRAS.356..925D}.

\subsubsection{Mar\v{c}enko-Pastur Model for DM and DE}
\label{sec:mpmcmc}

The first set of example constraints we show is the simplest both in model and computational terms. We take the Mar\v{c}enko-Pastur Law model, and tailor it to provide DE with fixed number of axions $n_{\rm ax}=20$. We fix the matter density to $\Omega_{\rm mat} h^2=0.148$, including dust-like CDM and baryons. 

\begin{figure}
    \centering
    \includegraphics[width=0.48\textwidth]{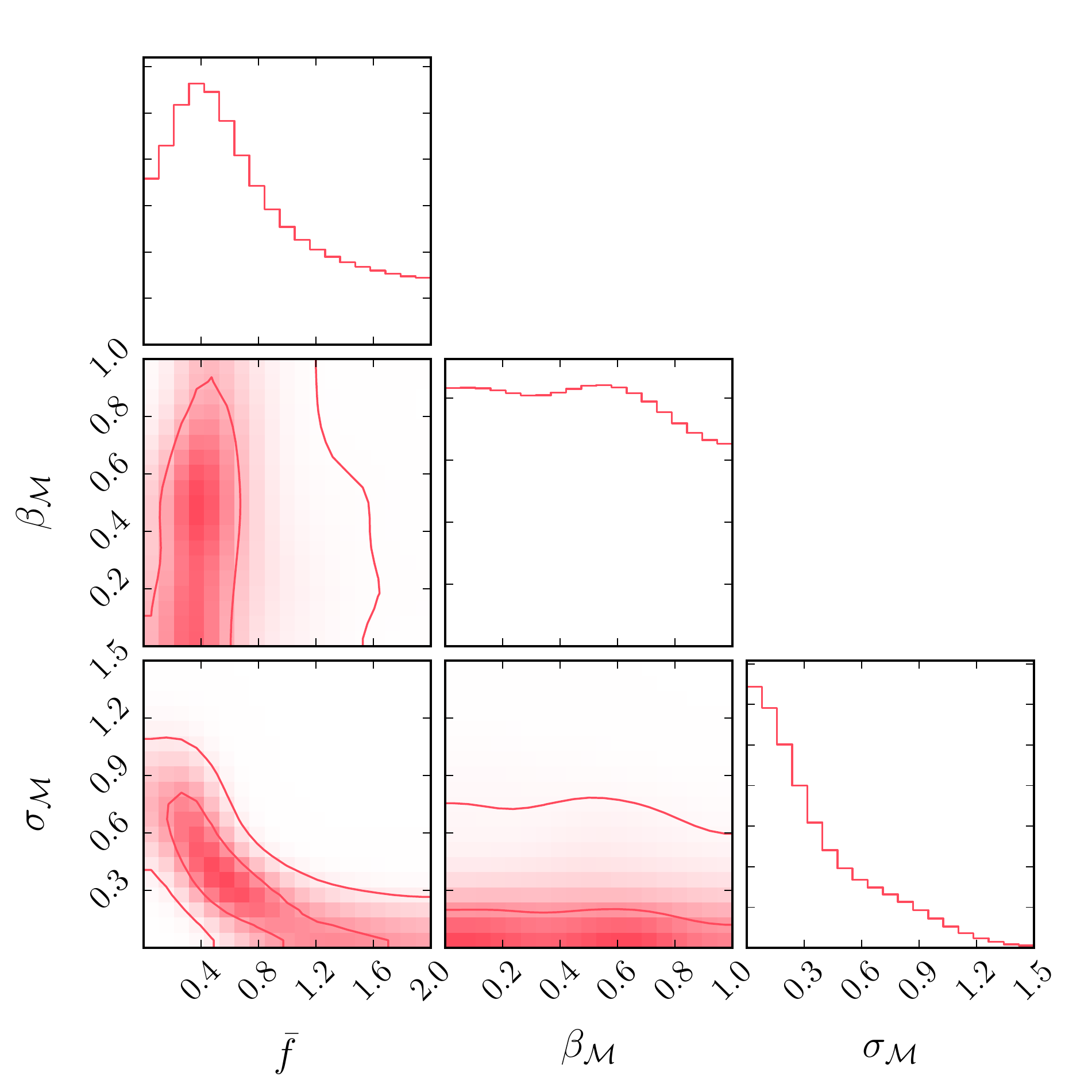}
    \caption[]%
    {{\bf Constraints on the Mar\v{c}enko-Pastur RMT model for DE:} Contours 1 and 2 $\sigma$ in the posterior distribution after imposing likelihoods and cuts on the quasi-observables. Demanding acceleration with $\Lambda=0$ gives the bound $\sigma_\mathcal{M}<0.9 M_H=1.9\times 10^{-33}\text{ eV}$ (95\% C.L.) from requiring the total equation of state $w<-1/3$ with the fields in slow roll at $z=0$.}    
    \label{fig:RMT_DE_triangle}     
\end{figure}

The L1 hyperparameters have the following priors (fixed L2 parameters):
\begin{align}
\bar{f}&\in \mathcal{U}[0.0,5.0]\, , \\
\sigma_\mathcal{M}&\in \mathcal{U}[0.0,10.0]\, , \\
\beta_{\mathcal{M}}&\in \mathcal{U}(0.0,1.0]\, .
\end{align}
After applying Gaussian likelihoods to $h$, $z_{\rm eq}$ and $\Omega_m$, and a cut for $\ddot{a}>0$, we find the constraints shown in Fig.~\ref{fig:RMT_DE_triangle}. The mass parameter and $\bar{f}$ are constrained to values consistent with the DE density. The cut on acceleration with the requirement $\Lambda=0$ leads to a maximum allowed value of $\sigma_\mathcal{M}$. This model shows no preference on $\beta_{\mathcal{M}}$: with a linear prior on $\sigma_\mathcal{M}$ near $M_H$ the width of the mass distribution is not important.

\begin{figure}
    \centering
    \includegraphics[width=0.48\textwidth]{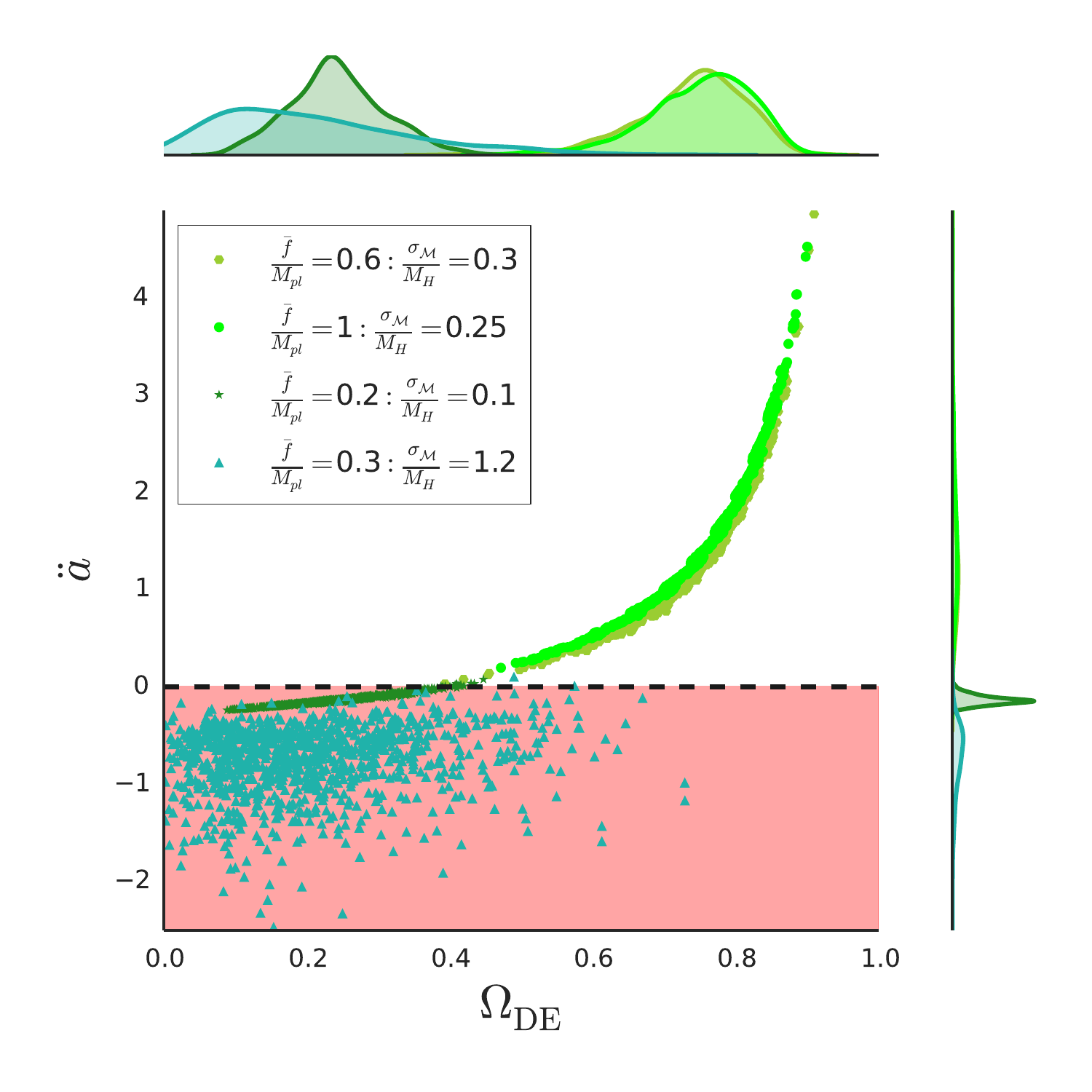}
    \caption[]%
    {{\bf Degeneracies in the Mar\v{c}enko-Pastur RMT model for DE:} We show random samples form four locations in the $(\bar{f},\sigma_\mathcal{M})$ plane at fixed $\beta_{\mathcal{M}}=0.5$, marked in Fig.~\ref{fig:RMT_DE_triangle}. Along the degeneracy direction the quasi observable distributions do not change much. Across this direction, models are disfavoured, with the quasi-observables distributions moving in opposite directions on either side.}    
    \label{fig:RMT_DE_degen}     
\end{figure}

The degeneracies in the MP-DE are demonstrated in Fig.~\ref{fig:RMT_DE_degen}. We show random samples drawn four different values of $(\bar{f},\sigma_\mathcal{M})$ with $\beta_{\mathcal{M}}=0.5$ and demonstrate how the quasi-observable distributions shift. The models moving along the degeneracy direction give accelerated expansion and consistent values of $\Omega_{\rm DE}$ which change relatively little. Perpendicular to this direction, the DE density is too low if the mass is too large (oscillations begin before $z=0$) or the decay constant is too low (field displacement too small). This has a knock-effect of making the acceleration parameter negative in these models.

\begin{figure}[t!]
    \centering
    \includegraphics[width=0.48\textwidth]{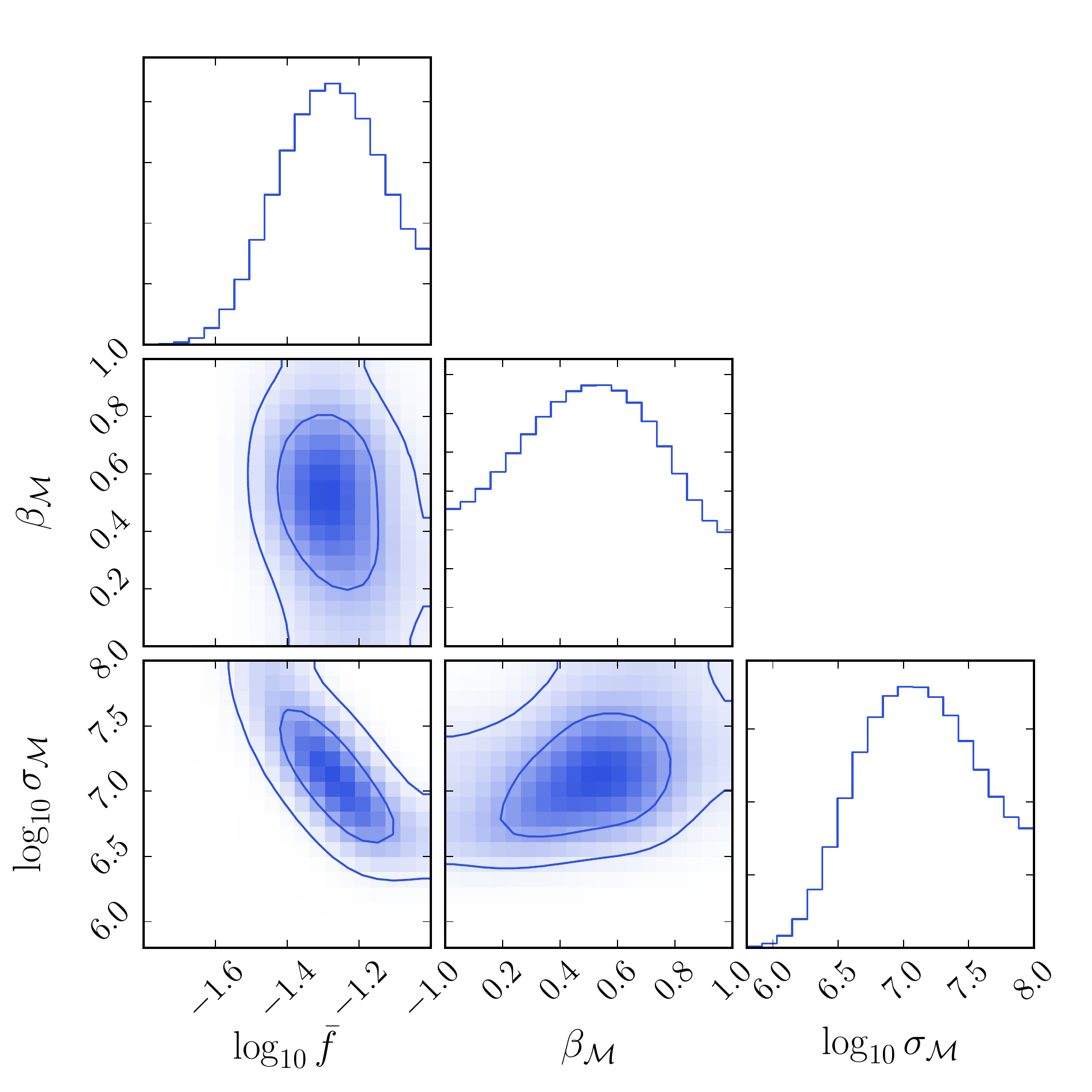}
    \caption[]%
    {{\bf Constraints on the Mar\v{c}enko-Pastur RMT model for DM:} Contours 1 and 2 $\sigma$ in the posterior distribution after imposing likelihoods and cuts on the quasi-observables. Fixing $z_{\rm eq}$ with only baryons as additional matter leads to the constraint $\log_{10}\sigma_\mathcal{M}>6.6\Rightarrow \sigma_\mathcal{M}>4\times 10^{-27}\text{ eV}$ (95\% C.L.) from requiring the fields to be oscillating with $w=0$ prior to this epoch. There is a mild preference for $\beta=0.5$.}    
    \label{fig:RMT_DM_triangle}     
\end{figure}

Next, we consider the computationally more challenging but physically more interesting case of the Mar\v{c}enko-Pastur Law model for DM. The model is more computationally challenging than the DE model due to the required switch in the equations of motion and following of axion field oscillations before the switch (an average run of \textsc{AxionNet} for this model takes $\mathcal{O}(20{\rm s})$ in wall-clock time). We fix the (non-axion) matter density to the baryon density, $\Omega_b h^2=0.022$, and we fix the physical cosmological constant density to $\Omega_\Lambda h^2=0.31$ (this gives the central \emph{Planck} value for $\Omega_\Lambda=1-\Omega_m$ when $h=0.673$). 

The L1 hyperparameters have priors:
\begin{align}
\log_{10}\bar{f}&\in \mathcal{U}[-9.0,-1.0] \, , \\
\log_{10}\sigma_\mathcal{M}&\in \mathcal{U}[0.0,8.0] \, , \\
\beta_{\mathcal{M}}&\in \mathcal{U}(0.0,1.0]\, .
\end{align}
The posterior distributions are shown in Fig.~\ref{fig:RMT_DM_triangle}. The constraint on the matter density parameter, $\Omega_m$, fixes a direction in the $(\bar{f},\sigma_\mathcal{M})$ space. The constraint on $z_{\rm eq}$ leads to a minimum allowed value of $\sigma_\mathcal{M}$. Interestingly, this model shows a mild preference for $\beta_{\mathcal{M}}=0.5$. The preference for $\beta_{\mathcal{M}}=0.5$ is possibly driven by the preference for a \emph{not-too-wide} mass distribution. Preventing the occurence of axions with $m_a<H(z_{\rm eq})$ selects against $\beta_{\mathcal{M}}=1$ and a wide distribution. There is no strongly preferred mass for DM above this scale, and so $\beta_{\mathcal{M}}=0$ is disfavoured to keep the distribution from becoming singular. The minimum value of $\bar{f}$ depends on the maximum value of $\sigma_\mathcal{M}$, fixed by obtaining the relic density.

In both the above considered Mar\v{c}enko-Pastur models we observe a constraint on the characteristic axion mass and decay constant. The location of the constraint on the mass is fixed by the quasi-observables, and the problem under consideration: either by the condition on $\ddot{a}$ for $h\approx 0.7$ for axion DE, or by the conditions on $z_{\rm eq}$ and $\Omega_m$ for axion DM. The modal value of $\bar{f}$ in the Mar\v{c}enko-Pastur model is determined by the required energy density in axions, and is thus dependent on our \emph{fixed} parameter $n_{\rm ax}=20$. In the DE example, the modal value (after binning on the linear prior) is $\bar{f}=0.3 M_{pl}$, reduced from the naive value $\bar{f}=M_{pl}$ in a single axion model by the ``N-flation'' $1/\sqrt{n_{\rm ax}}$ effect (c.f. constraints on axions as quintessence~\cite{2017JCAP...01..023S}). There is a similar effect in the DM example, where $\bar{f}$ is lowered from the value needed in a single field $m^2\phi^2$ model for the DM relic density (e.g. Ref.~\cite{Marsh:2015xka}). A model with varying $n_{\rm ax}$ would display a degeneracy in the $(\bar{f},n_{\rm ax})$ plane. 

\subsubsection{Dark Matter from the M-theory Axiverse }
\label{sec:mtmcmcdm}
\begin{figure*}[t!]
    \centering
    \includegraphics[width=0.8\textwidth]{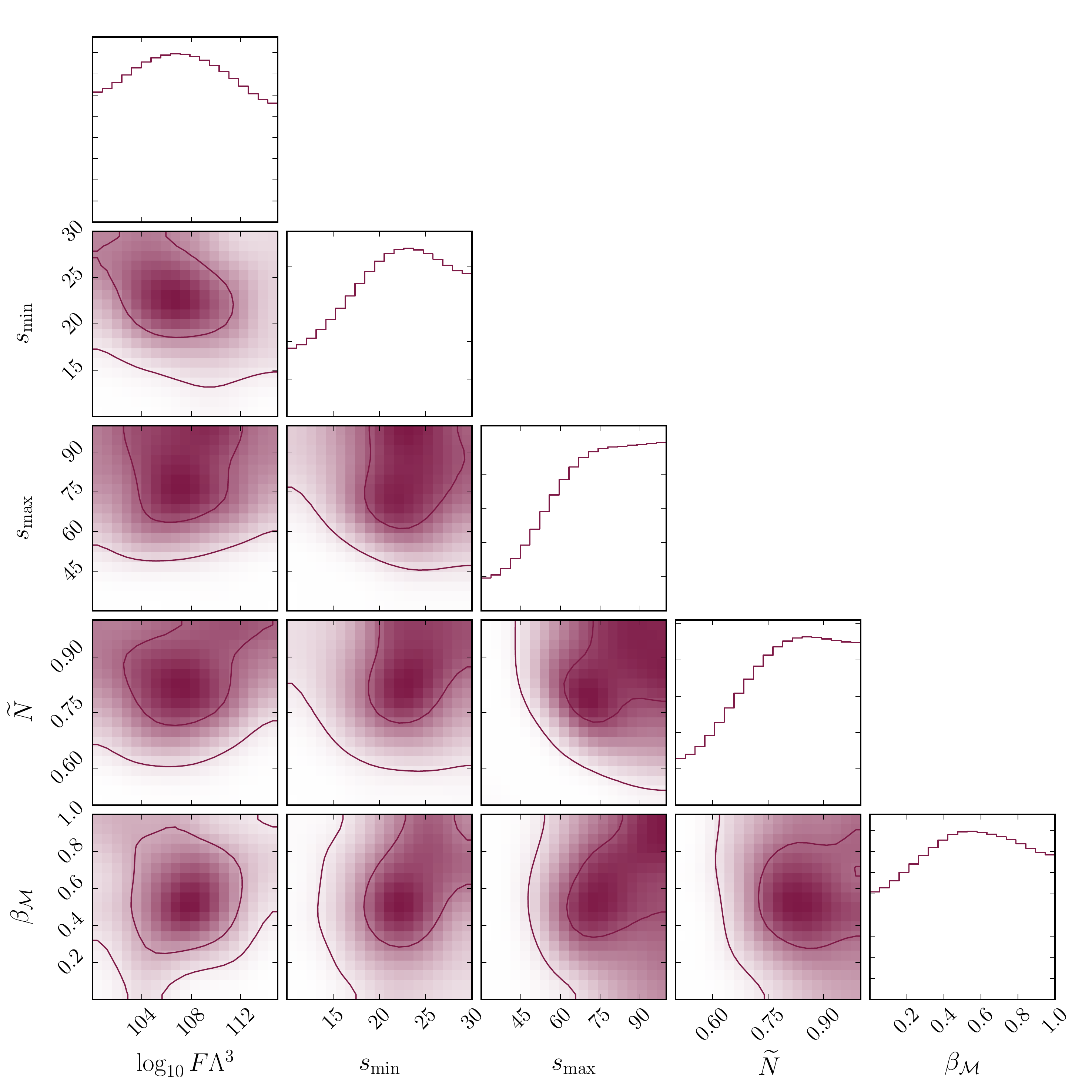}
    \caption[]%
    {{\bf Constraints on M-theory RMT model DM:} Contours 1 and 2 $\sigma$ in the posterior distribution after imposing likelihoods and cuts on the quasi-observables. One sided constraints on parameters are driven by the simultaneous requirements of not overproducing DM and maintaining an accelerating Universe at $z=0$. The constraints are one-sided due to the best-fot region being very narrow, with a plateau in the likelihood away from this region where the axion density drops to zero, $z_{\rm eq}$ is fixed by the baryons alone, adn acceleration is guaranteed by the cosmological constant.}    
    \label{fig:mtheory_bayes}     
\end{figure*}

The M-theory axiverse, with it's log-normal mass distribution and very wide spread, mean that the constraints must be read carefully (in a preliminary investigation, we found the same considerations apply to the log-flat matrix elements model.). The constraints on the the M-theory model parameters for the case of uniform distributions in $s$ and $\tilde{N}$ are shown in Fig.~\ref{fig:mtheory_bayes}. 

The constraints on the M-theory model primarily derive from not over-producing DM. With decay constants typically of order the GUT scale, axions with masses $m_a\gtrsim 10^{-18}\text{ eV}$ typically provide too much DM density (``anthropically constrained''~\cite{Arvanitaki:2009fg}). This leads to minimum values of $s_{\rm min}$ and $s_{\rm max}$, with large moduli giving large instanton actions, low axion masses, and correspondingly lower relic densities. There is also a lower bound on $\tilde{N}$, which sets the scale of the instanton charges, and also leads to lower axion masses. 

The vast majority of the M-theory DM models within the 2$\sigma$ allowed region in Fig.~\ref{fig:mtheory_bayes} produce a cosmology with quasi-observables: $(h,\Omega_m,z_{\rm eq})\approx (0.57,0.06,520)$, with $\ddot{a}>0$ provided by the cosmological constant, and the matter density provided by the baryons. While this is a bad fit to the data, it is a better fit than a model with, for example, total DM domination at $z=0$, $\ddot{a}<0$, and $z_{\rm eq}\approx 10^5$, which results if heavy axions ``overclose'' the Universe by providing too much DM. This is not to say that there are not examples of M-theory models that do provide good fits to the data. For example, it is easy to find a model in our chains with hyperparameters $(\log_{10}F\Lambda^3,s_{\rm min},s_{\rm max},\tilde{N},\beta_\mathcal{M})\approx(105,26,54,0.7,0.9)$ and quasi-observables $(h,\Omega_m,z_{\rm eq})\approx (0.7,0.3,3000)$. We have checked that this general trend also applies in the alternative Gaussian priors on $s$ and $\tilde{N}$, and also using the alternative quasi-observable $\Omega_d h^2$ for the axion DM instead of the total matter content including baryons.

This one-sided behaviour in the constraints, and with many samples being poor fits, can be understood by considering the results of grid-based sampling in a simplified model. We took the Gaussian priors model for $s$ and $\tilde{N}$, holding $\sigma_s=1$, $\sigma_N=0.01$ fixed and varying $\bar{s}\in [20,21]$, $\bar{N}\in [0.5,0.55]$ with $n_{\rm ax}=20$. We sampled each point in parameter space ten times, and interpolated the average quasi-observables on a linear grid.  

\begin{figure}[t!]
    \centering
    \includegraphics[width=0.48\textwidth]{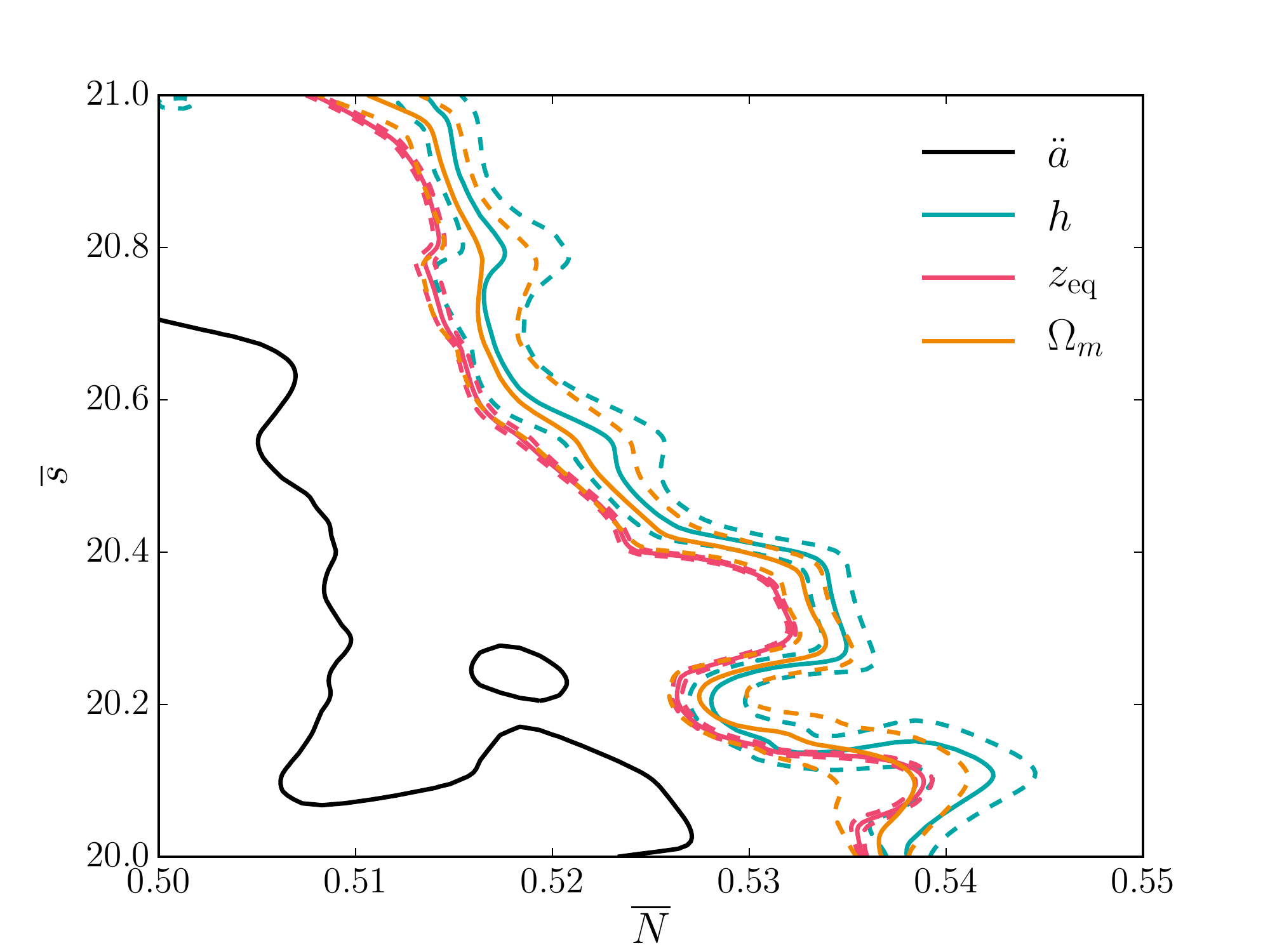}
    \caption[]%
    {{\bf Grid sampling of M-theory RMT model DM:} Solid (dashed) contours show the mean ($\pm 2 \sigma$) values of the quasi-observables on a grid based sampling of $(\bar{N},\bar{s})$ for $n_{\rm ax}=20$. For small $\bar{N}$, $\ddot{a}<0$ leading to zero likelihood (cut), while for large $\bar{N}$, $\ddot{a}>0$. For large $\bar{N}$ the axion density goes to zero, but the likelihood plateaus due to the inclusion of the baryons and the cosmological constant.}    
    \label{fig:mtheory_grid_contour}     
\end{figure}

Fig.~\ref{fig:mtheory_grid_contour} shows the results of this sampling. The contours show the location of $\bar{x}\pm 2\sigma_x$ for quasi-observable $x$, and the location of $\ddot{a}>0$. We see that there is only a very narrow region of parameter space where the quasi-observables all have values near the means. For small $\bar{N}$ the likelihood goes to zero due to the cut on $\ddot{a}$. On the other hand, for large $\bar{N}$ the likelihood plateaus. As the axion DM density drops to zero, the baryon contribution leads to minimum values of $z_{\rm eq}$ and $h$. Thus the whole region of parameter space with large $\bar{N}$ is equally disfavoured, and has large prior volume. This leads to a one-sided constraint on parameters driven by $\ddot{a}>0$, which is in turn driven by the requirement of not overproducing DM and having $z_{\rm eq}$ too large.

These observations highlight some limitations of our methodology when applied to a model with a larger number of parameters and a very small prior volume in the best-fit region. It also highlights how our use of quasi-observables does not equally disfavour all possibilities away from the best-fit.

\section{Discussion and Conclusions} \label{sec:conclusions}

The existence of a ``dark sector'' of particles largely decoupled from the Standard Model is necessary to explain the phenomenon of dark matter, and could also play a role in the accelerated expansion of the Universe as dark energy. String theory and M-theory predict the existence of a complex, multi-component dark sector containing (among other things) many axion fields. Making definite predictions in such a landscape of possibilities seems at present impossible. However, statistical tools enable us to explore these possibilities. In the context of inflationary theory, random matrix models have proven to be a useful simplification, owing to the universality of the eigenvalue distributions.

In the present work we have investigated random matrix models for the axion dark sector, and computed the spectrum of axion masses and initial field values. These quantities determine the resulting energy densities of dark matter and dark energy. By treating these as quasi-observables we have been able to constrain the parameters of the random matrix models. This is the first such investigation (that we are aware of) of random multi-field models applied to the problem of the dark sector. We have used the adaptable framework of Bayesian networks to perform a Monte Carlo investigation of this scenario.

We have chosen to investigate axion models for DM and DE separately. A model for axion DM \emph{and} DE \emph{together} requires a mass splitting at least of $\mathcal{O}(H(z_{\rm eq})/M_H)\sim 10^6$ so as not to generate too much energy density in light states~\cite{Hlozek:2014lca}. Such a hierachy cannot be generated in the models we have considered. The structure of the matrices we have assumed is that all the stable axions acquire their masses from similar sources. That is, the elements of the matrices are all drawn from the same distributions. There are no separate sectors, which would occur for matrices with mixed distributions and for block-diagonal  matrices. In our models, the only effect that can lead to hierarchies in the mass spectrum is the existence of large eigenvalues, and we have not found this to be sufficient to allow axions to simultaneously provide DM and DE. 

An interesting extension of our work would be to consider a \emph{hierarchical} model, constraining the $\{m_i\}$ and $\{\phi_i\}$ distributions separately for DM and DE. With this information one could design block-diagonal random matrix models for an entirely axionic dark sector. In a high energy physics context, such a model could be realised if part of the axion sector was protected from the leading order instanton effects and received its mass only at some higher order. 

Hierarchies can also be generated in multi-axion models with non-trivial potentials~\cite{2017PhLB..765..293D}, where isocurvature perturbations (see below) can also be suppressed. This highlights another major simplification and limitation of our work: the use of the mass term only in the potential. While it is technically trivial to replace the mass term with some general function (such as the instanton expansion), computationally it is more challenging. Firstly, by the simplification it is necessary to impose after oscillations (for a non-quadratic minimum, one cannot use $w_a=0$), and secondly by the possibility of meta-stable minima leading to dynamics on widely separated timescales.

We have found, in the case of DM models, data-driven lower bounds on axion mass distributions set by the matter density and $z_{\rm eq}$. Low mass scales for axions find theoretical and phenomenological motivation also. Theoretically, as discussed, the mass scale $m_a\approx 10^{-15}\text{ eV}$ emerges from fixing the GUT scale unified gauge coupling, $\alpha_{\rm GUT}\approx 1/25$, in the M-theory compactifications~\cite{Acharya:2010zx}, with a similar approximate relation in string models~\cite{Hui:2016ltb}. Generation of ultralight masses has been discussed extensively recently, in string theory and supersymmetry~\cite{Halverson:2017deq}, in QCD-related theories~\cite{Davoudiasl:2017jke}, and through use of discrete symmetries~\cite{Kim:2015yna}. Ultralight DM has distinctive effects on cosmic structure formation that allow it to be distinguished from cold DM, and it represents a frontier of DM research~\cite{Marsh:2015xka,Hui:2016ltb}. The idea of ``catastrophic boundaries''~~\cite{2009PhRvD..80f3510B} in the multiverse may lead to a preference for universes ``on the edge'' of such a frontier.

Phenomenologically, axion masses in the range we have constrained [approximately $H_0<m_a<H(z_{\rm eq})$], and up to $10^{-23}\text{ eV}$, are probed by the CMB power spectrum and large scale structure~\cite{Hlozek:2014lca,Urena-Lopez:2015gur,Hlozek:2016lzm}. Higher masses in the range $10^{-22}\text{ eV}\lesssim m_a\lesssim 10^{-20}\text{ eV}$ are motivated by their interesting effects on galaxy formation~\cite{Marsh:2013ywa,Schive:2014dra,Marsh:2015xka,2016PhRvD..94d3513S,Hui:2016ltb}, and are probed by high redshift galaxy formation~\cite{2015MNRAS.450..209B,2016ApJ...818...89S,2016JCAP...04..012S,2016arXiv161105892C} and the Lyman-alpha forest flux power spectrum~\cite{Armengaud:2017nkf,Irsic:2017yje}. Still more massive axions in the range $10^{-20}\text{ eV}\lesssim m_a\lesssim 10^{-18}\text{ eV}$ can be probed purely gravitationally by the 21cm power spectrum~\cite{2015PhRvD..91l3520M}. 

Constraints from quasi-observables cannot make contact to such detailed constraints as discussed above. To even begin such a task would require the perturbation theory of multi-axion models. While technically trivial, this is a computationally challenging task that we have not attempted to take on. However, even without perturbation theory the range of masses $10^{-18}\text{ eV}\lesssim m_a\lesssim 10^{-10}\text{ eV}$ are constrained by black hole superradiance~\cite{2011PhRvD..83d4026A,2012PhRvL.109m1102P,2012IJMPS...7...84K,2015PhRvD..91h4011A,2015CQGra..32u4001Y}.  Incorporating the superradiance constraints into the axion mass distribution will be a relatively simple task given the adaptability of the Bayesian networks approach.

As well as axion mass distributions, we have computed the distributions of decay constants, $f_a$, from the eigenvalues of the kinetic matrix. The ``weak gravity conjecture''~\cite{2007JHEP...06..060A} (WGC) can be used to place bounds on combinations of axion decay constants and masses, and broadly speaking can be said to constrain the existence of super-Planckian values for $f_a$ (without the alignment mechanism). Overcoming this apparent constraint is a prime motivation for the introduction of multi-field models of axion inflation, and has in part motivated the present work on DM and DE. 

We have held $n_{\rm ax}$ fixed in our example Bayesian Network constraints. It would be interesting to explore in a future work how imposing the (weak or strong forms of the) WGC as a prior could lead to a lower bound on $n_{\rm ax}$ required by providing the correct energy densities in a given DM or DE model. In the case of N-flation (and related models), the necessary minimum number of fields has been argued to be in conflict with entropy bounds in de Sitter space~\cite{Conlon:2012tz}, and a similar conclusion for axion DE or DM could have profound implications.

A notable multifield axion model for DE considered previously in the literature is Ref.~\cite{2014PhRvL.113y1302K}, with more detailed cosmological consequences computed in Ref.~\cite{2016PhRvD..93l3005E}. The model considered a simplified distribution for the axion masses and decay constants, equivalent to log-flat mass eigenvalues and uniform kinetic matrix eigenvalues. The model has a one in one hundred ``chance'' of providing the correct DE density. An interesting extension considered in Ref.~\cite{2014PhRvL.113y1302K} is the use of non-canonical multi-instanton potentials to facilitate the decay of problematic heavy axion fields that otherwise provide too large energy densities.

Our random matrix approach provides a more versatile, and realistic, approach to the distributions. Our Bayesian forward model is able to quantify and extend the estimates outlined in Ref.~\cite{2014PhRvL.113y1302K} for the mass and decay constant distributions. Ref.~\cite{2016PhRvD..93l3005E} consider the observables for DE models more thoroughly, such as the angular diameter distance to the CMB, and improvements from future baryon acoustic oscillation measurements by the Square Kilometer Array. It would be interesting to include these in our Bayesian network.  

The only concrete axiverse construction we have used to inform our random matrix models has been the M-theory model of Ref.~\cite{Acharya:2010zx}. An explicit axiverse model has also been realised in Type-IIB~\cite{2012JHEP...10..146C}, where models for N-flation and ``N-quintessence'' have also been constructed~\cite{2014JCAP...08..012C}, and our methodology could easily be applied to these models also. We note, however, that in the case where these models can have a low string scale, $M_s\sim 10^{12}\text{ GeV}$, the DM abundance from vacuum realignment of light axions will be hard to achieve.

Our discussion in this paper has been set entirely in the late Universe, in particular during radiation domination post-BBN and we have made no explicit connection between our models and inflationary theory. This neglects the very important constraints on axion DM coming from isocurvature perturbations (e.g. Ref.~\cite{2004hep.th....9059F}). High scale inflation, in particular with observably large tensor-to-scalar ratio, typically generates large amplitude number density perturbations in axions, which contribute to the CMB power spectrum acoustic peaks such as to shift their phase, inconsistent with observations~\cite{2009ApJS..180..330K,2016A&A...594A..20P}. The Hubble scale during inflation is constrained, for the QCD axion with typical $f_a$, to be $H_I\lesssim 10^8\text{ GeV}$. 

The requirements on $H_I$ are significantly loosened for ultralight axions, with isocurvature perturbations becoming negligible for $m_a\lesssim 10^{-26}\text{ eV}$~\cite{2013PhRvD..87l1701M,Marsh:2014qoa}. The constraints become multiplicatively worse, however, in the case of multiple axion fields~\cite{Mack:2009hs}. In Ref.~\cite{Acharya:2010zx} it was shown that the M-theory axiverse requires $H_I\lesssim 10^{10}\text{ GeV}$. The adaptability of the Bayesian networks approach means that including the isocurvature amplitude as a quasi-observable and $H_I$ as a model parameter is another easily tackled problem. Such an investigation would clarify the prior dependence in the results of Ref.~\cite{Mack:2009hs}.

The study of random matrix multi-axion models has been popular for some time in inflationary theory. While inflation is well-motivated by cosmological observations, it is unlikely to be possible to determine the theory precisely due to the limited information available. Dark matter, on the other hand, offers far greater prospects for precision measurement~\cite{Hlozek:2016lzm}, and so by studying multi-axion  models in the late Universe we might discover more about physics beyond the Standard Model. In this work we have presented the first exploration of a random matrix multi-axion model for dark matter and dark energy and have used statistical methods to bound the axion mass and decay constant distributions.


\section*{Acknowledgements}

We are thankful to Thomas Bachlechner, Jonathan Frazer, Eugene Lim, and M.C. David Marsh for helpful discussions and correspondence. The work of MJS is supported by funding from the UK Science and Technology Facilities Council (STFC). DJEM is supported by a Royal Astronomical Society Postdoctoral Fellowship. LCP is supported in part by the DOE DE-SC0011114 grant.

\appendix

\section{Computations} \label{appendix:numsim}

We numerically solve the equations of motion for $n_{\rm ax}$ axions with fixed initial conditions and evolve the solutions forwards in time using \textsc{scipy}. We rescale the fields in terms of Planck units. The cosmic time, $t$, is the independent variable measured in units of $M_H$. This naturally sets the axion mass scale in units of $M_H$, and the cosmological densities in component $X$ appear in the Friedmann constraint in terms of their density today as $\Omega_X h^2$.

\subsection{Energy Densities}

We define our initial and final conditions using the photon temperature as a clock. The total energy density for the relativistic degrees of freedom, $\rho_{r}$ is,  
\begin{equation}
\rho_r = \frac{\pi^2}{30}g_\star(T) T^4\,, 
\label{eq:rhorr}	
\end{equation}   
where $g_\star(T)$ counts the relativistic degrees of freedom (e.g. Ref.~\cite{1990eaun.book.....K}). We fix ``\emph{today}'' from the CMB temperature, $T_{\rm CMB}=2.725\text{ K}$~\cite{1996ApJ...473..576F}. Normalising the scale factor such that $a(T_{\rm CMB})=1$, the scale factor $a(t)$ is found by integrating the Friedmann constraint. 

For simplicity, we treat the total relativistic degrees of freedom as a constant, and thus we must begin our solutions after neutrino decoupling. Using the fits from Ref.~\cite{Wantz:2009it}, this occurs at $T_i\approx 23\text{ keV}$ when the scale factor is $a_i\approx 10^{-8}$. After this time, the radiation energy density evolves as:
\begin{equation}
\rho_r(a)=3 M_H^2 M_{pl}^2 \frac{\Omega_r h^2}{a^4}\, ,
\end{equation}
where, 
\begin{equation}
\Omega_r h^2=\rho_r(T_{\rm CMB})/(3M_H^2M_{Pl}^2)=4.16\times 10^{-5}\,.
\end{equation}
Assuming radiation domination at $T_i$ allows us to set the initial physical time, 
\begin{equation} 
t_i=(a_i^2/2)(\Omega_r h^2)^{-0.5}\,.
\end{equation}
We allow for the inclusion of a cosmological constant with fixed physical density $\Omega_\Lambda h^2$. The total (ordinary+CDM) matter density is
\begin{equation}
\rho_{\rm mat}(a) =3 M_H^2 M_{pl}^2 \frac{\Omega_{\rm mat} h^2}{a^3}\, .
\end{equation}
The minimum value for $\Omega_m h^2$ is given by the physical baryon density, $\Omega_b h^2=0.022$~\cite{Ade:2015xua}. 

In the homogeneous limit, the energy-momentum tensor for the axions is described by a perfect fluid with components, $T_0^0 = -\rho$ and $T^i_j = P\delta^i_j$. The energy density and pressure for a single axion are,
\begin{align}
\rho_a = \frac{1}{2}\dot{\phi}^2 &+ \frac{1}{2}m^2_a\phi^2	\, , \\
P_a = \frac{1}{2}\dot{\phi}^2 &- \frac{1}{2}m^2_a\phi^2	 \, . \label{eqn:axion_pressure}
\end{align}
The pressure of the matter, radiation, and cosmological constant are determined by the equations of state: $P_i=w_i\rho_i$ (no sum on $i$) with $w_r=1/3$, $w_m=0$ and $w_\Lambda=-1$. The total pressure appears in the acceleration equation,
\begin{equation}
	\dot{H}+H^2 = \frac{\ddot{a}}{a} = -\frac{1}{3}\sum_i\left(\rho_i+3P_i\right)\,,
\label{eqn:acceleration}
\end{equation}
with an accelerating universe satisfying the condition $\ddot{a}>0$. We do not solve the acceleration equation, but we compute $\ddot{a}$ using the right hand side on Eq.(\ref{eqn:acceleration}).

\subsection{Initial Conditions and Axion Mass Limits}
\label{appendix:heavy_states}

The Hubble parameter, $H$, provides a friction term in the Klein-Gordon equation, which, as long as the condition $H\gtrsim m_a$ is satisfied, the axion field velocity will remain small. In the limit that the mass can be entirely neglected, the attractor solution is $\dot{\phi}=0$. We assume this condition is met for our initial conditions. This assumption sets an upper limit for the axion masses that we can consistently consider for any given initial temperature. Demanding that $m_a<3H(T_i)$ fixed by neutrino decoupling, we find the upper limit for the axion mass:
\begin{equation}
m_a < 4 \times 10^{-19} \text{ eV}\, .	
\label{eqn:strict_mass}
\end{equation}

In principle we could extend to higher temperatures, and thus higher axion masses, by modelling the evolution of $g_\star$ above neutrino decoupling. We have chosen not to do this for a number of reasons. Firstly, the particle content is not known beyond a few TeV. Secondly, above about 1 MeV (BBN), the Universe need not have been radiation dominated, and there is no observational necessity to assume so. Thirdly, in string/M-theory, we expect a non-thermal cosmology at early times dominated by the energy density of moduli coherently displaced by vacuum fluctuations during inflation. The matter dominated phase is known to alter the relic densities of axions that begin oscillating during that period~\cite{Banks:1996ea,Acharya:2010zx}. 

Furthermore, when the moduli are displaced, and before they have decayed, our entire treatment of the axiverse effective theory is not valid, since the K\"{a}hler metric is dynamical. For our simple treatment to hold, we must consider axions still in slow-roll after the lightest modulus field $X_0$ has decayed: $m_a < \Gamma_{X_0} = \mathcal{O}(1)\sfrac{m^3_{X_0}}{M^2_{pl}}$. For $m_{X_0}\approx 30\text{ TeV}$, so as to avoid the cosmological moduli problem~\cite{1983PhLB..131...59C,Banks:1993en,deCarlos:1993wie}, we can extract a slightly higher maximum value for the axion mass we could consider: 
\begin{equation}
m_a < 1 \times 10^{-15} \text{ eV}\, .	
\label{eqn:modulus_mass}
\end{equation}
Axions violating these bounds must be removed from the spectrum for our treatment to be consistent. Numerically, this is simple to achieve: we locate axions in the spectrum violating the bound, and set the mass to zero. Our initial conditions then ensure that the realignment energy density in these fields remains zero.

A simple way to achieve this is to assume a large amount of entropy production and/or short period of inflation caused by the modulus-dominated epoch prior to BBN. This will dilute the population of heavy axions that begun oscillations prior to BBN. Such a scenario is relatively natural in the context of a string/M-theory cosmology with many moduli \cite{Lazarides:1990xp,Fox:2004kb,Kaplan:2006vm,Acharya:2010zx}.

A second possibility is that these heavier axions themselves decay rapidly prior to BBN, and simply contribute to setting the correct radiation content and baryon density. Theoretically, such axion decays are more problematic. Axion decays through the canonical two photon coupling are comparatively slow (see e.g. Ref.~\cite{2015PhRvD..92b3010M}), and decays before BBN require $m_a\gtrsim 1\text{ keV}$. For axions respecting our bounds to decay, one would require much larger than expected couplings and rapid decay channels. 

Alternatively, we could assume a gapped spectrum with any axions violating our bounds taken to have their masses lifted to a much higher scale to allow decays through standard channels. Another mechanism to remove heavy axions is via the multi-instanton potential, $U(\theta)\propto (1-\cos\theta)^3$, of Ref.~\cite{2014PhRvL.113y1302K}, which causes the misalignment population to redshift faster than $a^{-3}$ due to the non-quadratic potential minimum. Whether or not the appearance of such a multi-instanton potential occurs naturally in string/M-theory models is not clear.

All the above options (removing heavy axions from the spectrum) are covered by the simple command \texttt{remove\_masses=True} in \textsc{AxionNet}. We also allow for the option to reject outright (set zero likelihood) all models with large masses violating our bounds. This is controlled by a setting inside the likelihood function in \textsc{AxionNet} and is operative when \texttt{remove\_masses=False}.

Despite the construction of the mass matrix guaranteeing positive semi-definiteness mathematically, and thus mass eigenvalues $m^2_a\geq 0$, the huge spread in the elements of the mass matrix in the M-theory model leads to numerical precision errors and the existence of ``tachyonic'' $m^2_a<0$ eigenvalues. We have not been able to overcome this issue of numerical precision within the confines of \textsc{numpy}. We remove these tachyonic states from the spectrum just as we remove the heavy states, and they do not contribute to the energy density. Fortunately, the negative eigenvalues are guaranteed to be those for which the true values are smallest in absolute value. Since the true eigenvalue is $m_a\ll H_0$ and the field displacements $\phi_i^{\rm ini}\sim\mathcal{O}(M_{pl})$, even with the correct (positive) eigenvalue these states would not contribute significantly to the spectrum, and so removing them does not affect the results. 

Options for alternative thermal histories and evolution of $g_\star$ in \textsc{AxionNet} will be the subject of future developments. The two mass limits, in Eq.~(\ref{eqn:strict_mass}) and Eq.~(\ref{eqn:modulus_mass}), are both far exceeding axion masses probed by our simple DM constraints and thus the model of the Universe used above a few keV, the treatment of heavy axions, and use of constant $g_\star$ does not affect our results. These effects will be important for treatments going beyond considerations of the simple quasi-observables.

\subsection{Axion Oscillations}
\label{sec:axosc}
    
As the Universe expands $H$ decreases monotonically. When any individual field satisfies the condition $m_a \gtrsim H$, the field begins to roll towards its potential minimum and then begins coherent oscillations about it. The solution is given by:
\beq
\phi(a>a_{\rm osc})= \phi(a_{\rm osc})\left( \frac{a}{a_{\rm osc}}\right)^{\sfrac{-3}{2}} \cos (m_a t) \, ,
\eeq
where $a_{\rm osc}$ occurs at approximately $H(a_{\rm osc})\approx m_a$ (we define it more precisely shortly). As $H$ further decreases, the time scale of the oscillation induces a very small time step in the integrator of order $(\sim m_a^{-1})$ (much smaller than the dynamical time, $t_{\rm dyn}\approx H$. This is computationally prohibitive to integrate directly given the hierarchical nature of the axion mass distribution. 

Although the axion field oscillates, the energy density does not, and obeys a simple scaling:
\beq
\rho_a (a>a_{\rm osc}) =\rho(a_{\rm osc})\left(\frac{a_{\rm osc}}{a}\right)^3 \, .
\eeq
It is the well-known fact that fields oscillating in a quadratic potential will behave as non-relativistic matter (e.g. Ref.~\cite{1983PhRvD..28.1243T}). The pressure oscillates with a frequency $P\sim \cos (2m_a t)$, leading to a time-averaged equation of state $\langle w_a\rangle =0$, and can be safely neglected for our purposes.\footnote{For a selection of interesting astrophysical consequences of the pressure term, see Refs.~\cite{2014JCAP...02..019K,Porayko:2014rfa,Aoki:2016kwl,Blas:2016ddr}.}

The dynamical time scale in our integration is fixed to be of order the Hubble scale today, $M_H$. In order to be able to integrate models with $m_a\gg M_H$, we must approximate the axion evolution for time scales $t>t_{\rm osc}$. The method we choose is simply to set $w_a(t>t_{\rm osc})=0$ such that the energy density in heavy axions evolves exactly as $a^{-3}$ at late times. An alternative method uses a change of co-ordinates in the axion phase space, as implemented in Ref.~\cite{Urena-Lopez:2015gur}.

We define $t_{\rm osc}$ by allowing the equation of state in the full solution to oscillate (cross zero) a fixed number of times denoted by the parameter, $n_{\rm cross}$. We then define $t_{\rm osc}$ using $n_{\rm cross}$. This is an accuracy parameter in our numerical results, with larger values of $n_{\rm cross}$ leading to more accurate, but considerably slower, numerical computations. We find that results for the quasi-observables converge above $n_{\rm cross}=3$, and we use $n_{\rm cross}=5$ in the examples and constraints in the text. Care must be taken, however, as using a too large value of $n_{\rm cross}$, while improving the numerical integration accuracy, incorrectly assigns DM axions to the DE density in the quasi-observables (see below).

\subsection{Computing the Quasi-Observables}
\label{sec:comquasi}
Our quasi-observables are $(\Omega_m,z_{\rm eq},\ddot{a}, h)$. We compute in physical time, $t$, up to some maximum time $t_{\rm f}\approx \mathcal{O}(10)$ and output a fixed number of log-spaced time steps. We begin by locating $z=0$ in the output variables. If $z=0$ has not been reached in ten Hubble times (which may occur for extreme cosmologies) \textsc{AxionNet} outputs default quasi-observables which lead to very low likelihood (in particular, failing the acceleration cut). This is equivalent to a cut on the age of the Universe.

Having located $z=0$, computing $h$ is trivial as it is given by the Friedmann constraint evaluated at $z=0$. Computing the other variables relies on the separation of axions into DM and DE-like based on $n_{\rm cross}$.  The split at $z=0$ trivially gives the matter density: $\Omega_m=\Omega_b+\Omega_{\rm DM}$. The acceleration is computed from the total pressure and density as:
\beq
\ddot{a}=-\frac{a}{3}\sum_i (\rho_i+3P_i)\, ,
\eeq
where the index $i$ runs over axions and the ordinary cosmological components. The pressure for the axions with a number of crossings less than $n_{\rm cross}$ is computed directly from the fields using Eq.~(\eqref{eqn:axion_pressure}), while for those with crossings greater than $n_{\rm cross}$ we set $P_i=0$.

Finally we compute $z_{\rm eq}$. At all values of $z$ the axions are split into the energy density components, $\rho_{\rm DM}$ and $\rho_{\rm DE}$ by selecting those that have and have not passed the $n_{\rm cross}$ criterion. We are also in possession of the radiation energy density $\rho_r(z)$ and baryon energy density $\rho_b(z)$ at every value of the redshift. We locate $z_{\rm eq}$ by simply finding numerically the point where $\rho_{\rm DM}+\rho_b=\rho_r$. We do not include $\rho_{\rm DE}$ in the definition of equality. We also find equality using the list of output times, and not using interpolation. Therefore, the location of equality will depend on the number of output times used. In our numerical examples we use 1000 log-spaced times steps between $t_{\rm ini}=8\times 10^{-15}$ and $t_{\rm f}$.

\section{Connection to String Theory and M-theory}
\label{appendix:stringmot}
\subsection{The Superpotential in M-Theory}
\label{sec:mthethe2}
Axions generically arise in string compactifications as Kaluza-Klein modes of antisymmetric tensor fields which are present in all low energy string/M-Theory frameworks. The topology of such generic theories which can manifest realistic models in high-energy physics is complex, containing many cycles which in turn generate a `\emph{landscape}' of fields. This landscape provides a source to many axion-like fields which could, in the context of cosmology, potentially be of great interest given the hierarchical nature of their their associated physical parameter scales. The shift symmetries coming from the higher-dimensional gauge invariance of antisymmetric tensors leave the resulting scalar fields from string compactifications massless to any perturbative order. There are always plenty of instanton configurations arising in string theory models such as worldsheet, gauge, gravitational or membrane instantons that violate the shift symmetries. 

In the framework of four-dimentional supergravity, the superpotential is a holomophic function of the scalar part of the moduli superfield $z_i = t_i +is_i$ where $t_i$ denote the axion fields and $s_i$ denote the geometric moduli fields. We consider the following general form of the superpotential generated from non-perturbative effects,
\beq
W_{\text{inst}} = \sum_{i=1}^{N} \widetilde{\Lambda}_i^3 e^{i b_i F_i}\,,
\eeq
where $\widetilde{\Lambda}_i$ are the mass scales associated to each of the non-perturbative effects. $F_i$ represents the gauge kinetic functions which are linear combinations of the moduli superfields, 
\beq
F_i = \sum_k^{n_{\rm ax}} N_i^k z_k = \sum_k^{n_{\rm ax}} N_i^k (t_k + is_k)\,.
\eeq
The non-perturbative effects are assumed to be membrane instantons such that $b_{i} = 2\pi I_i$ where $I_i$ are positive integers. In general, the number of non-perturbative effects such as string/membrane instantons present in any compactification is larger than the number of axions which, in turn, allows for the possibility to stabilise the axion/moduli potential. Therefore, we will assume that the number of independent terms in the superpotential is always greater or equal to the number of axions, $N > n_{\rm ax}$. The supergravity potential is calculated using, 
\beq
V = e^{\mathcal{K}} \left(\mathcal{K}^{i\overline{j}}{DW\over Dz^i}{D\overline{W}\over D\overline{z}^j} - 3|W|^2 \right)\,, \label{sugrapot}
\eeq
where $\mathcal{K}$ is the K\"{a}hler potential and $\mathcal{K}^{i\overline{j}}$ is the inverse of the K\"{a}hler metric $\mathcal{K}_{ij} \equiv \frac{\partial^2 K}{\partial z_i \partial z_j}$. 

The periodic potentials arise from the interference of the instanton superpotential and the superpotential from other SUSY breaking sources, $W_{0}$. Assuming that the SUSY breaking scale is, 
\begin{equation}
F \sim \frac{DW_{0}}{Dz_i}\,,	
\end{equation}
this gives rise to the following form for the potential,
\begin{align}
V \approx & F\left( \sum_{i=1}^{n_{\rm ax}} \frac{\partial}{\partial z_i} \sum_{j=1}^{N} \widetilde{\Lambda}_j^3 e^{i b_j F_j}\right) + \text{c.c.} \nonumber\,, \\
\approx & \sum_{i=1}^{n_{\rm ax}} \sum_{j=1}^N \frac{2 F \widetilde{\Lambda}_j^3 b_j N_j^i}{M_{S}} e^{- b_j \sum_k^{n_{\rm ax}} N_j^k s_k} \cos{\left(\sum_{k=1}^{n_{\rm ax}} b_j N_j^k t_k \right)}\,,
\label{eq:mtheorysuper}
\end{align}
where $M_S$ is the string scale.

\subsection{The Superpotential in Type-IIB String Theory}
\label{appendix:nflation}

In this section we review the original arguments of Ref.~\cite{Easther:2005zr}, which provide a context for the Mar\v{c}enko-Pastur models in Kachru-Kallosh-Linde-Trivedi (KKLT)~\cite{2003PhRvD..68d6005K} comptactifications of Type-IIB string theory. In this set up, we find motivations for the relationship between the parameter, $\beta_{\mathcal{M}}$ and the ratio of axions to moduli, as well as highlighting the potential power random matrix theory might have in physical models.

N-flation models are proposed in order to solve the issue regarding the requirement of trans-Plackian displacements of inflatons. Given their symmetry properties, axions could potentially provide a very good candidate in these models. The original model for N-flation consisted of $n_{\rm ax} \gg 1$ decoupled axion fields each with identical masses that served to drive a period of inflation through the assisted inflation mechanism~\cite{1998PhRvD..58f1301L,Dimopoulos:2005ac}. The fields have periodic potentials as expressed in Eq.~(\ref{eq:b1}) where the scales $\Lambda_{a,i}$ can be significantly lower that the UV cutoff scale of the theory due to dimensional transmutation. The fields with identical masses undergo a common initial displacement $\phi'$ as they continue to roll in unison, providing an effective single field displacement of the order $\sqrt{n_{\rm ax}}\phi'$. 

Further expanding on these concepts Easter and McAllister incorporated the mathematics of random matrix theory in a more general framework in which the axion masses come from a distinct spectra in Ref.~\cite{Easther:2005zr}. In their framework the form of the matrix used to determine a spectrum of axion masses is only dependant on the basic structure the matrix possesses, which can be extracted by the supergravity potential, 
\beq
V=e^{k}\left(K^{AB}D_A W D_{\bar{B}}\bar{W}-3 |W|^2\right)\,.
\eeq
 The KKLT superpotential from nonperturbative effects which are generated from the associated moduli and axions is given as,
\beq
W_i = \widetilde{\Lambda}_i e^{-2\pi \rho_i} e^{2\pi i \phi_i} \equiv C_i e^{2\pi i \phi_i}\,,
\eeq
where $C_i$ are constants when the moduli are fixed at their minimum. A Taylor expansion about the origin at $\phi_i = 0$  along with the F-flatness conditions $D_A W |_{\phi_i = 0} = 0$, finds the mass matrix from quadratic order terms in the axion fields,
\beq
\mathcal{M}_{ij} = (2\pi)^2 e^K \left( K^{AB} D_A C_i D_B C_j - 3 C_i C_j\right)\,,
\eeq
where
\beq
V = \mathcal{M}_{ij}\phi^i \phi^j\,.
\eeq
Note that $i, j = 1, \ldots, N$ run over the K\"{a}hler moduli, where the terms $A, B = 1, \ldots , N+L$ run over the dilaton, complex moduli and K\"{a}hler moduli. After the kinetic terms are bought into their canonical form (see Section \ref{sec:effmodel}), the mass matrix can be expressed as,
\beq
\widetilde{\mathcal{M}}_{ij} = (2\pi)^2 \frac{e^K}{f_i f_j} U_i^k\left( K^{AB} D_A C_k D_B C_l - 3 C_i C_j\right)U_j^l\,.
\label{eq:complexm}
\eeq

Despite the complex form of $\mathcal{M}_{ij}$ in Eq.~(\ref{eq:complexm}) it can be shown that the characteristics of the N-flation model can be extracted from the eigenvalues of a random matrix with independent and identically distributed (i.i.d) entries. Numerically and semi-analytically it was shown that regardless of the input distributions for $K$, $f_i$, $U_i^k$, $C_i$, $D_A C_i$, and $K^{AB}$, the complicated structural form of the mass matrix above can be simplified by assuming that the leading contribution to $\mathcal{M}_{ij}$ takes the following form:
\begin{equation}
\mathcal{M}_{ij} = B_{iA}B_{Aj}\,, 
\label{eq:simplem} 
\end{equation}
where $B_{iA}$ is defined as,
\beq
B_{iA} = 2\pi \frac{e^{\sfrac{k}{2}}}{f_i}U^K_i \mathcal{Z}_{Ak}\,,
\label{eq:simplemcomp}
\eeq
with $\mathcal{Z}_{Ak}$ a matrix constructed of K\"{a}hler covariant derivatives. The approximation made in Eq.~(\ref{eq:simplem}) is subject to the arguments that the matrix $B_{iA}$ should be a $N \times (N+L)$ dimensional matrix constructed from i.i.d variables with zero mean and variance $\sigma^2$. The spectral properties of a matrix of this form are well known from the Mar\v{c}enko-Pastur limiting law in random matrix theory.
\begin{figure*}[t]
\centering
\begin{tabular}{cc}
    \includegraphics[width=0.5\textwidth]{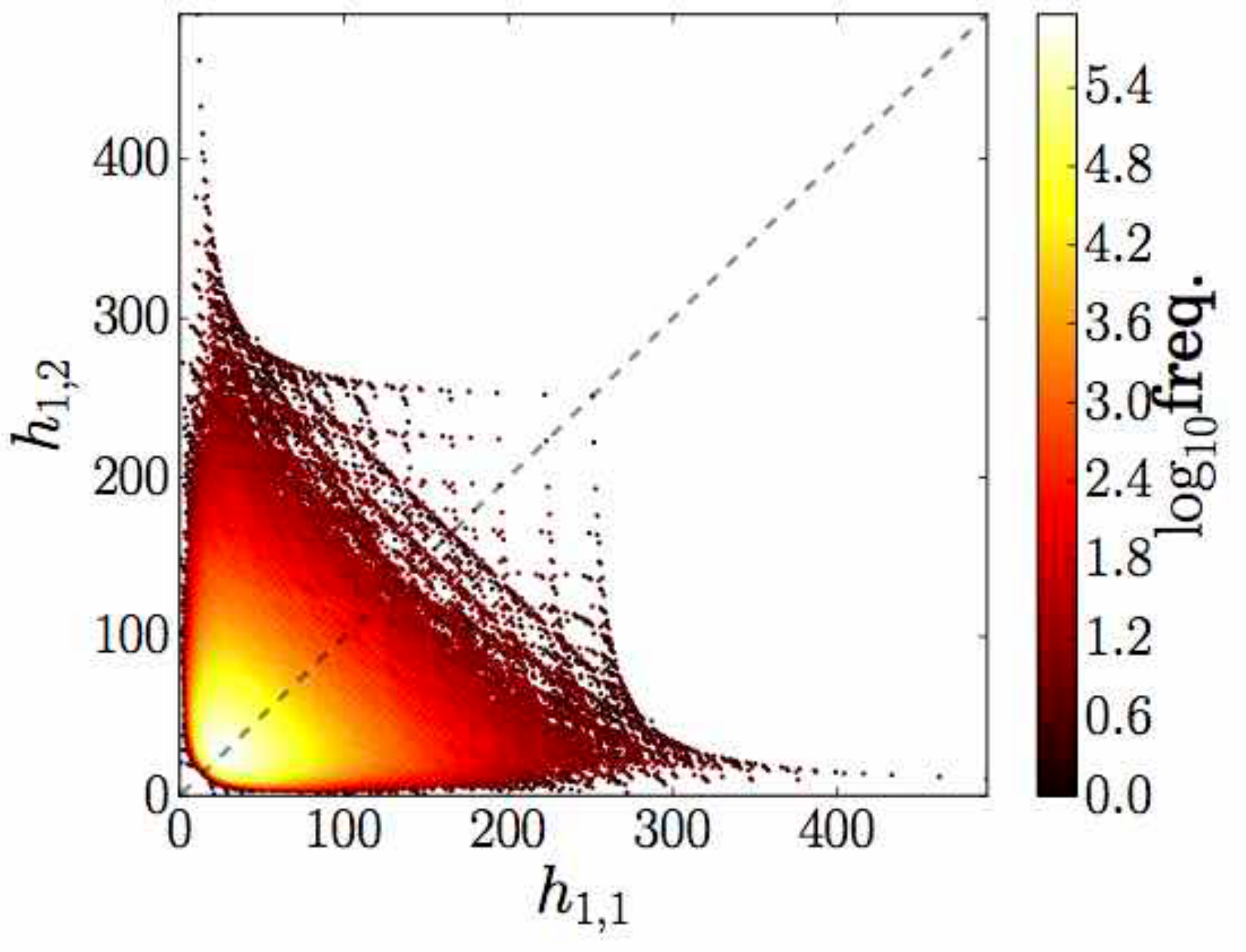}&
    \includegraphics[width=0.5\textwidth]{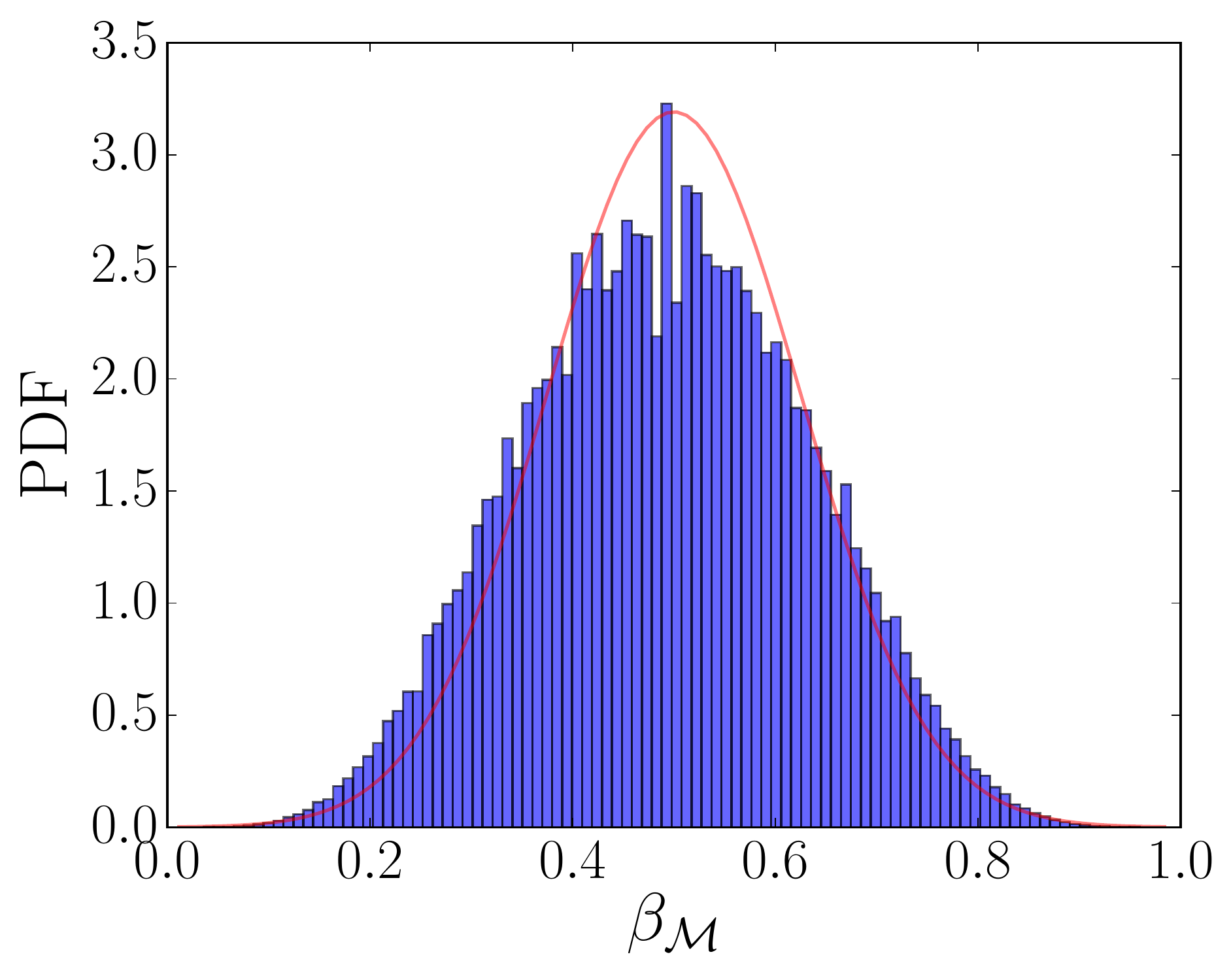}\\
\end{tabular}
\caption{{\bf Database of String Compactifications and associated ratio of Axions to Moduli:} {\emph{Left Panel:}}  Probability density of the hodge numbers $h_{11}$ (K\"{a}hler moduli) and $h_{12}$ (complex structure moduli) on Calabi-Yau manifolds from the construction of Ref.~\cite{Kreuzer:2000xy}. \emph{Right Panel:} Probability density of $\beta_\mathcal{M}$ on Calabi-Yau manifolds. This is reasonably well fit by a Gaussian distribution with mean $\bar{\beta}_\mathcal{M}=0.5$ and standard deviation $\sigma_\beta=0.125$ (solid line). (Since the distribution has exactly zero probability density at the boundaries, a Gaussian fit cannot be perfect.)}
\label{CY-plot}
\end{figure*}
 \subsection{Axions and Moduli in String Theory}
\label{sec:axmod}

In the context of string theory the parameter $\beta_{\mathcal{K},\mathcal{M}}$ in our study can be related to the relative number of axions to moduli appearing in the axion mass and kinetic matrices. In M-theory, $\beta_{\mathcal{K}}=1$ while, as discussed in the main text, $0<\beta_{\mathcal{M}}\leq 1$ is specified by the number of instantons. In (weakly coupled) Type-IIB string theories compactified on Calabi-Yau manifolds, $\beta_{\mathcal{K}}=1$ while $\beta_{\mathcal{M}}$ is specified by the ratio of the number of axions (from the K\"{a}hler moduli) to total number of moduli (K\"{a}hler plus complex structure plus axio-dilaton). 

The value $\beta_{\mathcal{K}}=1$ in Type-IIB comes from the large volume, tree level result for the K\"{a}hler potential, which is sum separable for the K\"{a}hler and complex structure moduli~(e.g. Eq.(10.104) in Ref.~\cite{2007stmt.book.....B}). Values of $\beta_{\mathcal{K}}\neq 1$ can arise when a mixing between the K\"{a}hler and complex structure moduli occurs. For example, in Ref.~\cite{Grana:2003ek} they introduce matter fields from D-branes leading to the non-trivial mixing of all the moduli fields. Such mixing can also come from quantum corrections in $\alpha'$ or $g_s$, and from non-Calabi-Yau compactification considerations.

The number of K\"{a}hler and complex structure moduli coming from Calabi-Yau threefolds are topologicaly invariant and are given by the hodge numbers $h_{1,1}$ (K\"{a}hler moduli) and $h_{1,2}$ (complex structure moduli) respectively, which can be used to define the value of $\beta_{\mathcal{M}}$,
\begin{equation}
\beta_{\mathcal{M}} = \frac{h_{1,1}}{(h_{1,1}+h_{1,2}+1)}\,.
\label{eq:hodgebeta}
\end{equation}

Following from a complete construction of reflexive polyhedra from Kreuzer and Skarke \cite{Kreuzer:2000xy}, the topological and geometrical information of the extra dimensions can be extracted~(e.g. Ref.~\cite{Altman:2014bfa}). The data on the Hodge numbers from the Kreuzer-Skarke database is shown in the left-hand panel of Fig.~\ref{CY-plot}. In the right-hand panel of Fig.~\ref{CY-plot} we show the probability density of $\beta_{\mathcal{M}}$ defined in Eq.~(\ref{eq:hodgebeta}) in KKLT compactifications. 

In Type-IIB string theory on a Calabi-Yau manifold, topologies with $\beta_{\mathcal{M}}$ close to zero or unity (Hodge numbers of zero or going to infinity) will be rare. Distributions with values close to $\beta_{\mathcal{M}} = 0.5$ are more expected in fitting with the Kreuzer-Skarke data base and mirror symmetry. In the M-theory limit, however, one has exactly $\beta_{\mathcal{M}}=1$, i.e. equal numbers of axions and moduli. A final point to note about the low energy theory is that Type-IIB requires orientifold projection in order to obtain chiral fermions leading to the Betti number relation for axions versus moduli. However some axions are also ``eaten'' by gauge bosons from Green-Schwarz anomaly cancellation, which alters the number of light axions~\cite{Grimm:2004uq}.

\section{Random Matrix Theory}
\label{appendix:rmt}

\subsection{Matrix Ensembles}
\label{appendix:maten} 
The study of the statistical properties of the spectral behaviour for sample covariance matrices in models involving high dimensional data structures has seen prolific advancements in both their theoretical and practical applications. The most well known random matrix ensembles consistent with modelling physical systems are the \emph{Wigner-Dyson} or so called \emph{beta} ensembles of hermitian matrices classified by the \emph{three-fold} way \cite{Stephanov:2005ff,2011arXiv1104.2272B} with elements distributed according to the probability function, 

\begin{equation}
P(X)DX = \mathcal{Z}e^{-\frac{N\beta}{4}Tr X^{\dagger}X}DX\,,	
\end{equation}
where $\mathcal{Z}$ is a distribution normalisation constant and $DX$ is the Haar measure. The parameter $\beta$ is the Dyson index determined by the symmetry group of the matrix with classical values $\beta \in \{1,2,4\}$ \footnote{The $\beta$ parameterisation should not be confused with the sub-matrix dimension parameters for $\mathcal{K}_{ij}$ and $\mathcal{M}_{ij}$ denoted as $\beta_{\mathcal{K}}$ and $\beta_{\mathcal{M}}$ used throughout this paper.} each with well defined eigenvalue distributions for full matrix wishart ensembles which give unitary, orthogonal or symplectic transformation invariance in matrix space \cite{james1960,RATNARAJAH2005399,2009arXiv0901.3379L}. These special orders of $\beta$ defining the three division algebras over the real numbers correspond to correlation functions which can be explicitly expressed in terms of polynomials orthogonal to the associated invariant measure.   

We are primarily interested in the class of symmetric, ($A \rightarrow O^{T} AO)$, positive-definite, real matrices with orthogonal invariance residing in the Wishart ensemble. The eigenvalue spectral distribution and limit properties of Wishart matrices play an important part in many aspects of multivariate analysis \cite{rmtmulti}. We begin with the basic matrix property that it is always possible to take any ensemble of non-hermitian matrices and construct a random matrix in the Wishart form,

\begin{align}
X_{ij} = H_{ih}^T H_{hj}\,,
\label{eq:generalmatrix}
\end{align}
where $H_{hj}$ is a $(n \times p)$ dimensional rectangular matrix with a shaping index, $\beta = \sfrac{n}{p}$ \footnote {see Appendix~\ref{sec:betahodge} for physical motivations in the context of string theory.}.  $X_{ij}$ is defined as a positive-definite Wishart matrix in the class $W_{\mathbb{R}}(n,\Sigma_{p})$ with n degrees of freedom and  population covariance matrix, $\Sigma_{p}$. Its eigenvalues, which are our derived physical quantities are the positive real values, $\lambda_i = m_i^2$ or $\lambda_i = f_i^2$ where $\{\lambda_i\in \mathbb{R}|\lambda_i >0\}$. When $\Sigma_{p} = \mathbb{1}$ this is referred to as the ``null'' case corresponding to the class of \emph{white}-Wishart matrices or the $\beta =1$ Laguerre ensemble. The limiting normalised eigenvalue spectral density function, $P(x)$, of a \emph{white} wishart matrix is given by the Mar\v{c}enko-Pastur distribution (see Appendix~\ref{sec:mplaw}). It has been shown in the limit that $\Sigma_{p} = \mathbb{1}$ the ensembles will reproduce the Mar\v{c}enko-Pastur distribution with a total invariance over $\beta=\{1,2,4\}$ \cite{MIT-2003}. The asymptotic distribution for the rescaled largest eigenvalues of a \emph{white} Wishart covariance matrix is determined by the \emph{Tracy-Widom} distribution \cite{johnstone2001}. Finite dimensional analysis of real Wishart matrices whereby the properties of the largest eigenvalues can be extracted, incorporate determining the hypergeometric functions of matrix arguments \cite{2014arXiv1401.3987C,2012arXiv1209.3394C}.

\subsection{The Mar\v{c}enko-Pastur Law}
\label{sec:mplaw}

The Mar\v{c}enko-Pastur Law, is but one of several limiting laws in random matrix theory used to described the asymptotic behaviour of empirical measures of sample covariance matrices \cite{0025-5734-1-4-A01}. Matrices of this form can find a purpose in many areas of physics and have recently found traction in the context of string theory models. We provide a brief review of the properties of the Mar\v{c}enko-Pastur Law  continuing the arguments made in \cite{Easther:2005zr} as discussed in Appedix~\ref{appendix:nflation} in the context of axion mass spectra arising in KKLT compactifications in Type-IIB string theory. 

The elements of $B_{iA}$ in Eq.~(\ref{eq:simplem}) are drawn from a standard Gaussian statistical distribution such that the axion mass matrix in the canonical basis is defined in the class of Wishart matrices. Given the mass matrix is sufficiently large the eigenvalue spectrum of $m^2_a$ values is governed by the Mar\v{c}enko-Pastur distribution law parametrised by the two quantities, $\beta_{\mathcal{M}} = \sfrac{n_{\rm ax}}{p}$ and $\sigma^2_{\mathcal{M}}$. The closed form density expression for the Mar\v{c}enko-Pastur distribution who's shape is encoded by $\beta_{\mathcal{M}}$ is given as,
\begin{equation}
    p\left(m^2_a\right) = 
\begin{cases}
    \frac{1}{2 \pi m^2_a \beta_{\mathcal{M}} \sigma^2_{\mathcal{M}}}\sqrt{\left(\gamma_+ - m^2_a\right)\left(m^2_a - \gamma_-\right)},\\
    0,  
\end{cases} \,,
\label{eq:MPprobden}
\end{equation}
where $\gamma_+$ and $\gamma_-$ are defined as,
\begin{align}
\gamma_+ = \sigma^2_{\mathcal{M}} \left(1+\sqrt{\beta_{\mathcal{M}}}\right)^2\,,\\
\gamma_- = \sigma^2_{\mathcal{M}} \left(1-\sqrt{\beta_{\mathcal{M}}}\right)^2\,.	
\end{align}
\label{sec:betahodge}

The density function in Eq.~(\ref{eq:MPprobden}) is defined on the compact interval $[\gamma_-,\gamma_+]$ such that $\gamma_-\geq m^2_a \geq \gamma_+$ where the probability density drops out to zero outside this region. The eigenvalues will surely converge to the compact interval bounds in the asymptotic limit. The rate of convergence for the real case was found in the work by Johnstone \cite{johnstone2001} where he found approximations satisfactory up to dimensions as low as $n,p \approx 10$. We will therefore treat our distributions as ``safe'' and within the asymptotic understanding of their spectral convergence when using axion population numbers, $n_{\rm ax} \geq \mathcal{O}(10)$.  
 
It follows that the overall scale of eigenvalues is controlled by the  variance, $\sigma^2_{\mathcal{M}}$ where,
\beq
\langle m^2_{a} \rangle = \sigma^2_{\mathcal{M}}\,.
\eeq
In the original axiverse models for N-flation, the overall scale for the axion masses are fixed for inflationary concerns, i.e., $\sigma \sim 10^{-5}M_{Pl}$ in order to enforce that density perturbations from inflation are consistent with observational limits. However, such constraints are of no interest for cosmological concerns for axions in the dark sector.

\subsection{Eigenvalue Spectra: Non-Universality and Free Multiplicative Convolution}
\label{sec:eigspect}

\subsubsection{Generalised Wishart Matrices}
We have currently only considered the construction of $\mathcal{K}_{ij}$ and $\mathcal{M}_{ij}$ in our effective model residing in Eq.~(\ref{eq:effL}) involving matrix products to first order with no level of decomposition or  considerations of the free convolution of matrix ensembles.~\footnote{See Ref.~\cite{Marsh:2011aa} for the potential uses of additive free convolutions of matrix ensembles in the context of random Hessian construction in supergravity.} This study is only concerned with a focus on the products of fixed ensembles with the entries constructed from some predefined statistical distribution, however there are several areas which could be of interesting for further study in this regard. More general considerations of the the construction of Wishart matrices such as those appearing in our MP RMT and WW RMT models involve the product of random independent gaussian matrices. See Refs.~\cite{PhysRevE.92.012121,2015JSP...158.1051F,PhysRevE.83.061118,
2012arXiv1211.7259M,2011PhRvE..83f1118P} for detailed work regarding this subject.

Following the approach in this work, a generalised construction of random matrices residing in the Wishart ensemble involves the product of $\mathcal{S}$ non-hermitian sub-matrices residing in the Ginibre ensaumble, $X_{ij} = H^{(1)}_{ih}H_{hl}^{(2)}H^{(\mathcal{S})}_{lj}$. The study of the singular values of these products corresponding to the root of the associated eigenvalues is of interest in generalisations of these random matrix ensembles. The spectral density functions, $P_{\mathcal{S}}(x)$ (where $\mathcal{S} \in \mathbb{N}=\{1,2,3,...\}$) for these ensembles involving the product of an arbitrary number of matrices are asymptotically described by the \emph{Fuss-Catalan} distributions with their moments defined by the \emph{Fuss-Catalan} numbers \cite{2010arXiv1012.2743A}. These distributions can be expressed as the multiplicative free convolution of the Mar\v{c}enko-Pastur spectral density limit, of the order $\mathcal{S}$ such that,
\begin{equation}
 P_{\mathcal{S}}(x) = [P_1 (x)^{\boxtimes \mathcal{S}}]\,.	
\end{equation}

 A powerful two-dimensional parameterisation of the \emph{Fuss-Catalan} numbers comes in the form of the \emph{Raney} sequences. An explicit density $W_{p,r}(x)$ characterised by the indices $p,r \in \mathbb{R}$ defines a family of measures incorporating the multiplicative free measures of the Mar\v{c}enko-Pastur distribution reproducing the both the \emph{Fuss-Catalan} densities and Wigner semi-circle distribution for specific values of $r$ and $p$. 

\subsubsection{Spiked Wishart Matrices}

Recent work involving so called \emph{non-white} Wishart matrices or spiked population models has yielded an interesting insight into the effects of a phase-transition phenomena \cite{2004math......3022B} in the fluctuations of the largest eigenvalues of the population covariance matrix \cite{2007arXiv0711.2722W,2008PhDT.......217W,2010arXiv1011.5404M,2011arXiv1109.3704B,2012arXiv1209.3394C}. These models can make predictions beyond the traditional ensembles found in the literature. The presence of large eigenvalues in the population covariance matrix can have a significant effect of the total spectral width and limiting distributions of the sample covariance matrix in the limit $n,p \rightarrow \infty$ and have been incorporated into many interesting areas such as finance \cite{2009arXiv0910.1205B}. 

Fig.~\ref{spikedis} shows the eigenvalue spectrum for a mass matrix displaying the features of these models. Singular eigenvalues in these models will leave the support of the Mar\v{c}enko-Pastur bulk with a value $\sim \mathcal{O}(N)$ for $(N\times N)$ dimensional data structures. The determination of the true values of the largest eigenvalues in these models can be analysed using various methods such as the stochastic operator method \cite{2010arXiv1011.1877B} or using the Painleve formula \cite{2011arXiv1101.5144M}. In general the effects in these models will be most prevelant when considering high dimensional data structure or in our case high a population number of axions.

\section{Outlying Cosmologies}

\label{sec:rejec}
 
\begin{figure}
    \centering
    \includegraphics[width=0.48\textwidth]{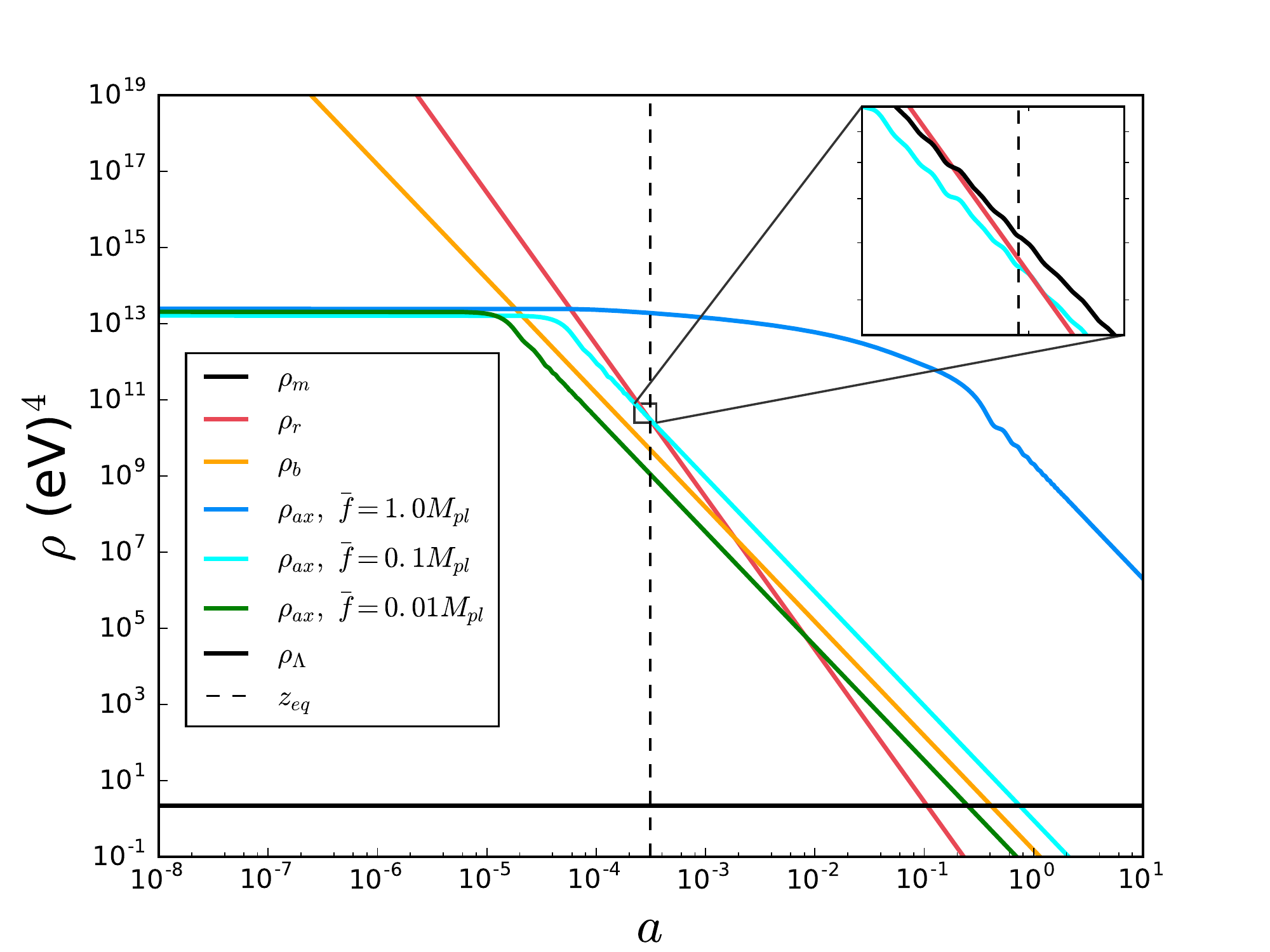}
    \caption[]%
    {{\bf MP-DM example outlier cosmology density evolution.} Evolution of the collective axion density, $\rho_{\rm ax}$ using $n_{\rm ax}$ = 20. We highlight the effect of using different initial field condition scales set by $\bar{f}$ where values of $\bar{f} \rightarrow 1$ returning cosmologies which don't fulfil the criterion for acceptable values of $z_{\rm eq}$.}    
    \label{nfdmexcosmo1}     
\end{figure}

 In this section we provide a picture of the evolution of the cosmological densities in the context of example cosmologies which would not pass the cuts outlined in Section~\ref{sec:results2}. In Fig.~\ref{nfdmexcosmo1} we show the cosmological evolution for three example configurations using the MP RMT model for a population of axions behaving as dark matter. We allow the equal field condition scaling parameter $\bar{f}$ to approach the high scale limit, $\bar{f} \rightarrow M_{pl}$ (\emph{blue} line). The large value for $\bar{f}$ causes the population of axions to collectively ``inflate'' the Universe for a period ($10^{-4} \lesssim a \lesssim 10^{-1}$) with the collective energy density overshooting the expected value of $z_{\rm eq}$ before it has entered the scaling regime behaving as non-relativistic matter. The evolution of the collective axion field density as dark matter begins to scale accordingly at an approximate time of $z \approx 0$ with a value of $z_{\rm eq}$ far too early in the cosmic history. Such cosmologies return axion DM domination with $\Omega_{\rm DM} \approx 0.9999$. 
 
Decreasing the scale of $\bar{f}$ to $0.1M_{pl}$ (\emph{cyan} line) causes the axions to account for the correct total dark matter density at the current time where $\Omega_{\rm DM} = 0.2528$. The reduced initial field conditions cause the axions to enter the correct scaling regime with a significantly reduced redshift. The inset of Fig.~\ref{nfdmexcosmo1} shows the value of $z_{\rm eq}$ falling within acceptable bounds (crossing of black ($\rho_{b}+\rho_{\rm ax}$) and red ($\rho_{r}$) lines). Further decreasing $\bar{f} = 0.01 M_{pl}$ (\emph{green} line) corresponds to an example configuration in which the total matter density is insufficient for the Universe to reach redshift zero within ten Hubble times according to our numerical configurations. The lowest value of $z$ reached corresponded to an axion dark matter density parameter value of $\Omega_{\rm DM} = 0.0119$.
     \begin{figure}
    \centering
    \includegraphics[width=0.48\textwidth]{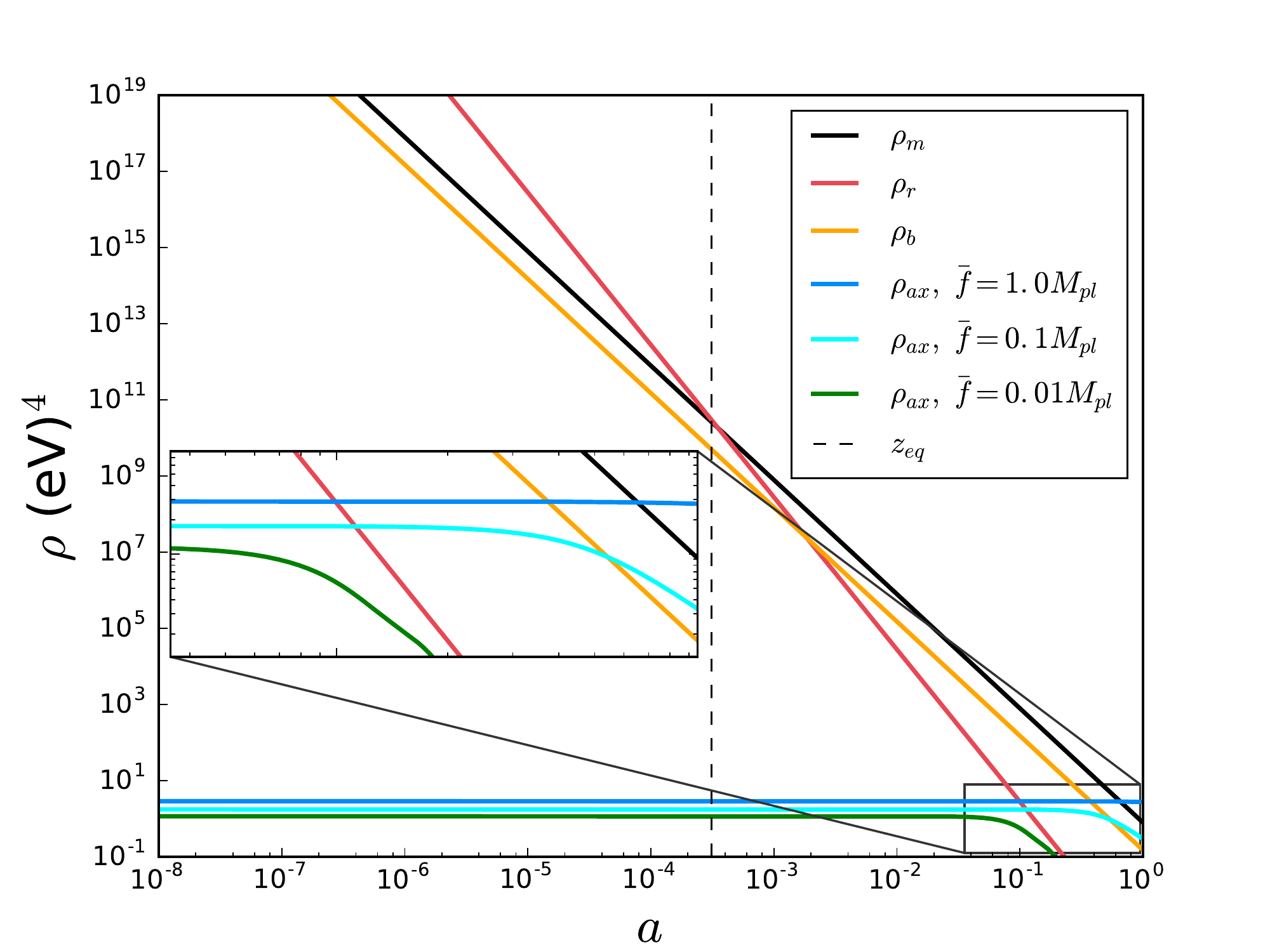}
    \caption[]%
    {{\bf MP-DE example outlier cosmology density evolution.} Evolution of the collective axion density, $\rho_{\rm ax}$ for $n_{\rm ax}$ =20 axion for example dark energy cosmologies in the MP RMT model. We highlight the effect of using different scales for $\bar{f}$ where insufficient values of $\bar{f}$ lead to cosmologies outlier by the acceleration criterion, $\ddot{a}>0$.}    
    \label{nfdmexcosmo2}     
\end{figure} 
In Fig.~\ref{nfdmexcosmo2} we show potential configurations which do not pass the acceleration criterion, $\ddot{a}>0$ or give dominant contributions to the critical density at $z=0$ for MP-DE cosmologies. The axion density is set by the initial field displacement and axion mass, $m_a^2 \phi^2$. Without a sufficient scaling of the initial field displacements (\emph{light blue} and \emph{green} line), the axion masses need to be higher to account for the acceptable amount of dark energy density. However, this generally causes the axion to start oscillating earlier following the condition $m_a \leq H$, which returns smaller values of $\Omega_{\rm DE}$. Increasing the value of the scaling $\bar{f}$ in this configuration would satisfy an accelerating universe with sufficient dark energy density. The increased value of $\bar{f}=0.1M_{pl}$ enhances the final dark energy density at $z=0$ returning a value of $\Omega_{\rm DE} = 0.1979$. Finally the configuration (\emph{blue} line) with $\bar{f} = 1.0M_{pl}$ is sufficient for an effective dark energy cosmology returning a value of $\Omega_{\rm DE} = 0.7732$.

\bibliography{bib}
\bibliographystyle{h-physrev3.bst}


\end{document}